\documentclass[reprint,superscriptaddress,preprintnumbers,
amsmath,amssymb,aps,floatfix,pra,twocolumn,
tightlines,nofootinbib]{revtex4-2}
\usepackage{dcolumn}
\usepackage{bm}
\usepackage{amsmath,amssymb}
\usepackage{graphics}
\usepackage{amsfonts}
\usepackage{epsfig}

\usepackage{hyperref}
\usepackage{adjustbox}
\usepackage{rotating}

\usepackage{dblfloatfix}
\usepackage{blindtext}
\usepackage{multirow}
\usepackage{graphicx}
\usepackage{dcolumn}
\usepackage{makecell}
\usepackage{bbold}
\usepackage{bm}
\usepackage{amsmath}
\usepackage{import}
\usepackage{physics}
\usepackage{verbatim}
\usepackage{listings}
\usepackage{xcolor}
\usepackage{xr-hyper}
\usepackage{hyperref}
\usepackage{algorithm}
\usepackage{algpseudocode}
\usepackage{cleveref}
\usepackage{newfloat}
\usepackage[table]{xcolor}
\usepackage{caption}
\usepackage{subcaption}
\DeclareFloatingEnvironment[name={Fig S},fileext=lof]{suppfigure}
\begin{document}
\title{Fractional Topological Phases, Flat Bands, and Protected Zero-Energy Edge States in Small Cyclic Quantum Walks}

\author{Dinesh Kumar Panda$^{\divideontimes}$}
\email{dineshkumar.quantum@gmail.com}
\author{Aditi Rath$^{\divideontimes}$}
\email{aditi.rath@niser.ac.in}

\author{Colin Benjamin}
\email{colin.nano@gmail.com}
\affiliation{School of Physical Sciences, National Institute of Science Education and Research, Bhubaneswar, Jatni 752050, India}
\affiliation{Homi Bhabha National Institute, Training School Complex, Anushaktinagar, Mumbai 400094, India}

\begin{abstract}
\textcolor{blue}{} 
We report the first realization of a fractional topological phase in a fully unitary, noninteracting discrete-time quantum walk implemented on finite cyclic graphs. Using a single-coin split-step cyclic quantum walk (SCSS-CQW), we uncover topological phenomena that are inaccessible within conventional cyclic quantum-walk dynamics. The protocol enables controlled engineering of quasienergy spectra, flat bands, and topological phase transitions through the step-dependency parameter and coin-rotation angle. We show that cyclic graphs with even and odd numbers of sites exhibit qualitatively different band structures, while rotational flat bands arise exclusively in even cycles; a general analytic condition for their emergence is derived. The SCSS-CQW produces fractional winding numbers $\pm \frac{1}{2}$ (Zak phases $\pm \frac{\pi}{2}$), in sharp contrast with the integer invariants in standard cyclic quantum walks. These fractional invariants lead to an unconventional bulk-boundary correspondence and support protected zero energy edge states beyond the usual integer topological classification. In the step-dependent protocol, transitions between distinct fractional winding sectors generate robust edge modes pinned at zero quasienergy. Numerical simulations show that these states remain stable in the presence of both dynamic and static coin disorder as well as phase-preserving perturbations, while survival-probability analysis demonstrates their persistence over the simulated evolution window. Requiring only a constant number of detectors independent of the evolution time and fewer operators, the proposed scheme offers a minimal-resource and experimentally accessible platform for realizing fractional topology, flat bands, and protected zero energy edge states in small-scale synthetic quantum systems.
\end{abstract}
\maketitle
\begingroup
\renewcommand{\thefootnote}{$\divideontimes$}
\footnotetext{These authors contributed equally to this work.}
\endgroup
\textit{\textcolor{brown}{Introduction.--}}Topological phases, characterized by band invariants, protected edge states, and flat bands, constitute a central theme in condensed matter physics and play an important role in quantum memory and topological quantum computation (TQC)~\cite{pan16,pan23,pxue}. Since the discovery of the integer quantum Hall effect~\cite{hall1,hall2}, followed by the theoretical~\cite{kita21,kita22,kita23} and experimental~\cite{kita24,kita25} realization of topological insulators~\cite{kita25,kita26}, topology has been understood to arise from gapped band structures, where symmetry-protected edge states emerge at the interfaces between distinct phases through the bulk–boundary correspondence~\cite{panahiyan20,pan-2,pxue}. Band-gap closings may appear in different forms, including linear dispersions (Dirac cones), nonlinear structures (Fermi arcs), or dispersionless modes (flat bands). In particular, gapped flat bands can host strong correlation effects such as Mott phases and fractional quantum Hall states, while gapless flat bands are associated with quantum criticality and unconventional transport phenomena~\cite{fqh,correl,mott,semimetal,criticalsyst}. Experimental realizations of topological phases have been demonstrated in platforms such as photonic lattices~\cite{kita28} and ultracold atoms and molecules~\cite{kita42,kita40}. Despite their potential for robust quantum information storage and noise-resilient state transfer~\cite{qmemory,qmemory2,edge1,edge2,edge4}, naturally occurring topological materials are relatively rare and often constrained by symmetry requirements~\cite{Schnyder2008,Zhang2019,Hasan2010,Halperin1982,chandra-topo}. This limitation has motivated the exploration of synthetic platforms, including discrete-time quantum walks (QWs). Quantum walks describe the coherent lattice evolution of particles with internal degrees of freedom, governed by interference and entanglement~\cite{QW93,pxue,ina2023,joshua2025,Cinthia}. Implementations in one, two, and three dimensions, particularly photonic realizations, have enabled the simulation of a wide range of topological phenomena~\cite{kita-exp,pan16,filippo17,pan21,pan22,pan23,pxue,karimi19,panahiyan20,pan-2,kita-pra,asboth}, providing tunable control over symmetries and access to regimes that are difficult to realize in conventional condensed-matter systems.\\
Despite these advances, the realization of fractional topological invariants within finite, fully unitary quantum walk architectures remains largely unexplored. In particular, the simulation of topological phenomena using quantum walks on finite cyclic graphs, cyclic quantum walks (CQWs), has received little attention, with only two studies reported so far~\cite{p6,karimi19}. Due to their finite-dimensional Hilbert space, CQWs are particularly resource-efficient for photonic implementations, requiring only a small number of particle detectors and passive optical components (e.g., waveplates or Jones plates)~\cite{bian,karimi19,p2}, in contrast to conventional QWs defined on extended 1D–3D lattices. Ref.~\cite{p6} demonstrated integer winding numbers $\pm1$ together with gapped flat bands and topological edge states, whereas Ref.~\cite{karimi19} considered the Zak phase for a Hadamard-coin QW on a six-site cycle. Fractional winding numbers $\pm\frac{1}{2}$ (corresponding to Zak phases $\pm\frac{\pi}{2}$), previously reported mainly in interacting many-body systems~\cite{frac-interact}, scattering geometry~\cite{frac-asboth,frac2-asboth,frac3-asboth} or non-Hermitian settings~\cite{frac1,frac2,frac3}, encode exotic phase transitions and unconventional edge states beyond those associated with integer topological invariants~\cite{frac1,frac2,frac3}. These features indicate richer band topology and anomalous bulk-boundary correspondence~\cite{frac1,frac2,frac3}. Notably, the fractional invariants in Refs.~\cite{frac1,frac2,frac3} arise through non-Hermitian mechanisms while Refs.~\cite{frac-asboth,frac2-asboth,frac3-asboth} use a combination of a finite and two semi-infinite lattices in a scattering junction to generate the same. In contrast, here we generate fractional winding numbers within a fully unitary framework governed by a Hermitian Hamiltonian in finite, non-interacting systems (small cyclic graphs). Moreover, the emergence of topological flat bands and fractional winding numbers in cyclic graphs, as well as engineering robust zero energy edge states from them, has not been previously investigated. In this work we address this gap by analyzing the effects of dynamic and static disorder, as well as phase-preserving perturbations, modeling the decay of survival probabilities, and demonstrating the long-time stability of the resulting zero energy edge states. Our protocol further requires only a constant number of particle detectors, independent of the evolution time.
\begin{figure}[!]
\includegraphics[width = 9.0cm,height=3.3cm]{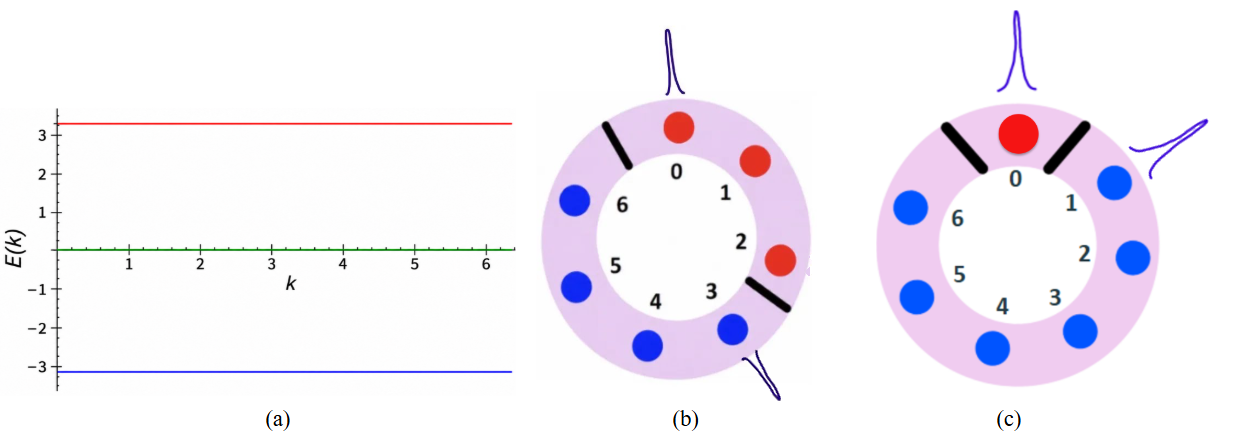}
\caption{\textcolor{black}{Schematics for (a) symmetric flat bands (red, blue) and zero-energy edge states (green) in the gap at $E = 0$; (b) a seven-site cyclic graph with two bulk domains (sites 0–2 and sites 3–6), whose interfaces host zero-energy edge states at sites 0 and 3: a walker started at $x = 0$ ($x = 3$) remains localized there, whereas one started at any other site $x \notin \{0, 3\}$ spreads; (c) a single-site domain at $x= 0$ embedded in the remaining sites, whose zero-energy edge states sit at sites 0 and 1: a walker
started at $x = 0$ ($x = 1$) remains localized there (blue peak), whereas one started at $x \notin \{0, 1\}$ spreads.}}
\label{c6-f-f1}
\end{figure}

\begin{figure*}[!t]
  \centering
         \begin{minipage}{\textwidth}
        \centering
\includegraphics[width=\textwidth,height=5.5cm]{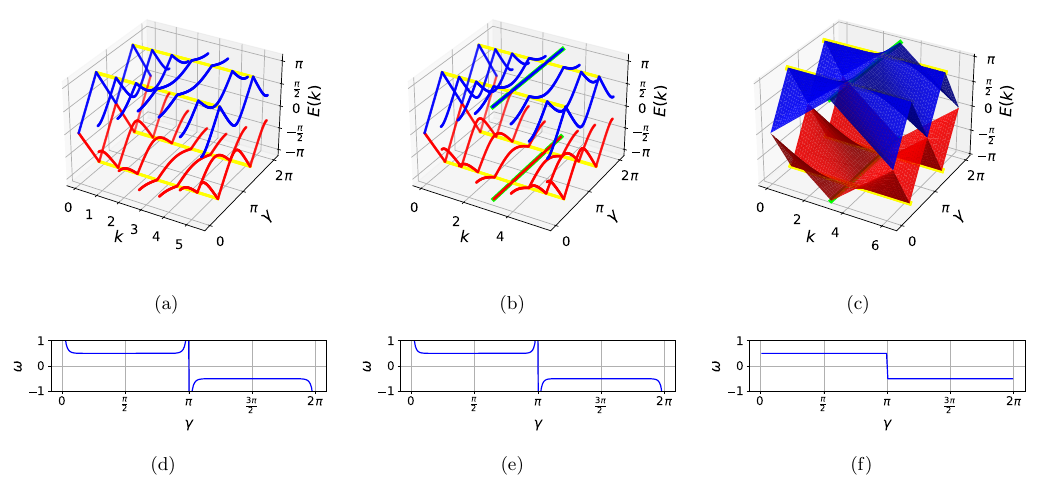}
\captionof{figure}{Dependence of quasienergy $E(k)$ on momentum $k$ and coin angle ($\gamma$) for: (a) $7$-cycle; (b) $8$-cycle; (c) $1000$-cycle. The red (blue) curves represent the upper (lower) quasienergy branches. (d)–(f) The corresponding winding number $\omega$ as a function of $\gamma$ for the $7$-cycle, $8$-cycle, and $1000$-cycle, respectively, obtained using the SCSS-CQW protocol with $D=2$. Fractional winding numbers $\pm \frac{1}{2}$ occur with a phase transition at $\gamma=\pi$. Yellow (sea green) lines indicate $k$-flatbands (rotational flatbands).}
\label{c6-f-f78scsst2}
\end{minipage}
    \hfill
    \begin{minipage}{\textwidth}
        \centering
\includegraphics[width=\textwidth]{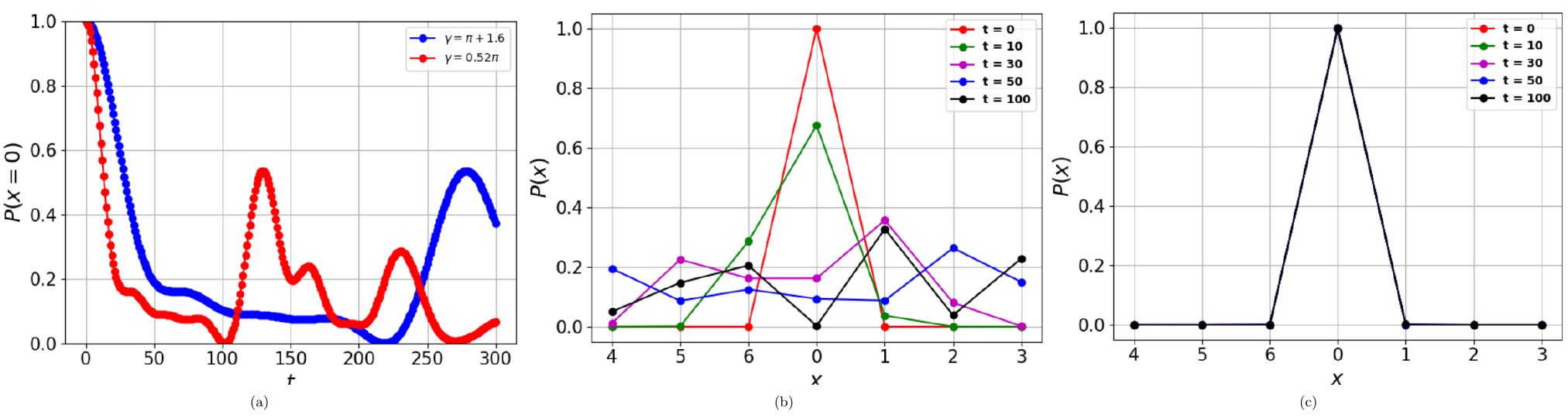}
\captionof{figure}{\textcolor{black}{(a) Probability $P(x =0)$ of the walker executing a SCSS-CQW ($D=2$) vs time steps $t$, for coin with $\gamma=0.52\pi$ in red and coin with $\gamma=\pi+1.6$ in blue, depicting chaotic (non-periodic) evolution. 
(b) A uniform topological phase with $\omega=\frac{1}{2}$ for coin $\gamma= 0.52\pi$ at all sites, produces no topological phase boundary and therefore no edge state, seen in the probability profile $P(x)$ versus $x$ for increasing time steps $t$. (c) Introducing an interface between two distinct topological phases, $\omega=-\frac{1}{2}$ via coin $\gamma=\pi+1.6$ at site 0 and $\omega=\frac{1}{2}$ via coin $\gamma=0.52\pi$ at other sites, $x\ne 0$, results in robust edge states localized at site $x=0$, persisting over time. The probability at site 0 is almost constant ($\approx1$) even at time step $t=100$.}}
\label{c6-f-edge7scss}
\end{minipage}
\end{figure*}

\textit{\textcolor{brown}{Single coin split-step CQW protocol.--}}
A CQW describes the evolution of the spatial distribution of a single quantum particle (e.g., an ion, photon, or electron) on a benzene-like ring, i.e., an $N$-cycle graph ($N$ being the number of sites)~\cite{p6,karimi19}. The walker state evolves in composite Hilbert space $\mathcal{H}=\mathcal{H}_P\otimes\mathcal{H}_C$, composed of an $N$-dimensional position space $\mathcal{H}_P$ (with basis $\{\ket{x}\,|\,x\in\{0, 1,\ldots,N-1\}$) and a two-dimensional coin space $\mathcal{H}_C$ (with basis $\{ \ket{g}=\ket{1_c},\ket{0_c}\}$).
Due to the spatial symmetry of the cyclic graph, the CQW dynamics can be diagonalized using the discrete quantum Fourier transform (QFT),
\begin{equation}
|k'\rangle=\frac{1}{\sqrt{N}}\sum_{x=0}^{N-1}e^{i\frac{2\pi}{N}xk'}|x\rangle,
\end{equation}
where $k',x\in\{0,1,2,\dots,N-1\}$. At each time step, the particle may move anticlockwise or clockwise depending on the coin state $\ket{1_c}$ or $\ket{0_c}$. This motion is governed by the shift (translation) operator $\hat{S}$, conditioned on the action of a single-qubit coin (gate) $\hat{C}$. The time-evolution operator of a standard CQW is given by $
\hat{U}=\hat{S}\,[I_N\otimes\hat{C}],$
and the quantum state after $t$ steps evolves as $\ket{\psi(t)}=\hat{U}^{\,t}\ket{\psi(0)}$, starting from the initial state $\ket{\psi(0)} = \ket{x}\otimes\ket{g}$.
By appropriately modifying the shift and coin operations, the CQW dynamics can be extended to a richer protocol known as the single-coin split-step CQW (SCSS-CQW), which enables distinct topological phenomena that are inaccessible within the standard CQW framework~\cite{p6}.
The topological invariant~\cite{karimi19,filippo17,panahiyan20,p6} characterizing the topological phases of our protocol is the  winding number~\cite{p6}, i.e.,
\begin{equation} \resizebox{0.5\linewidth}{!}{$ \omega= \frac{1}{2\pi} \sum_{k'=0}^{N-1} \left( \vec{n} \times \frac{\partial \vec{n}}{\partial k'} \right) \cdot \hat{A} . \label{c6-e-wind0} $} \end{equation}
which is obtained via the stroboscopic evolution, $\hat{U}_{evo} = e^{-i\hat{H}},\; \hat{H} = \vec{\sigma} \cdot \hat{n}E(k),$ which by definition is Hermitian, where, $\hat{A}$ is a vector of unit magnitude (in a quantum Bloch-sphere), and it is orthogonal to the winding vector $\vec{n}$ for all $k\in[0,2\pi]$ (entire Brillouin zone) and Eq. (\ref{c6-e-wind0}) is the discretized form of the continuous winding-number integral, with $k=\frac{2\pi k'}{N}$. A phase is topological if $\omega\ne 0$ and trivial if $\omega=0$.

The temporal evolution of a quantum walker particle evolving via SCSS-CQW dynamics with a single coin $\hat{C}_\gamma$, is governed by the operator, \begin{equation}
\hat{U}_{evo} = \hat{S}_{+} \hat{C}_\gamma \hat{S}_{-} \hat{C}_\gamma,
\end{equation}
involving alternating two conditional shifts and coin rotations, where the shift operators are:
$\hat{S}_{+} = \sum_{x=0}^{N-1} \left( \ket{0}_c \bra{0}_c \otimes  \ket{(x+1)\mod N} \bra{x} + \ket{1} \bra{1} \otimes \ket{x} \bra{x} \right)$, $
\hat{S}_{-} = \sum_{x=0}^{N-1}\left(  \ket{0}_c \bra{0}_c \otimes  \ket{x} \bra{x} + \ket{1} \bra{1} \otimes \ket{(x-1)\mod N} \bra{x} \right),$ 
and the coin,
$
\hat{C}_{\gamma} = e^{-i \frac{D \gamma}{2} \sigma_y}$.
$\gamma \in [0,2\pi]$ is the coin rotation-angle and the step-dependency parameter $D$ is an integer, i.e., \( D \ge 2 \) refers to a step-dependent (SD) coin while \( D = 1 \) refers to a step-independent (SI) coin. In the SCSS-CQW evolution, $D$ controls the step-dependent coin rotation through the angle $D\gamma/2$. In momentum space, $\hat{S}_{+} = e^{-i \left( \frac{2\pi k}{2N} \right) (\sigma_z + \mathbb{1})},\;\;\hat{S}_{-} = e^{-i \left( \frac{2\pi k}{2N} \right) (\sigma_z -\mathbb{1})}$. Under stroboscopic evolution $\hat{U}_{evo} = e^{-i\hat{H}}$, see for more details Supplementary Material (SM)~\cite{SM} Sec.~I, we calculate the energy dispersion in SCSS-CQW as,
\begin{equation}
\resizebox{0.8\linewidth}{!}{$
E_{\pm} = \pm \cos^{-1} \left[\cos k (\cos \Delta.\gamma)^2 - (\sin \Delta.\gamma)^2  \right],
\label{c6-e-en1}
$}
\end{equation} 
with $\Delta = D/2$ and winding number (Eq.~(\ref{c6-e-wind0})) for a $N$-cycle graph is,
\begin{equation}
\resizebox{0.92\linewidth}{!}{$
\omega_{\gamma,D,N}=  \sum_{k'=0}^{N-1} -\frac{2 \sin \left(\Delta.\gamma\right)}{N \left(2 \cos ^2\left(\frac{k}{2}\right) \cos (2\Delta.\gamma)+\cos \left(k\right)-3\right)},\;\text{with } k=\frac{2\pi k'}{N}.
$}
\label{c6-e-wind1}
\end{equation}

When $D=1$, i.e.,  SI-SCSS-CQW, and for $\gamma=\frac{\pi}{2}$ and $N=7,8,15$, the winding numbers are, $\omega_{\frac{\pi}{2},1,N}=0.500004,0.500001,0.500000$ which are almost same as with $N\rightarrow$ very large, i.e., $N=1000$ and $\omega_{\frac{\pi}{2},1,1000}=\frac{1}{2}$. For $D = 2$, i.e., SD-SCSS-CQW, with the coin $\gamma = 0.52\pi$ used below, $\omega_{0.52\pi,2,N} = 0.500000$ for $N = 7, 8$ and $15$ (deviations from $1/2$ below $10^{-21}$), identical to $N = 1000$.  We show that the fractional winding number converges to exactly $\frac{1}{2}$ as system size $N$ increases and a finite cyclic graphs with $N\geq6$ is sufficient to generate fractional topology, without going to the infinite-lattice limit, see also details in SM Sec.~I (SM Fig.~8). Earlier studies on fractional topological invariants ~\cite{frac1,frac2,frac3,frac-asboth,frac2-asboth,frac3-asboth} involve non-Hermitian Hamiltonians or interactions or complex scattering configuration in 1D infinite lattices, whereas in our work we uniquely realize the fractional invariants via an unitary evolution that also obeys Hermitian physics in finite cyclic graphs and operates with minimal resources, see SM~\cite{SM}, Sec. VII.

Below, we discuss the generation of topological phases, energy gap closing, flat bands, and topological winding numbers, edge states for step-dependent variants of SCSS-CQW dynamics, on finite even and odd cycle-graphs. We also test robustness of the generated edge states against dynamic-static coin disorder~\cite{p5,sasha,pxue,p6} and phase preserving perturbations~\cite{pxue,p6} in SM Secs. II and III~\cite{SM}.

\textit{\textcolor{brown}{Energy dispersion, topological phases and edge states.--}}
Quasienergy as a function of momentum $k$, and the winding number (topological invariant) as a function of the rotation angle $\gamma$ [Eqs.~(\ref{c6-e-en1}), (\ref{c6-e-wind1})], are shown in Fig.~\ref{c6-f-f78scsst2} for a SD (step-dependent) coin ($D=2$) (see, SM Figs~11 and 12). Both the step-dependent ($D\ge2$) and independent ($D=1$) SCSS-CQW exhibit conducting (energy gap closing) and insulating phases.
\begin{figure*}[!t]
         \centering
          \begin{minipage}{\textwidth}
        \centering
     	\includegraphics[width=\textwidth,height=4cm]{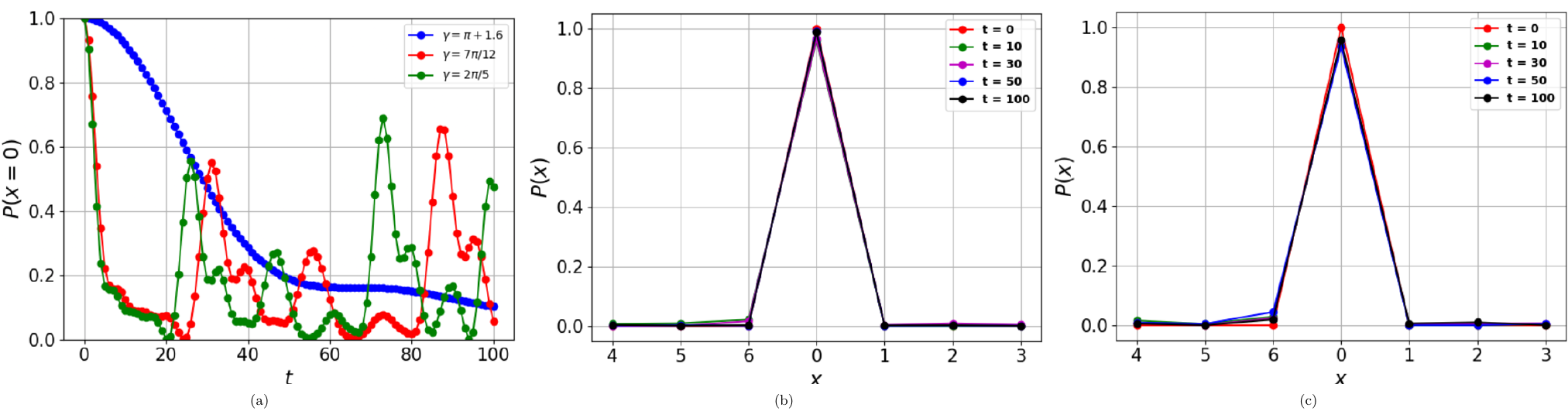}
     	\captionof{figure}{\textcolor{black}{(a) Probability $P(x =0)$ of the walker executing a SCSS-CQW ($D=2$) vs time steps $t$, for coin with $\gamma=\pi +1.6$ in blue, with $\gamma=7\pi/12$ in red, and with $\gamma = 2\pi/5$ in green depicting chaotic (non-periodic) evolution. Introducing an interface  on a 7--cycle graph between two distinct topological phases, $\omega=-\frac{1}{2}$ via coin $\gamma=\pi+1.6$ at site 0 and $\omega=\frac{1}{2}$ via coin (b)$\gamma=7\pi/12$, (c) $\gamma=2\pi/5$ at other sites, $x\ne 0$, results in a robust edge state localized at site $x=0$, persisting over time.} }
	\label{other} \end{minipage}
         \begin{minipage}{\textwidth}
        \centering
     	\includegraphics[width=\textwidth,height=4cm]{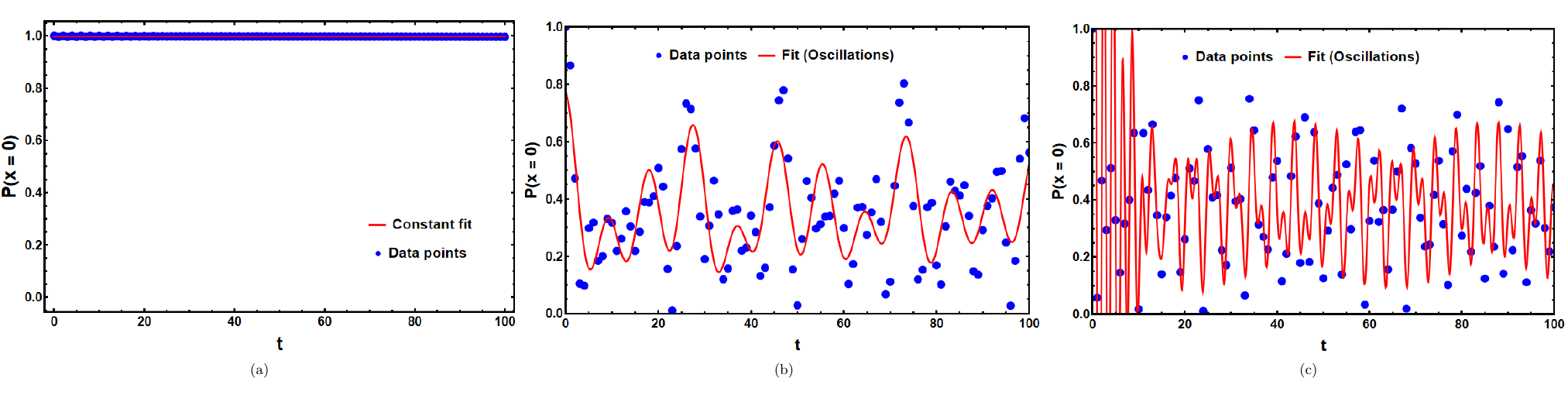}
     	\captionof{figure}{\textcolor{black}{(a) Edge-state survival probability $P(x=0)$ versus time-step $t$ for the SCSS-CQW protocol ($D=2$) on a $7$-cycle without disorder, using coin $\gamma=\pi+1.6$ ($\omega=-\frac{1}{2}$) at site 0 and $\gamma=0.52\pi$ ($\omega=\frac{1}{2}$) at the remaining sites $x\ne0$. The numerical fit shows a constant probability ($\approx 1$) (red), demonstrating the long-time stability of the localized edge state. (b) Localized-state survival probability $P(x=0)$ versus $t$ without a phase boundary, using coin $\gamma=\tfrac{\pi}{3}$ ($\omega=\frac{1}{2}$) at site 0 and $\gamma=\tfrac{3\pi}{5}$ ($\omega=\frac{1}{2}$) at the remaining sites $x\ne0$. (c) The corresponding case with a phase boundary, using $\gamma=\pi+0.2$ ($\omega=\frac{-1}{2}$) at site 0 and $\gamma=\tfrac{3\pi}{4}$ ($\omega=\frac{1}{2}$) at the remaining sites $x\ne0$. In both (b) and (c), $P(x=0)$ shows rapid, large-amplitude oscillations, in contrast to the stable edge state in (a).} }
	\label{c6-f-lsq-clean}
	\end{minipage}
    
    \begin{minipage}{\textwidth}
        \centering
	\includegraphics[width=\textwidth,height=6cm]{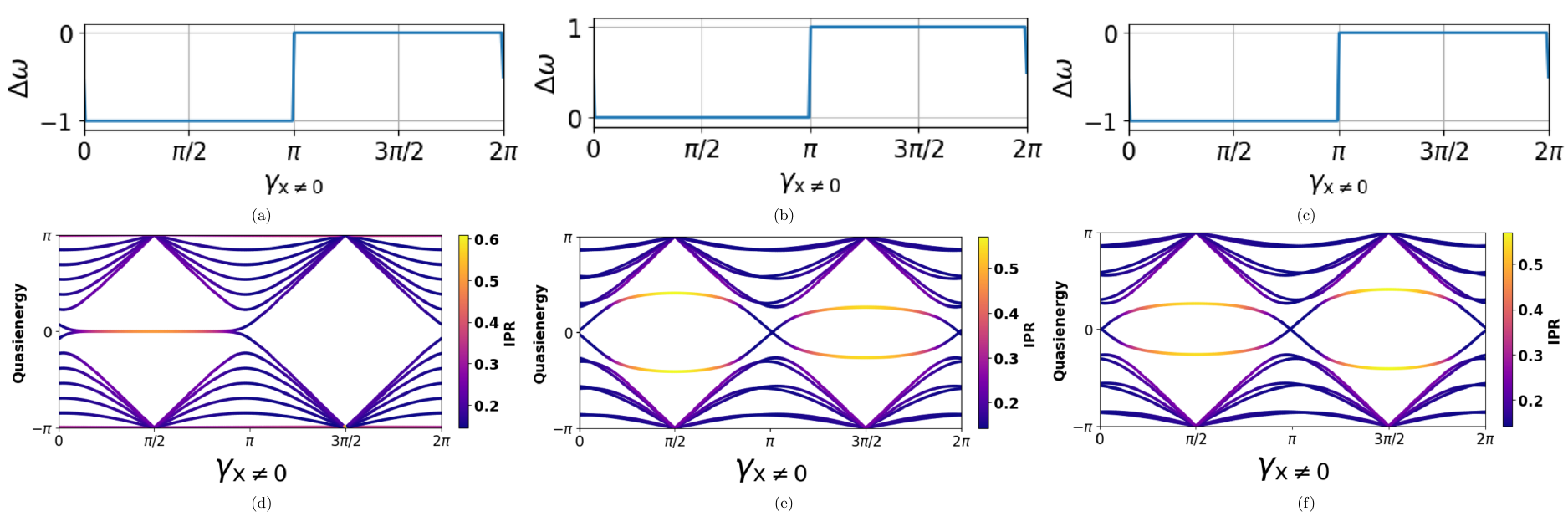}
\captionof{figure}{\textcolor{black}{Comparison of the winding-number difference and quasienergy spectra for FELS and topological ZEES on a 7-cycle graph. The winding-number difference $\Delta\omega$ versus coin angle $\gamma_{x\ne 0}$ for a single interface at site $x = 0$ with (a) $\gamma_{{x=0}} = \pi + 1.6$, (b) $\gamma_{{x=0}} = \pi - 0.2$, (c) $\gamma_{{x=0}} = \pi + 0.2$, (d)-(f) present the corresponding quasienergy spectra which yield (d) topological ZEES, and (e--f) FELS. The color of each eigenstate denotes its inverse participation ratio (IPR), with larger IPR indicating stronger spatial localization. In (d) the topological ZEES occur only for $\Delta \omega \ne 0$, and hybridize with the bulk when $\Delta \omega = 0$.}}
\label{compare}
\end{minipage}
     \end{figure*}

SCSS-CQW ($D\ge 2$) supports two distinct topological phases ($\omega=\pm \frac{1}{2}$) and exhibits topological phase transitions that can be induced by tuning the rotation angle $\gamma$. The number of topological phase transitions increases linearly with $D$ (See, Sec. I of SM~\cite{SM}). \\  In Fig.~\ref{c6-f-edge7scss}, the walker is initialized in the state $\ket{\psi(0)} = \ket{0} \otimes \ket{0_c},$
on a seven-site cyclic graph. \textcolor{black}{The coin angle at $x =0$ is $\gamma = \pi+1.6$ (corresponding to $\omega = -\tfrac{1}{2}$), whereas at all other sites, coin angle is fixed at $\gamma = 0.52\pi$ (i.e., $\omega = \tfrac{1}{2}$) in Fig.~\ref{c6-f-edge7scss}(c). Both coins ($\gamma = \pi+1.6$ or $0.52\pi$) beget a chaotic probability distribution (i.e., the walker does not return to the initial site with probability 1), so as to avoid cases of periodicity masking the edge state formation (see, Fig. \ref{c6-f-edge7scss}(a)). Following Refs.~\cite{treg,dukes}, we call a walk chaotic when it is non-periodic, i.e. $U^{N} \neq I$ for every $N$. As a reference case, Fig.~\ref{c6-f-edge7scss}(b) considers a homogeneous seven-site cycle in which the coin $\gamma=0.52\pi$, corresponding to $\omega=\tfrac{1}{2}$, is applied at every site. Since the entire system belongs to the same topological phase, no topological phase boundary is formed.
Consequently, the probability profile $P(x)$ remains delocalized over the cycle as the evolution proceeds, and no edge state occurs.} This spatial modulation of the coin angle induces a topological phase boundary localized at site $0$. A pronounced probability amplitude at site 0 serves as a hallmark of the edge state~\cite{kita-pra,pxue,chandra-topo,p6} and we observe resilient edge-states, which are stable over a long time ($t$), see Fig.~\ref{c6-f-edge7scss}(c). We emphasize that, for the seven-site cyclic graph considered, the site \(x=0\) constitutes a one-site topological domain embedded in another bulk six-site domain. Owing to the periodic boundary condition, this domain has two interfaces, namely between \(x=6\) and \(x=0\), and between \(x=0\) and \(x=1\). In the full site-plus-coin Hilbert space, which is 14-dimensional for a seven-site cycle, these two interfaces host a pair of zero-energy eigenstates with coin states $\ket{0}$. These are topological zero energy edge states (ZEES) (see, Figs.~\ref{c6-f-lsq-clean}(a) and ~\ref{compare}(d)).  Their chiral components sit at $x = 0$ and $x = 1$, one per interface (see Sec. VI of SM~\cite{SM}). A walker started at $x = 0$ remains there throughout the simulated window, giving the single peak in Fig.~\ref{c6-f-edge7scss}(c). \textcolor{black}{We avoid coin angles at $\gamma= \pi/2 (mod \pi),$ where each coin on its own makes the evolution trivial ($U = -1$); all coin angles used here lie away from these points, so the walks are non-periodic. Moreover, periodic quantum-walk dynamics can produce localized states irrespective of the winding-number mismatch, making it difficult to isolate the topological contribution~\cite{p6}. To distinguish these effects, we consider coin parameters for which the individual homogeneous walks exhibit chaotic dynamics, rather than periodic evolution, as shown in Fig.~\ref{c6-f-edge7scss}(a). The coin angles employed in  Fig.~\ref{c6-f-edge7scss} are $\gamma_{x=0}=\pi+1.6$ and $\gamma_{x\ne0}=0.52\pi$, which yield an angular difference of $0.989\pi$. Further, Fig.~\ref{other} shows examples of topological ZEES with $\gamma_{x=0}=\pi+1.6$, $\gamma_{x\ne0}=7\pi/12$ and $\gamma_{x\ne0}= 2\pi/5$ (see, Fig. \ref{compare} (d)), that occur for coin angle differences $0.926\pi$ and $1.1\pi$ respectively. More importantly, in SM Fig.~30 (b,e) we show examples with a phase boundary defined via coin angles ($\pi + 0.2, 3\pi/4$) for $N = 51, W = 25$, i.e., a difference of $0.31\pi$. }\\
  The algorithm used to generate the edge state is given in Sec VIII A of SM~\cite{SM}, and the Python code is available in GitHub~\cite{AR-git}. We model the decay of edge state probabilities vs time $t$, in Fig.~\ref{c6-f-lsq-clean} (a), and it displays a constant fit \textcolor{black}{($0.99743$)}, i.e., almost no decay. In comparison, the edge state generated via SS-QW's on 1D lattice shows a rapidly decaying power law fit~\cite{kita-pra,kita-exp}. This shows that the generated edge states of this letter remain stable over the simulated time window, see SM Sec. III for more details. \textcolor{black}{However, spatial localization alone does not establish the topological origin of a state. In an inhomogeneous quantum walk, a local modification of the coin parameter can itself generate a localized state at the modified site. Such defect-induced localized states (see, Figs.~\ref{c6-f-lsq-clean} (b) and (c)), referred to as finite energy localized states(FELS), can occur both without ($\Delta \omega = 0$) and with a topological phase boundary ($\Delta \omega \ne 0$) ~\cite{ptdef-1, ptdef-2, ptdef-3}. FELS may exhibit enhanced localization and a finite survival probability even in the absence of topological protection. This motivates a systematic comparison of  FELS with ZEES generated at the boundary between distinct topological phases. A detailed comparison of the topological ZEES and trivial FELS due to point-defects is discussed below and in Sec. IV of SM~\cite{SM}.} Our numerical simulations demonstrate that the ZEES generated in finite cyclic graphs using the SCSS-CQW protocol are robust under both moderate dynamic/static coin disorder and phase-preserving perturbations; see SM Sec.~II. This resilience to noise and disorder underscores the strong potential of these edge states for fault-tolerant quantum information processing and topological quantum computing applications.

\textcolor{brown}{\textit{Topological ZEES and Trivial FELS --}}
\textcolor{black}{As discussed previously, one can generate localized states even in the absence of a topological phase boundary. Fig.~\ref{c6-f-lsq-clean} shows that the introduction of a topological phase boundary ($\Delta \omega \ne 0$) is not a sufficient condition for topological edge states. Therefore, it is important to establish clear criteria for distinguishing genuine topological ZEES from trivial FELS.}

\textcolor{black}{Fixing the coin parameter at site $x= 0 (\gamma_{{x =0}})$ while continuously varying the bulk coin parameter $\gamma_{{x \ne 0}} $ (coin parameter at $x\ne 0$) causes the winding-number difference, $\Delta \omega = \omega (\gamma_{{x =0}}) - {\omega({\gamma_{x \ne 0}})}$, to change whenever the bulk undergoes a topological phase transition. Distinguishing topological ZEES from FELS requires examining both their spatial localization and quasienergy. We therefore diagonalize the evolution operator $\hat{U}_{evo}$ to obtain its eigenvalues and eigenvectors, $\hat{U}_{evo}\ket{\psi}=\lambda_i\ket{\psi}$. The quasienergy of each eigenstate ($i= 1,2...2N$) is obtained from $E_i=-\arg(\lambda_i)$, while its localization is quantified by the inverse participation ratio (IPR), calculated from the position-resolved probability as $\mathrm{IPR}=\sum_x[P(x)]^2$. The quasienergy determines whether a localized state lies within the symmetry-protected gap, whereas the IPR measures the degree of spatial localization. The calculation details of quasienergy and IPR are given in Sec. IV B of SM~\cite{SM}, and the algorithm is provided in Secs. VIII B and C of SM~\cite{SM}, with the corresponding Python code available in Ref.~\cite{AR-git}. We apply this procedure to three representative configurations: (i) the topological edge-state configuration  ($\gamma_{{x=0}} = \pi + 1.6 $ $(\omega = -1/2)$), (ii) a point-defect state with $\gamma_{{x=0}} = \pi - 0.2$ $(\omega = 1/2)$, and (iii) a point-defect state with $\gamma_{{x=0}} = \pi + 0.2 $ $(\omega = -1/2)$. Fig.~\ref{compare} shows the winding-number difference $\Delta \omega$ ((a)--(c)) together with the corresponding quasienergy spectra ((d)--(f)) as functions of the bulk coin parameter $\gamma_{x\ne0}$. A topological edge state is expected to be localized within the bulk spectral gap~\cite{asboth}. Although the second and third configurations exhibit a non-zero winding-number difference over part of the parameter range, only the edge-state configuration (see Figs. \ref{compare} (a) and (d)) supports a localized quasienergy mode pinned at the gap center $E=0$ whenever $\Delta \omega \ne 0$~\cite{asboth, kita-pra}. There are two zero-energy eigenmodes in the full 14-dimensional site$+$coin Hilbert space, corresponding to the two interfaces, their chiral components sit at $x=0$ and $x=1$, one per interface. The zero-energy states exist only in the topological regime ($\Delta\omega \ne0$) and merges with the bulk spectra in the trivial regime ($\Delta\omega =0$). By contrast, the point-defect configurations exhibit finite-energy localized states (FELS), which occur irrespective of the winding-number difference. These results demonstrate that a mismatch in the winding number is a necessary condition for the existence of a topological edge state, while the presence of an $E= 0$ mode provides the sufficient condition. A detailed comparison between topological ZEES and trivial FELS is presented in Sec. IV of SM~\cite{SM}. }

\textit{\textcolor{brown}{Finite width topological domains.--}}\textcolor{black}{To establish that the single-site domain considered in this work is not an artificial construct, we extend the SCSS-CQW protocol to finite-width bulk domains. Specifically, a contiguous domain of width (W) wherein the coin angle $\gamma_{\text{bulk}_1}$ is applied, while the remaining sites coin angle $\gamma_{\text{bulk}_2}$ is employed. By diagonalizing the corresponding evolution operator for different values of (W), we determine the number of localized zero-quasienergy modes, $N_{\text{loc}}$. Here, $N_{\text{loc}}=0$ indicates absence of topological ZEES, while $N_{\text{loc}}=2$ corresponds to two topological ZEES associated with the two boundaries of the finite-width domain. Further details of this analysis are provided in Sec. V of SM~\cite{SM}}.
\begin{figure*}
    \centering
\includegraphics[width=\textwidth]{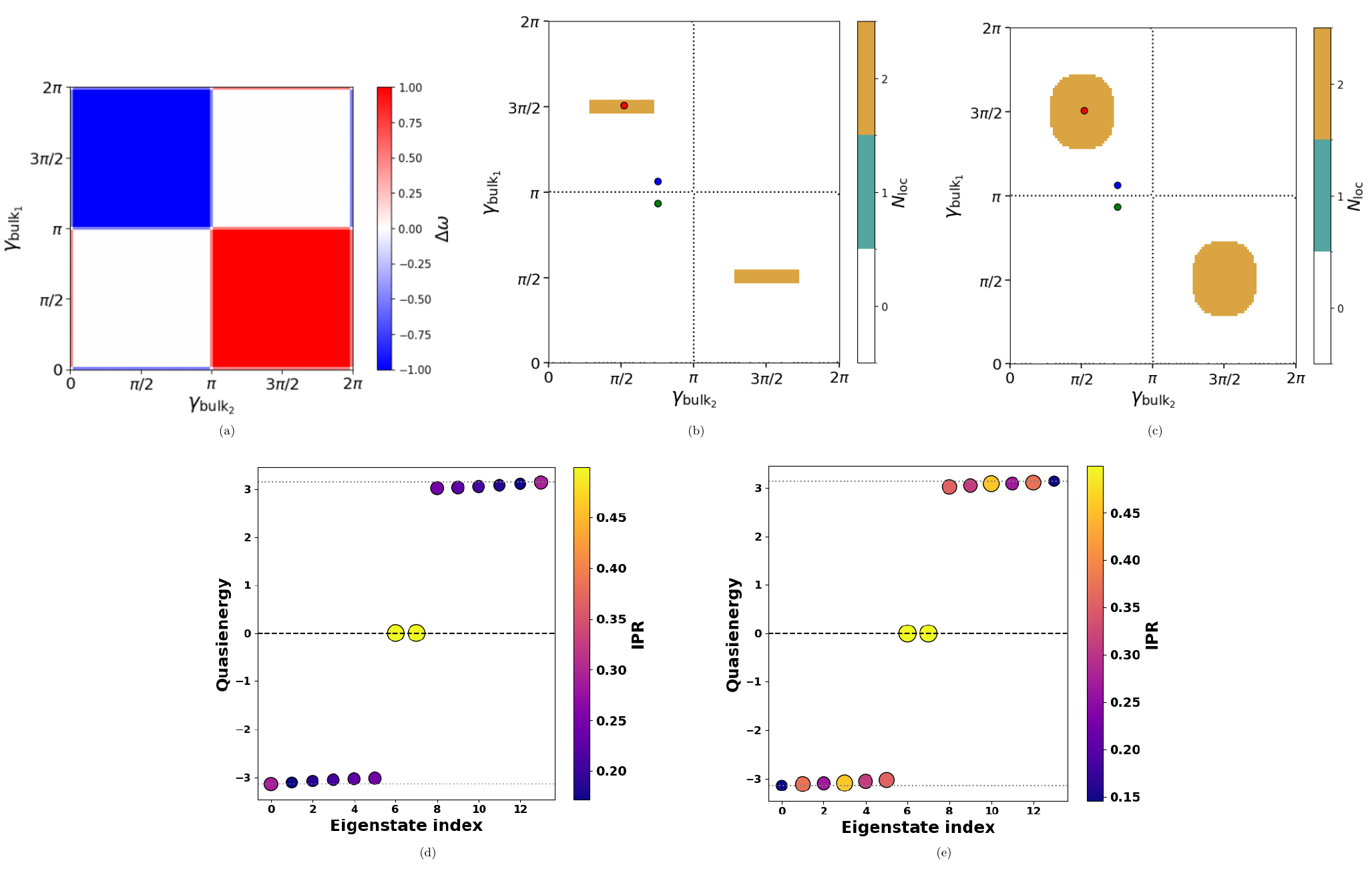}
\caption{\textcolor{black}{Topological phase diagram and number of topological zero-energy states both for a single-site and finite-width bulk domains. (a) Winding-number difference $\Delta\omega$, as a function of the bulk coin parameters. The two nontrivial quadrants $(\Delta \omega = \pm1)$ correspond to topological phase boundaries between the two bulk regions.  Number of topological zero-quasienergy modes, $N_{\text{loc}}$, for bulk domains of width (b) $W=1$ and (c) $W=3$ on a 7-site cyclic graph. In both cases, topological ZEES remain confined to the topologically nontrivial regions. (d-e) Quasienergy spectra for representative parameter values selected from the corresponding topological regions in (b)-(c) (for coin angles identified with red dot). The color of each eigenstate denotes its inverse participation ratio (IPR), with larger IPR indicating stronger localization. The representative spectra in panels (d)–(e) show two isolated eigenstates pinned at $E=0$, confirming that the localized modes are genuine topological edge states.  The red, blue, and green dots mark the three selected coin-angle pairs $\gamma_{\text{bulk}_1},\gamma_{\text{bulk}_2}=(\pi+1.6,0.52\pi)$ (ZEES), $(\pi+0.2,3\pi/4)$, and $(\pi-0.2,3\pi/4)$ (FELS), respectively. Topological edge states are observed for the choice of red dot but are absent for the blue and green dots, which lie outside the yellow region corresponding to $N_{\mathrm{loc}}=2$. As the bulk domain width increases from a single site to a finite number of sites, the range of bulk coin angles supporting topological ZEES expands in the presence of a winding-number mismatch, see SM~\cite{SM} for N = 15 and 51 cycles with bulk widths W = 7 and 25, where the $N_\text{loc} =2 $ region encompass nearly all coin angles for which winding number mismatch is seen.}}
\label{width2}
\end{figure*}

\textcolor{black}{Fig. \ref{width2} considers a 7-site cyclic graph to examine the evolution from a single-site bulk to a finite-width domain. For $W=3$, sites $0,1,2$ are with the coin angle $\gamma_{\text{bulk}_1}$ while sites $3$--$6$ are with coin angle $\gamma_{\text{bulk}_2}$ (see, Fig. 1(b)), corresponding to different topological phases. }\\
\textcolor{black}{Fig. \ref{width2} (a) shows the winding-number difference, $\Delta \omega = \omega_{\text{bulk}_1} - \omega_{{\text{bulk}_2}} $, which identifies the topological phase boundaries separating the bulk regions. According to the bulk-boundary correspondence, localized interface states are expected only when the two regions possess different topological invariants ($\Delta \omega \ne 0$). The $N_{\text{loc}}$ map for the single-site bulk ($W=1$) shown in Fig.~\ref{width2} (b) is consistent with this prediction, demonstrating that even the narrowest possible bulk region supports localized zero-quasienergy interface modes only when the two bulks belong to distinct topological phases. Fig. \ref{width2}(c) depicts the phase diagram for the finite-bulk case ($W=3$). The topological ZEES remain confined to parameter regions with $\Delta\omega\ne 0$, while the region supporting $N_{\text{loc}}=2$ expands as the domain width increases. This trend indicates that finite-width domains support topological ZEES over a broader range of coin parameters than the single-site bulk. The corresponding quasienergy spectra in Figs. \ref{width2}(d)–\ref{width2}(e) further illustrate this behavior. The color of each eigenstate represents its inverse participation ratio (IPR), with larger values indicating stronger localization. In the representative cases shown, two eigenstates remain pinned at the quasienergy ($E=0$) and are strongly localized, consistent with the presence of topological edge states. Together, the phase diagrams and quasienergy spectra demonstrate that the emergence of these modes is associated with the topological distinction between the two domains and that the parameter region supporting them expands with increasing domain width. For larger cyclic graphs (N = 15, 51) with finite-width bulk domain (see, N = 15, W = 7 and N = 51, W = 25 in Sec. V of SM~\cite{SM}), we observe that the coin parameters associated with the topological ZEES encompass nearly the entire region where winding number mismatch occurs. In this case, a topological ZEES implies that there is a winding number mismatch, and the reverse also holds true for nearly all coin angles, i.e., winding number mismatch also implies the presence of a topological zero-energy edge state.}\\
\textcolor{black}{The pinned $E=0$ modes also have a chiral symmetry signature. The unit vector $\hat A $ used in the definition of the winding number in  Eq.~(\ref{c6-e-wind0}) defines a site-dependent chiral operator $\hat\Gamma=\sum_x\ket{x}\bra{x}\otimes\hat A\!\cdot\!\vec\sigma$ with $\hat\Gamma\hat U_{\rm evo}\hat\Gamma=\hat U_{\rm evo}^{\dagger}$~\cite{asboth}, which maps a state at quasienergy $E$ to one at $-E$. Chiral density therefore vanishes for all bulk states and FELS, while for the states pinned at $E=0$ have finite values $\pm1$. We map the chirality in real space to the local chiral density $\rho_\Gamma(x)=\operatorname{Tr}_c\bra{x}\hat P_{\rm gap}\hat\Gamma\ket{x}$, where $\hat P_{\rm gap}$ projects onto the eigenstates in the zero-quasienergy gap. For the 7-cycle with $\gamma=\pi+1.6$ on the first $W$ sites and $0.52\pi$ elsewhere, $\rho_\Gamma(x)$ has two peaks of opposite sign at the two phase boundaries, $\rho_\Gamma(0)=-0.9988$ and $\rho_\Gamma(1)=+0.9998$ for $W=1$, and $\rho_\Gamma(0)=-0.9990$ and $\rho_\Gamma(3)=+0.9998$ for $W=3$, with negligible values at the remaining bulk sites. The chiral density is invariant with respect to both time frames. Further details are in SM Sec.~VI~\cite{SM}.}\\

\textit{\textcolor{brown}{Photonic realization.--}}Our SCSS-CQW protocol can be experimentally implemented using single photons, which effectively simulate cyclic quantum walkers, together with passive optical components such as waveplates, Jones plates, and polarizing beam splitters (PBS) to realize the shift and coin (gate) operations~\cite{bian,pxue,karimi19}. The coin degree of freedom is encoded in the polarization of the photon, while the position space can be represented either by spatial modes or by the orbital angular momentum (OAM) of photons~\cite{bian,pxue,karimi19}. \textcolor{black}{A sequence of PBS and Jones plates can implement the shift operators. The PBS conditionally routes photons into distinct spatial modes depending on their polarization, while the waveplates (or Jones plates) rotate the polarization to prepare for subsequent steps. Together, these elements implement the $S_+$ and $S_-$ conditional shifts required in the SCSS-CQW protocol~\cite{yasir2023}.} Furthermore, the site-dependent rotation angles defining the interface between different topological phases can be locally controlled by appropriately adjusting the orientation of the waveplates or Jones plates. The resulting probability distribution can be detected using single-photon detectors; in particular, a pronounced peak at the boundary site provides a clear experimental signature of the edge states generated by our protocol.

\textit{\textcolor{brown}{Summary.--}}We introduced a new quantum walk protocol, the SCSS-CQW on small finite cyclic graphs, which generates exotic topological phases characterized by fractional winding numbers that are inaccessible using standard CQW schemes. Moreover, these topological effects are achieved with significantly reduced resource consumption compared with existing approaches (see SM Sec VII~\cite{SM}). Even and odd cycles exhibit distinct energy dispersions, and rotational flat bands arise exclusively in even cyclic graphs. We analytically and numerically demonstrate that the protocol generates topological flat bands, and we derive a general condition for the rotation angle ($\gamma$) and $D$ that yields flat bands (i.e., with zero group velocity); see also SM Sec.~I.B. The proposed protocol produces fractional winding numbers $\{\pm \frac{1}{2}\}$, in contrast to the integer winding numbers $\{\pm1\}$ obtained in standard CQW. The emergence of distinct winding numbers (i.e., phase transitions) for $D\ge2$ enables the creation of a phase boundary in real position space, leading to the formation of topological ZEES. 

\textcolor{black}{We distinguish topological ZEES from trivial FELS arising from point defects by examining their quasienergy spectra and inverse participation ratios (IPR). The topological configuration is characterized by a localized mode pinned to the center of the quasienergy gap at $E=0$, whereas point defects produce localized states at finite quasienergies. Therefore, a mismatch in winding numbers is necessary but not sufficient for identifying a topological ZEES; the additional presence of a localized state pinned at the gap center, $E=0$, provides the sufficient criterion for its topological character.} This letter demonstrates that miniaturized synthetic quantum systems can serve as versatile and experimentally feasible platforms for the realization and control of fractional topological phases, nontrivial topological phase transitions, flat bands, and robust edge states. The ability to realize these phenomena in compact systems has potential implications for fault-tolerant quantum computing, the simulation of nontrivial topological-insulator physics, robust quantum-memory architectures, quantum state engineering, and topological quantum cryptography, while also providing a route toward resource-efficient fault-tolerant quantum architectures.

\newpage
\onecolumngrid
\section*{\underline{Supplementary Material}}

\vspace{1cm}
This material provides additional details and results for single-coin split-step cyclic quantum walks  (SCSS-CQW), supporting the key findings of the main text. In Sec.~\textcolor{blue}{I}, for the SCSS-CQW evolution, we derive energy dispersion, group velocity, and generic winding number which characterizes its topological phases. We quantitatively demonstrate that the convergence of fractional winding number with system size $N$ and small cyclic graphs (with $N\geq6$) are sufficient to generate fractional winding numbers without going to the infinite-lattice limit.  We then analytically derive the conditions on rotation angles and step-dependency parameters, for the occurrence of flat-bands. These are then validated with vanishing group-velocity (both analytically \& numerically). The numerical results of energy dispersion as well as topological phases (with fractional winding-numbers) with step-dependent variant and step-independent variant of SCSS-CQW for 7,8-cycle graphs are then put forth. We numerically demonstrate the resilience of the generated edge-states, \textcolor{black}{which remain pinned to zero quasienergy}, against static, dynamic disorder as well as phase-preserving perturbations in Sec.~\textcolor{blue}{II}. In Sec.~\textcolor{blue}{III}, we model the decay of survival probabilities vs time, and show the long-time stability of the generated \textcolor{black}{zero energy} edge states. 
\textcolor{black}{In Sec.~\textcolor{blue}{IV}, we show that the topological edge states generated via SCSS-CQW are completely different from the localized or point-defect states which are generated with or without any phase boundary, spread significantly into the bulk, and show rapid  large-amplitude oscillations in time, unlike stable edge states created at a topological phase boundary. Besides comparing their real-time dynamics in clean and disordered systems, we also include a comprehensive quasienergy analysis, inverse participation ratio (IPR) and robustness under local coin perturbations establishing that only topological edge states support localized zero-quasienergy modes pinned to the symmetry-protected gap. In Sec.~\textcolor{blue}{V}, we further investigate finite-width bulk domains and demonstrate that the single-site bulk considered in the main text represents the limiting case of an extended topological domain, with the parameter regime supporting localized edge states increasing with bulk width in accordance with the bulk-boundary correspondence. \textcolor{black}{Sec. \textcolor{blue}{VI} provides chiral signature of the topological edge states in comparison to the bulk states.} \textcolor{black}{Sec. \textcolor{blue}{VII} provides the resource overhead required by the protocol, with a comparison with other existing methods. }Finally, in Sec.~\textcolor{blue}{VIII} provides the algorithms for plotting probability distribution, edge state generation, obtaining quasienergy spectrum and IPR via the SCSS-CQW protocol.}

\section{Energy dispersion and topological phases with single-coin split-step cyclic quantum walk  (SCSS-CQW)}

\subsection{Energy dispersion and topological winding numbers}
The time-evolution of a quantum particle with a single-coin split-step CQW  (SCSS-CQW) is governed by the evolution operator,
\begin{equation}
\hat{U}_{evo} = \hat{S}_{+} \hat{C}_\gamma \hat{S}_{-} \hat{C}_\gamma,
\label{evo3}
\end{equation}
where the shift operators are:\begin{equation}
    \hat{S}_{+} = \sum_{x=0}^{N-1}\left(  \ket{0}_c \bra{0}_c \otimes \ket{(x+1)\text{ mod }N} \bra{x} + \ket{1}_c \bra{1}_c \otimes \ket{x} \bra{x} \right),
\label{s+}
\end{equation}
\begin{equation}
    \hat{S}_{-} = \sum_{x=0}^{N-1} \left( \ket{0}_c \bra{0}_c \otimes \ket{x} \bra{x} + \ket{1}_c \bra{1}_c \otimes \ket{(x-1)\text{ mod } N} \bra{x} \right),
\label{s-}
\end{equation}

where $N$ is the number of sites of the cyclic graph, and the coin operator:
\begin{equation}
\hat{C}_{\gamma} = e^{-i \Delta . \gamma .\sigma_y}.
\label{cgamma}
\end{equation}
In Eq.~(\ref{cgamma}), $\Delta=\frac{D}{2}$ and $D$ being an integer. \( D= 1 \) refers to step independent coin while \( D \ge 2 \), i.e., $D=2,3,4,...$ refers to a step-dependent coin. 
We consider the Fourier transform (FT) of the position basis, that is, $
        |k'\rangle = (\frac{1}{\sqrt{N}} )\sum_{x=0}^{N-1} e^{i(\frac{2\pi}{N}) x k' } |x\rangle_p.$ The parameters $k',x \in \{0, 1,2,...,N-1\}$. Substituting \(\omega_N = e^{\frac{2\pi i}{N}}\), we can write,

\begin{equation}
\begin{split}
&\hat{S}_{+} \ket{0}_c \ket{k'} = \omega_N^{-k'} \ket{0}_c \ket{k'},\;\hat{S}_{+} \ket{1}_c \ket{k'} = \omega_N^{0} \ket{1}_c \ket{k'},
\\&
\text{i.e.,    }\hat{S}_{+} = \sum_{k'} \begin{pmatrix}
\omega_N^{-k'} & 0 \\ 
0 & 1
\end{pmatrix} \otimes \ket{k'} \bra{k'}.
\end{split}
\label{sssstart}
\end{equation}


Similarly, 
\begin{equation}
\begin{split}   
& \hat{S}_{-} \ket{0}_c \ket{k'} = \omega_N^{0} \ket{0}_c \ket{k'},\;\hat{S}_{-} \ket{1}_c \ket{k'} = \omega_N^{k'} \ket{1}_c \ket{k'},\\& \text{i.e., }
\hat{S}_{-} = \sum_{k'} 
\begin{pmatrix}
1 & 0 \\
0 & \omega_N^{k'} 
\end{pmatrix} \otimes \ket{k'} \bra{k'} .
\end{split}
\end{equation}

In terms of Pauli matrices, we arrive at, 
\begin{equation}
\hat{S}_{+} = e^{-i \left( \frac{2\pi k'}{2N} \right) (\sigma_z + \mathbb{1})},\;\;\hat{S}_{-} = e^{-i \left( \frac{2\pi k'}{2N} \right) (\sigma_z -\mathbb{1})}.
\label{sssend} 
\end{equation}

With the stroboscopic evolution, $\hat{U}_{evo} = e^{-i\hat{H}},\; \hat{H} = \vec{\sigma}\cdot \hat{n}E(k),\;k=\frac{2\pi k'}{N}$; we obtain,
\begin{equation}
    \hat{U}_{evo} = \hat{S}_{+} \hat{C}_\gamma \hat{S}_{-} \hat{C}_\gamma = e^{-i E(k)  \vec{\mathbf{\sigma}} \cdot \vec{n}},
    \label{evo33}
\end{equation} from which we calculate energy dispersion to be,
\begin{equation}
\begin{split}    
&\cos E = \left[ \cos k (\cos (\Delta. \gamma))^2 - (\sin (\Delta. \gamma))^2 \right],\\&
\text{or, } E_{\pm} = \pm \cos^{-1} \left[\cos k (\cos (\Delta. \gamma))^2 - (\sin (\Delta. \gamma))^2  \right].
\end{split}
\label{en3}
\end{equation}

Clearly, one can draw the following conclusion from the energy dispersion (Eq.~(\ref{en3})):
\begin{enumerate}
    
      \item $(\cos (\Delta. \gamma))^2 = 0$, implies that the energy bands become $k$-independent and we see symmetric flat bands. 
      
    \item -$(\sin (\Delta. \gamma) )^2 = 0$, leads to the possibility for linear energy gap closing, for SCSS-CQW in $N\rightarrow$ large limit. This is similar to what is seen in CQW, see Ref.~\cite{p6}.

\end{enumerate}

The group velocity ($V_{gr}$)  can be calculated from Eq.~(\ref{en3}), i.e.,
\begin{equation}
V_{gr} =\pm \frac{\sin k (\cos (\Delta. \gamma))^2}{\sqrt{1 - \left[ \cos k (\cos (\Delta. \gamma))^2 - (\sin (\Delta. \gamma))^2 \right]^2}},
\label{c6s-e-v3}
\end{equation}
and the effective mass ($m^* = \frac{\hbar^2}{\partial_k V_{gr}}$) of the walker in the units of $\hbar = 1$ is,

\begin{equation}
   m^* =\pm
\frac{16 (\sec \left(\gamma \Delta\right) )^2 \left(1-\left(\cos (k) (\cos \left(\gamma \Delta\right))^2 -(\sin \left(\gamma \Delta\right))^2 \right)^2\right)^{3/2}}{\cos (k) (8-8 (\cos (\gamma D))^2)+2 (\cos (2k)+3) (\sin (\gamma D))^2}, 
\label{m3}
\end{equation}

where,$+ (-)$ signs correspond to the upper (lower) energy bands.

As shown in Eqs.~(\ref{c6s-e-v3})-(\ref{m3}), the group velocity and effective mass can be controlled by two parameters: coin rotation-angle $\gamma$ and step-dependency parameter $D$. These quantities provide a check on flat band formation, i.e., the group velocity vanishes for flat bands, also see Sec.~ I B below for further details. These are also useful in determining diffusion rates, carrier mobility 
and wave-packet-spreading in the case of a solid state quantum-system, see Refs.~\cite{effmass1,effmass2,effmass3}. From the stroboscopic evolution, given in Eq.~(\ref{evo33}), we get the components of the winding vector $\vec{n}(k)$,

\begin{equation}
    \vec{n}(k) =
\frac{1}{\sin E(k)}
\begin{bmatrix}
- \sin k \cos (\Delta. \gamma) \sin (\Delta. \gamma) \\
\cos k \cos (\Delta. \gamma) \sin (\Delta. \gamma) + \sin (\Delta. \gamma) \cos (\Delta. \gamma) \\
\sin k (\cos (\Delta. \gamma))^2
\end{bmatrix}
\label{eq:n_vector}
\end{equation}

Finally, the winding number (see Eq.~(2) of Main) for a discretized system of a $N$-cycle graph reads,
\begin{equation}
\omega_{\gamma,D,N}= \frac{1}{2\pi} \sum_{k'=0}^{N-1} \left( \vec{n} \times \frac{\partial \vec{n}}{\partial k'} \right) \cdot \hat{A} \;,
\label{eq:zak_phase}
\end{equation}
where the unit vector, $\hat{A}(\gamma,D)=\left( \cos(\Delta.\gamma), 0, \sin(\Delta .\gamma) \right)$ and $\Delta=\frac{D}{2}$.
Thus, we get the winding numbers,  $\omega_{\gamma,D,N}=\frac{Z_{\gamma,D,N}}{\pi}$, from the Zak phase $Z_{\gamma,D,N}$, to be,

\begin{equation}
\omega_{\gamma,D,N}=  \sum_{k'=0}^{N-1} -\frac{2 \sin \left(\gamma \Delta\right)}{N \left(2 \cos ^2\left(\frac{k}{2}\right) \cos (\gamma D)+\cos \left(k\right)-3\right)}\;\;.
\label{wind3}
\end{equation}


\begin{figure}
\includegraphics[width=\linewidth]{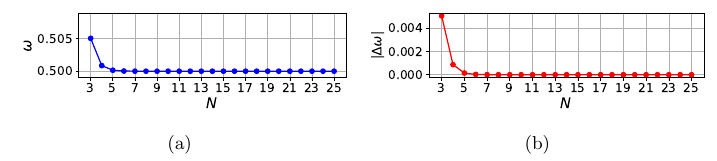}
\caption{(a) Winding number $\omega$ vs number of sites $N$ in cyclic graphs. (b) Deviation in winding number ($|\Delta \omega|$) vs $N$, showing convergence of $w$ value even for small $N$. }
\label{c6s-f-convergence}
\end{figure}

\begin{figure}
\includegraphics[width = 17cm,height=5cm]{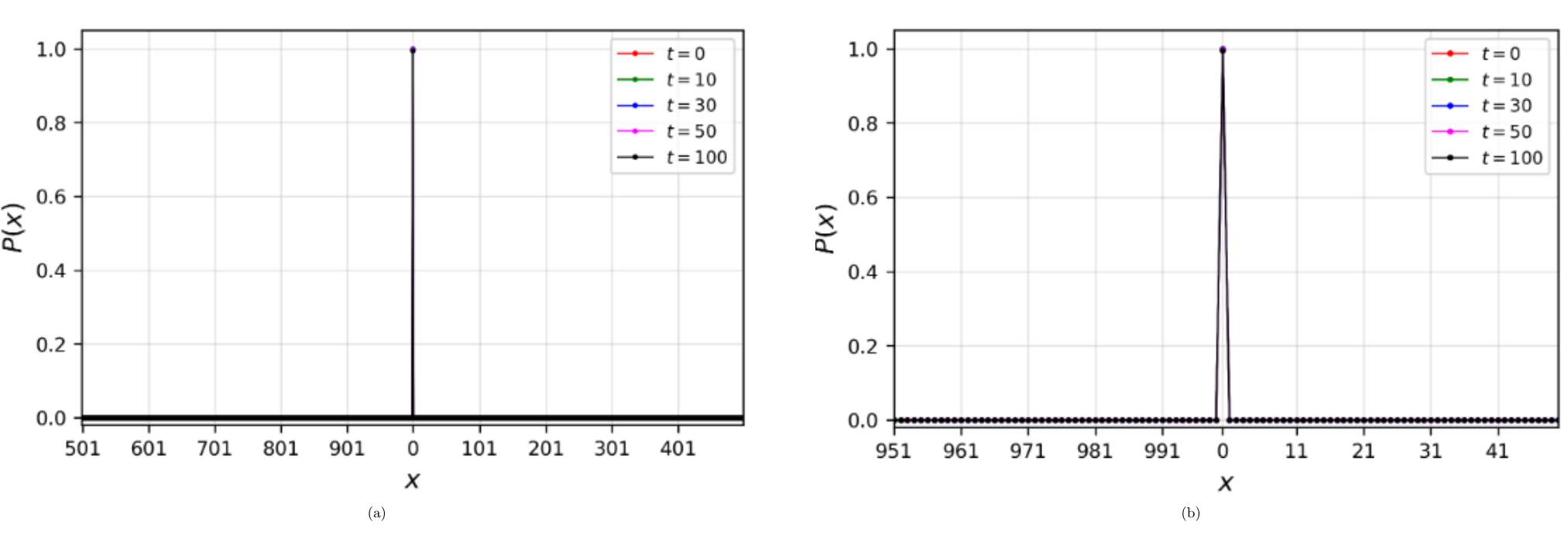}
\caption{\textcolor{black}{(a) Probability $P(0)$ of the walker at site $x=0$ as a function of time steps $t$. An interface is created between two distinct topological phases ($\omega=-\frac{1}{2}$ for coin $\gamma=\pi+1.6$ at site 0 and $\omega=\frac{1}{2}$ for coin $\gamma=0.52\pi$ at other sites $x\ne 0$), using SCSS-CQW protocol (with $D=2$) on a $1000$-site cyclic graph (large $N$ limit). This results in an edge state localized at site $x=0$, persisting over time. (b) Zoomed-in version of (a), displaying 50 sites on either side of $x=0$.}}
\label{edge-on-1000cycle}
\end{figure}

We achieve fractional winding numbers $\omega=\pm 1/2$ in small cyclic graphs, i.e., with finite $N$ value such as $N=7,8$ without going to the infinite $N$ limit. Computational and experimental complexities are much more reduced for finite $N$-cycles~\cite{bian, karimi19}. 
To quantitatively address the convergence of the fractional winding number with increase in the graph size $N$, we compute the deviation $|\Delta \omega|$ from the asymptotic value of $\omega_{\infty}=1/2$ for $N=1000$ (for coin rotation angle $\gamma=\pi/2$ and step-dependency $D=1$) as a function of number of sites $N$,
\begin{equation}
|\Delta \omega|= \left| \omega - \omega_{\infty} \right|= \left| \omega - \frac{1}{2} \right|.
\end{equation}
Fig.~\ref{c6s-f-convergence} shows the behavior of $|\Delta \omega|$ for $D=1$ and $\gamma=\pi/2$, with $N$ ranging from $3$ to $25$.
We observe that $|\Delta \omega|$ decreases rapidly with increasing $N$, approaching zero for relatively small cycles ($N \gtrsim 6$), wherein $|\Delta \omega|\rightarrow 0$ and $|\frac{\Delta \omega}{\omega_{\infty}}| \rightarrow 0$. This demonstrates that the winding number converges to the fractional value $1/2$ in a $N=6$-site cyclic lattice (lower limit). From the plot, the deviation becomes negligibly small beyond a threshold value $N_{\mathrm{min}} \approx 6$, indicating that the fractional characteristics can be reliably resolved for $N \geq N_{\mathrm{min}}$. 
These results confirm that the observed fractional winding number (characterizing the topological phases) is a finite cycle manifestation that converges to the asymptotic value not just in the thermodynamic limit, but also for $N\ge 6$. In Main, our use of small cycle graphs with $N=7,8$ sites is thus reasonable as it gives identical results to larger graphs but with much less resource expenditure (see, Sec. VII). \textcolor{black}{Furthermore, in Fig.~\ref{edge-on-1000cycle}, using SCSS-CQW protocol (with, $D=2$) on a $1000$-site cyclic graph, we show the probability of the walker at site $0$, $P(x=0)$ as a function of time steps $t$ by creating an interface between two distinct topological phases ($\omega=-\frac{1}{2}$ for $\gamma=\pi+1.6$ at site $x=0$ and $\omega=\frac{1}{2}$ for $\gamma=0.52\pi$ at $x\ne0$). This results in an edge state localized at site $x=0$, persisting over time, as in the case of 7-cycle shown in Fig.~3(c) in Main. This also supports the fact that results on small cyclic graphs agree with results in the thermodynamic limit.}

\subsection{Flat bands and group velocity}
We show flat-bands can be generated via SCSS-CQW by tuning the rotation angles and the step-dependency parameter $D$ in the SCSS-CQW evolution. 

As derived in main Eq.~(3) and Eq.~(\ref{en3}), the energy dispersion for SCSS-CQW in the cyclic graph of site $N$ is,
\begin{equation}
E_{\pm}= \pm \cos^{-1} \left[\cos k (\cos (\Delta. \gamma))^2 - (\sin (\Delta. \gamma))^2  \right]
\end{equation}

From the above dispersion, it is evident that symmetric flat bands are when the coefficient of $\cos k$ vanishes, giving
\begin{equation}
\cos\left(\Delta. \gamma\right) = 0,
\end{equation}
 \text{ i.e., energy $E_{\pm}$ becomes $k$ or $k'$ independent, i.e.,}
\begin{equation}
\begin{split}
\Delta. \gamma = \left(2n+1\right)\frac{\pi}{2}, \text{ or, } \gamma= \frac{(2n+1)\pi}{D}
\label{c6s-e-flat0}
\end{split}
\end{equation}
where $n \in \mathbb{Z}_{+} \cup \{0\}$.
The condition in Eq.~(\ref{c6s-e-flat0}) makes the quasienergy independent of $k$.
Substituting $\cos(\Delta.\gamma)=0$, i.e.\ $\sin^2(\Delta.\gamma)=1$, into Eq.~(14) gives
\begin{equation}
\cos E = -\sin^2(\Delta.\gamma) = -1, \qquad \text{i.e.}\qquad E_{\pm}=\pm\pi .
\end{equation}
Since quasienergy is defined modulo $2\pi$, $E_+=\pi$ and $E_-=-\pi$ are the same
quasienergy: the two flat bands coincide. Indeed, for $\Delta.\gamma=(2n+1)\pi/2$ the coin
is $\hat C_\gamma=\mp i\sigma_y$ and $\hat S_+\hat C_\gamma\hat S_-\hat C_\gamma=-\hat{1}$
for every $k$, so every state acquires only a sign at each step. The flat bands are
therefore degenerate at $E=\pi$: the quasienergy gap at $E=\pi$ is closed, while the gap
around $E=0$ is maximal. Such flat bands occur for $\gamma=\frac{(2n+1)\pi}{D}$ with
$n=0,1,\dots,D-1$; for example, for $D=1$ at $\gamma=\pi$ and for $D=2$ at
$\gamma=\frac{\pi}{2},\frac{3\pi}{2}$.

\begin{figure}
\includegraphics[width = 18cm,height=6cm]{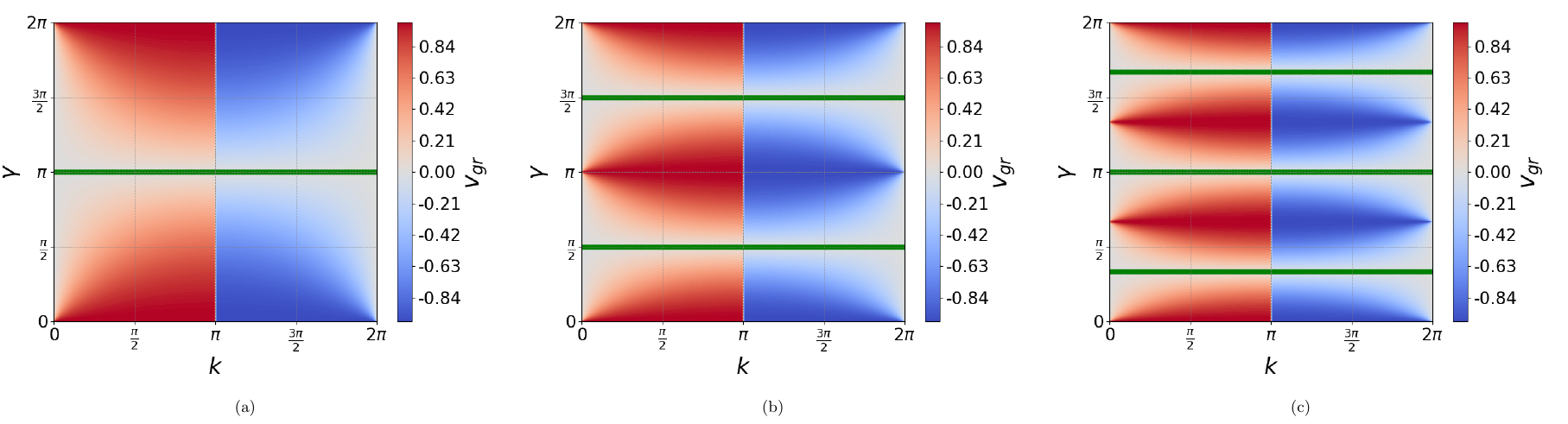}
\caption{Group velocity (see, Eq. (\ref{c6s-e-v3})) as a function of quasimomentum $k$ (for infinite site cyclic graphs) and coin rotation-angle $\gamma$ for SCSS-CQW evolution with (a) coin with $D= 1$, and coins with (b) $D= 2$ and (c) with $D= 3$. \textbf{Flat bands}, wherein the energy $E(k)$ is not dependent on momenta $k$, and are characterized by zero group velocity (that is, $V_{gr} = 0,$ for all $ k$), consistent with condition of Eq.~(\ref{c6s-e-flat0}) are marked by solid green-lines. The flat bands forms at (a) $\gamma= \pi$; (b) $\gamma= \frac{\pi}{2}, \frac{3\pi}{2}$; and (c) $\gamma= \frac{\pi}{3}, \pi, \frac{5\pi}{3}$.}
\label{c6s-f-figv0}
\end{figure}

Because $E$ is independent of $k$ at these angles, the group velocity
$V_{gr}=\partial E/\partial k$ vanishes identically. Equation~(\ref{c6s-e-v3}) cannot be
evaluated directly at these points, since its numerator and denominator both vanish;
approaching them, with $c=\cos(\Delta.\gamma)\to 0$, one finds
$E\simeq\pi-2|c|\,|\cos(k/2)|$ and hence $|V_{gr}|\le |c|\to 0$. This is confirmed
numerically in Fig.~\ref{c6s-f-figv0}, where the group velocity vanishes at the predicted
$\gamma$ values, indicated by the green lines.

\begin{figure}
\includegraphics[width=18cm,height=6.8cm]{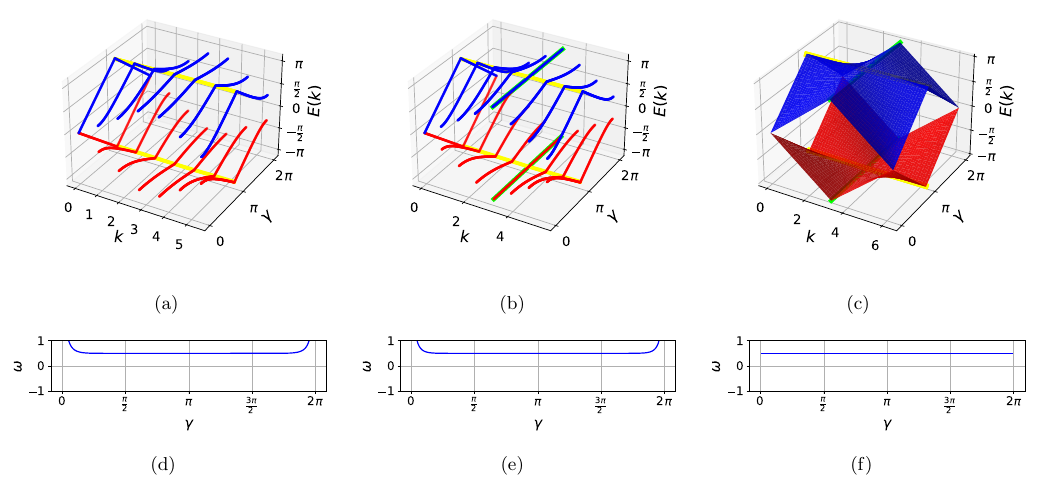}
\caption{Quasienergy spectrum $E(k)$ as a function of $k$ and coin-angle ($\gamma$) with: (a) $7$-cycle; (b) $8$-cycle; (c) $1000$-cycle. Red (blue) curves denote the upper (lower) quasienergy branches. 
(d)–(f) Corresponding winding number $\omega$ plotted vs $\gamma$ for $7$-cycle, $8$-cycle, and $1000$-cycle, respectively, via the SCSS-CQW protocol with $D=1$.}
\label{f78scsst1}
\end{figure}

\begin{figure}[H]
\includegraphics[width = 18cm,height=6.8cm]{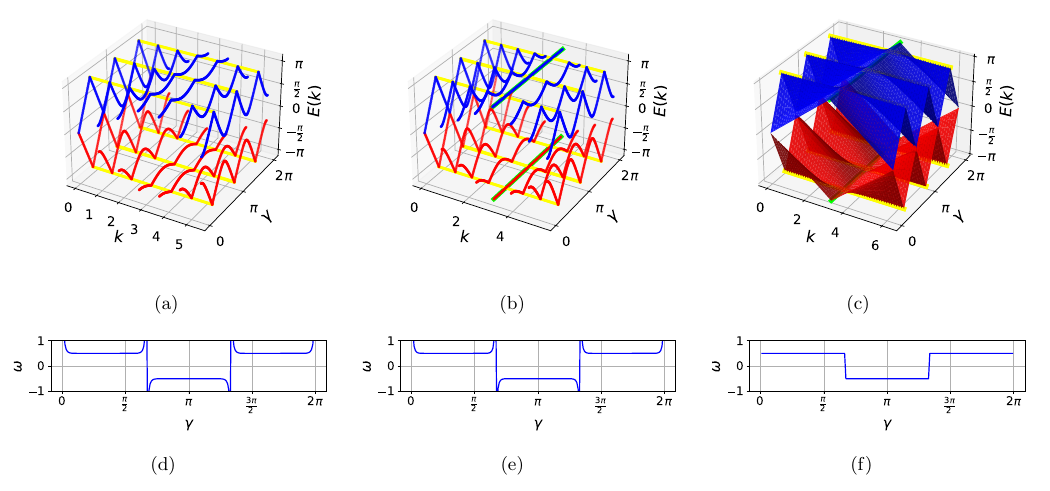}
\caption{Quasienergy spectrum $E(k)$ as a function of $k$ and coin-angle ($\gamma$) with: (a) $7$-cycle; (b) $8$-cycle; (c) $1000$-cycle. Red (blue) curves denote the upper (lower) quasienergy branches. 
(d)–(f) Corresponding winding number $\omega$ plotted vs $\gamma$ for $7$-cycle, $8$-cycle, and $1000$-cycle, respectively, for the SCSS-CQW protocol with $D=3$.}
\label{c6s-f-f78scsst3}
\end{figure}

\subsection{Energy Spectrum and topological phase structure of the SCSS-CQW}
Band dispersion and topological invariant (winding number, $\omega$) as a function of the coin rotation angle $\gamma$ (Eqs.~(\ref{en3}), (\ref{wind3})), are shown in Fig.~\ref{f78scsst1} for SI SCSS-CQW ($D=1$), and Fig.~\ref{c6s-f-f78scsst3} for step-dependent SCSS-CQW ($D=3$) for $N=7,8$-site cyclic graphs. Both SD and SI SCSS-CQW host conducting (energy gap closing) and insulating phases. Analytical calculations (Eqs.~4-5 of the Main and Eqs.~14-19 of SM) and numerical simulations (Figs. 2-4 of the Main and Figs. 9-15 of SM) show that SD ($D\ge 2$), and SI ($D=1$) SCSS-CQW on finite cycle graphs show topological effects. We see topological phases (with fractional winding-numbers), rotational flatbands for even cycles  for all $D$ and topological flatbands, along with resilient edge states generation. The SI version of SCSS-CQW shows energy band closing in even and odd cycles. Gap closing occurs in SD versions of SCSS-CQW, too. Even cycles show more number of gap closing points compared to odd cycles. Rotational flat-bands appear for even cycles and for all $D$  in SD version of SCSS-CQW with $E=\pi$ for every $\gamma$. This implies that the energy bands of the SCSS-CQW evolution become independent of the coin rotation angles. The SD version of SCSS-CQW shows  flat-bands (topological) and are linked to studies on quantum criticality and exotic transport in semimetals~\cite{semimetal,criticalsyst}. We have demonstrated the generation of topological zero-energy edge states (ZEES) via SD SCSS-CQW, which remain pinned to quasienergy $E=0$ (see Figs.~3,6 of Main). From numerical simulations (Figs.~13-15 of SM), we show that these topologically protected ZEES generated via the SCSS-CQW protocol are robust against small to moderate levels of dynamic and static coin disorder as well as against topological-phase preserving perturbations, see SM Sec.~II. We model the decay of the edge state probabilities vs time (with and without disorder) and show the long-time stability of the generated edge states (also, see SM Sec.~III). Table~\ref{c6-tab-t1} juxtaposes our key findings on the SD and SI versions of SCSS-CQW on even and odd cycles. 
\begin{table}
    \centering
    \renewcommand{\arraystretch}{1.1}
    \begin{tabular}{|c|c|c|c|}
        \hline
        \textbf{Model} & \textbf{Feature} & \textbf{Odd (7) Cycle} & \textbf{Even (8) Cycle} \\
        \hline
        \multirow{5}{*}{\makecell{SI-SCSS-\\CQW\\$(D=1)$}}
        & \makecell{Gap\\closing}
        & \makecell{At $E=0$: at $k=0$,\\ for $\gamma=0$ only.\\ At $E=\pi$: at all $k$,\\ for $\gamma=\pi$ (flat band).}
        & \makecell{As for odd cycles;\\ in addition, at $E=\pi$\\ at $k=\pi$ for every $\gamma$.} \\ \cline{2-4}
        & Flat band
        & \makecell{Yes, at $\gamma=\pi$: both bands\\ at $E=\pi$ ($\hat U_{evo}=-\hat 1$),\\ degenerate (not gapped).}
        & \makecell{Yes, at $\gamma=\pi$: both bands\\ at $E=\pi$ ($\hat U_{evo}=-\hat 1$),\\ degenerate (not gapped).} \\ \cline{2-4}
        & \makecell{Rotational\\flat band}
        & \makecell{Not possible\\ ($k=\pi$ absent).}
        & \makecell{Yes, at $k=\pi$:\\ $E=\pi$ for every $\gamma$.} \\ \cline{2-4}
        & \makecell{Winding\\number}
        & \makecell{Fractional, $\omega=\frac{1}{2}$\\ for $0<\gamma<2\pi$ (one value,\\ no phase transition).}
        & \makecell{Fractional, $\omega=\frac{1}{2}$\\ for $0<\gamma<2\pi$ (one value,\\ no phase transition).} \\ \cline{2-4}
        & Edge states
        & \makecell{Not possible\\ (no winding mismatch).}
        & \makecell{Not possible\\ (no winding mismatch).} \\
        \hline
        \multirow{5}{*}{\makecell{SD-SCSS-\\CQW\\$(D\ge2)$}}
        & \makecell{Gap\\closing}
        & \makecell{At $E=0$: at $k=0$,\\ for $\gamma=2m\pi/D$.\\ At $E=\pi$: at all $k$,\\ for $\gamma=(2n+1)\pi/D$.}
        & \makecell{As for odd cycles;\\ in addition, at $E=\pi$\\ at $k=\pi$ for every $\gamma$.} \\ \cline{2-4}
        & Flat band
        & \makecell{Yes, at the $D$ angles\\ $\gamma=(2n+1)\pi/D$: both bands\\ at $E=\pi$, degenerate.}
        & \makecell{Yes, at the $D$ angles\\ $\gamma=(2n+1)\pi/D$: both bands\\ at $E=\pi$, degenerate.} \\ \cline{2-4}
        & \makecell{Rotational\\flat band}
        & \makecell{Not possible\\ ($k=\pi$ absent).}
        & \makecell{Yes, at $k=\pi$:\\ $E=\pi$ for every $\gamma$\\ (for every $D$).} \\ \cline{2-4}
        & \makecell{Winding\\number}
        & \makecell{Fractional, $\omega=\pm\frac{1}{2}$;\\ $D-1$ phase transitions\\ at $\gamma=2m\pi/D$.}
        & \makecell{Fractional, $\omega=\pm\frac{1}{2}$;\\ $D-1$ phase transitions\\ at $\gamma=2m\pi/D$.} \\ \cline{2-4}
        & Edge states
        & \makecell{Yes: zero-energy states\\ pinned at phase boundary\\ where $\Delta\omega\neq 0$.}
        & \makecell{Yes: zero-energy states\\ pinned at phase boundary\\ where $\Delta\omega\neq 0$.} \\
        \hline
    \end{tabular}
    \caption{Comparison of the spectral and topological features of the step-independent
    (SI-SCSS-CQW, $D=1$) and step-dependent (SD-SCSS-CQW, $D\ge2$) single-coin split-step
    CQW on odd and even cycles. Gaps can close at $E=0$ or at $E=\pi$; here
    $m=0,1,\dots,D-1$ and $n=0,1,\dots,D-1$, and the phase transitions in $0<\gamma<2\pi$ are
    those with $m\ge1$. At the flat-band angles the evolution is trivial
    ($\hat U_{evo}=-\hat 1$) and the winding number is defined only by continuity from
    neighbouring angles.}
    \label{c6-tab-t1}
\end{table}

\section{Resilience of zero energy topological edge states to coin disorder (both dynamic and static) and minor phase-preserving-perturbations}
Here, we numerically investigate the effect of coin (gate) disorder (both dynamic and static), and phase-preserving perturbations on the generated topological edge states generated using the SCSS-CQW evolution, see Fig.~3 of main.

\subsection{Dynamic coin disorder}
\begin{figure*}[!t]
         \centering
         \begin{minipage}{\textwidth}
        \centering
     	\includegraphics[width = 18cm,height=4.5cm]{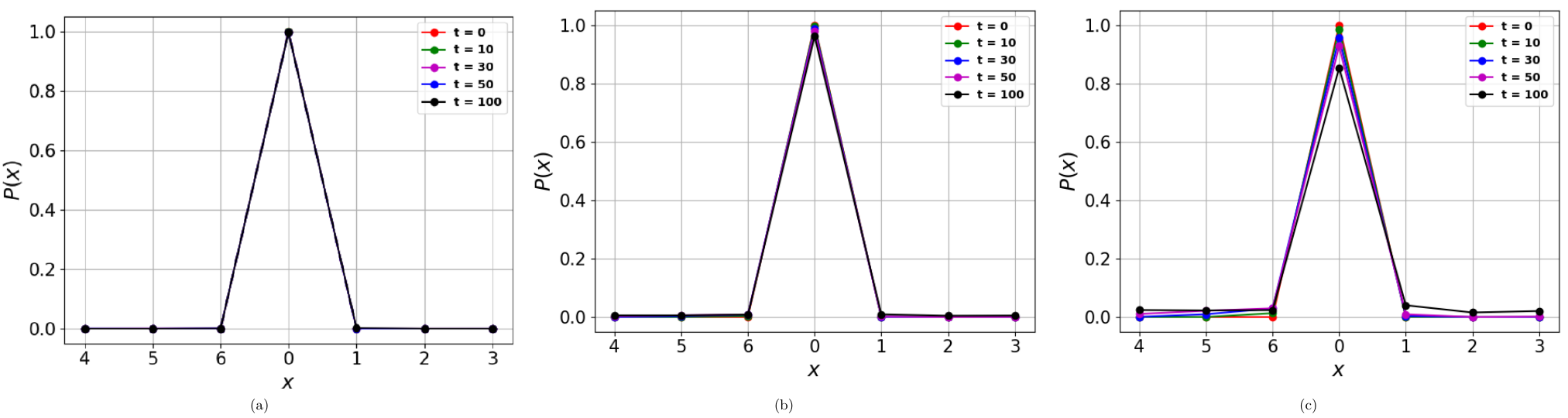}
\caption{\textcolor{black}{(a)~Probability profile ($P(x)$) versus position sites showing an edge-state localized at $x=0$, generated at the junction of two topological phases characterized by winding numbers $\omega=-\tfrac{1}{2}$ (for coin $\gamma=\pi+1.6$) and $\omega=\tfrac{1}{2}$ (for coin $\gamma=0.52\pi$), using SCSS-CQW protocol (with $D=2$) on a 7-site cycle in absence of disorder. 
(b) Effect of dynamic coin disorder with strength $\alpha=0.01$ on the edge state at $x=0$. 
(c) Corresponding behavior for stronger dynamic-coin disorder with strength $\alpha=0.02$. 
In panels (b) and (c), results are obtained by ensemble averaging over $500$ statistically independent disorder instances; dynamic-coin simulations require greater computational effort compared to the static-coin scenario.}}
\label{c6-f-edge7scss-dyncoin}
	\end{minipage}
    
    \begin{minipage}{\textwidth}
        \centering
	\includegraphics[width = 18cm,height=4.5cm]{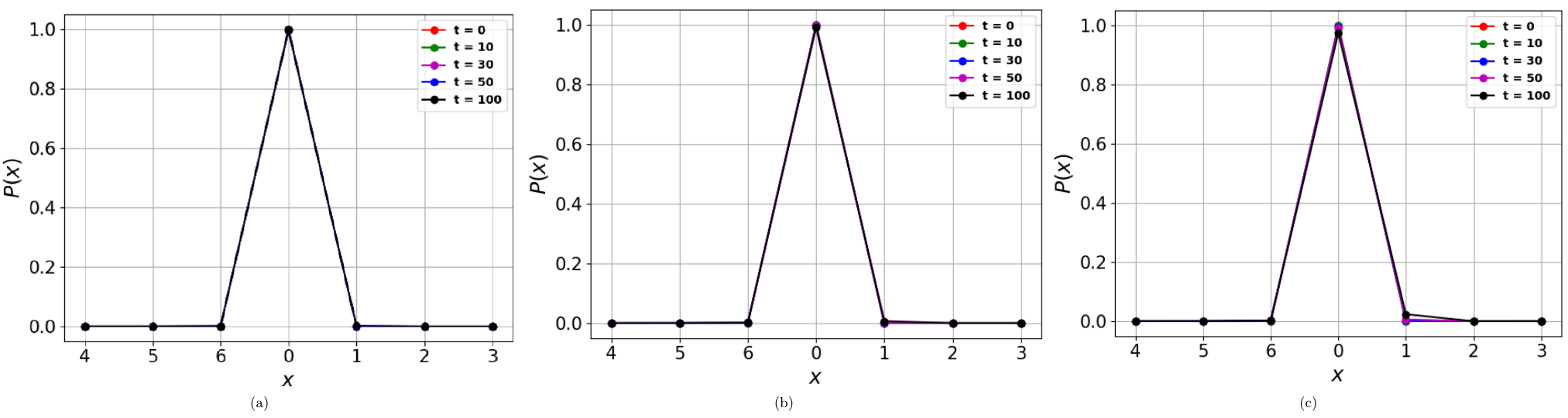}
\caption{\textcolor{black}{(a) The probability profile $P(x)$ versus position showing edge state formation at $x=0$, induced via an interface between 2 topological phases characterized by $\omega=-\tfrac{1}{2}$ (coin with $\gamma=\pi+1.6$) and $\omega=\tfrac{1}{2}$ (coin with $0.52\pi$), for SCSS-CQW protocol (with, $D=2$) on a 7-site cycle in the absence of disorder. 
(b) Effect of static coin disorder with strength $s=0.01$ on the edge state. 
(c) Corresponding behavior for strong static-coin disorder with $s=0.02$. 
In panels (b) and (c), results are obtained for an ensemble of $1000$ independent disorder realizations, and $P(x)$ represents the ensemble-averaged distribution.}}
\label{c6-f-edge7scss-statcoin}
\end{minipage}
\begin{minipage}{\textwidth}
        \centering
	\includegraphics[width = 18cm,height=4.5cm]{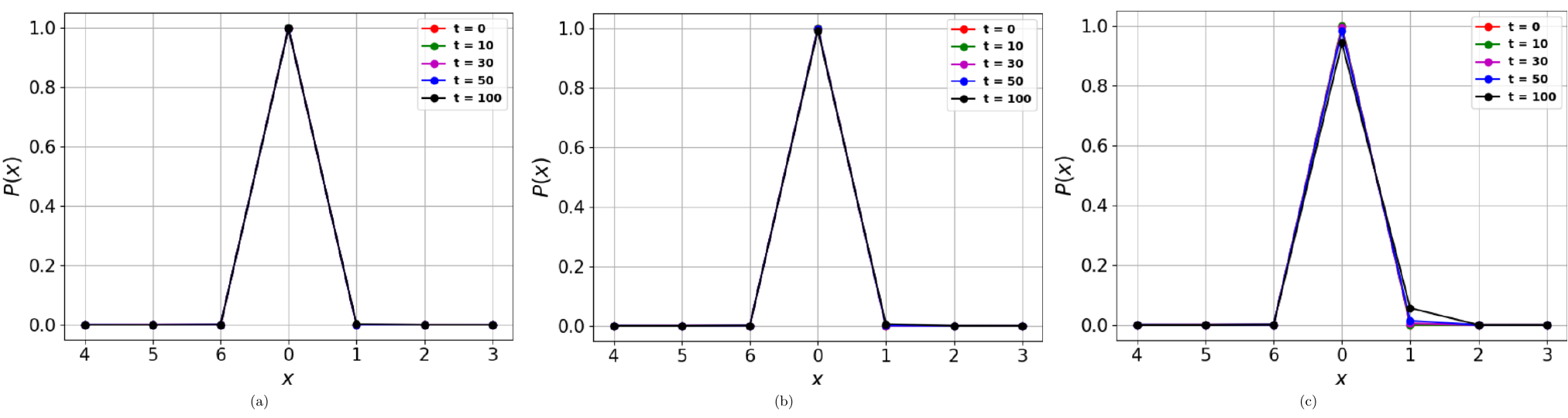}
\caption{\textcolor{black}{(a)~The probability profile $P(x)$ versus position, highlighting the edge state localized at site $x = 0$, appearing at the boundary between two different topological phases ($\omega = -\frac{1}{2}$ for coin with $\gamma=\pi+1.6$ and $\omega = \frac{1}{2}$ for coin with $\gamma= 0.52\pi$), for SCSS-CQW protocol (with, $D=2$) in a $7$-cycle graph. 
(b) Robustness of the topological edge state in presence of phase-preserving perturbations, i.e., edge state at phase boundary between the two topological phases: $\omega = -\frac{1}{2}$ ( $\gamma=\pi+1.61$) and $\omega = \frac{1}{2}$ ($\gamma= 0.52\pi$). 
(c) Persistence of the topological edge state under phase-preserving perturbation, i.e., edge state at the boundary between 2 topological phases: $\omega = -\frac{1}{2}$ (for coin with $\gamma=\pi+1.64$) and $\omega = \frac{1}{2}$ (for coin with $\gamma= 0.52\pi$). In (b) and (c), the coin at site $x=0$ are respectively $\gamma=\pi+1.61$ and $\gamma=\pi+1.64$, which are perturbation to the initial coin $\gamma=\pi+1.6$ used in (a), and all three coins correspond to the same winding number $\omega = -\frac{1}{2}$.}}
\label{c6-f-edge7scss-phasepres}
\end{minipage}
     \end{figure*}

We examine the role of dynamic coin (gate) disorder in the SCSS-CQW dynamics and on the generated zero energy edge states with dynamic disorder of strength \( \alpha\). Here, the site dependent rotation-angle $\gamma$ required to generate topological phases, and topological edge states is modified as~\cite{p5,p6}, $$\gamma \rightarrow \gamma + \alpha \, \delta(t),$$ the random fluctuations (site-independent) $\delta(t) \in [-\pi,\pi]$ are chosen at each time step, with the total number of random values matching the evolution time. For a specific realization, these random values are drawn from a uniform distribution. To evaluate the probability of finding the walker at a given site $x$, i.e., $P(x)$, we average over 500 such realizations. For more technical details on the implementation of dynamic coin disorder in quantum walk dynamics, see Refs.~\cite{p5,p6}. The case $\alpha = 0$ corresponds to a clean system without any dynamic coin disorder. As discussed in Fig.~3 of main, a topological edge state emerges localized at the lattice site $x = 0$, due to a phase boundary created by 2 different coin rotation angles: \textcolor{black}{$\gamma=\pi+1.6$ (applied at site $0$ with winding number: $\omega =-\frac{1}{2}$) and 
$\gamma= 0.52\pi$ (applied at the other sites with winding number: $\omega =\frac{1}{2}$)}, see Fig.~\ref{c6-f-edge7scss-dyncoin}(a), where the probability at position site 0 remains at $0.99743\approx 1$ and persists over time.
Figs.~\ref{c6-f-edge7scss-dyncoin}(b) and (c) illustrate how dynamical coin disorder with degrees of disorder: $\alpha \rightarrow 0.01$, and $\alpha \rightarrow 0.02$, influences the emergent localized edge state (Fig.~\ref{c6-f-edge7scss-dyncoin}(a)). The results indicate that the generated edge states remain stable under weak dynamic-coin disorder within the range $0 < \alpha \lesssim 0.02$.

\subsection{Static coin disorder} 
We now examine the role of static coin disorder~\cite{p6}, in SCSS-CQW dynamics and on the edge states, with the strength of disorder
\( s \). In this case, the site-dependent coin rotation-angle \( \gamma_x \), where the subscript $x$ labels the position sites, responsible for generating the topological phases and topological edge states, is modified as~\cite{p6}, 
\[
\gamma_x \;\to\; \gamma_x + s \, \delta_x.
\]
$\delta_x$ takes random values in \( [-\pi, \pi] \) and is sampled uniformly with the sample size equal to the site number in the cycle graph. Each realization of disorder is taken independently. 
\( s = 0 \) corresponds to a completely disorder-free SCSS-CQW evolution. In Fig.~3 of Main, we discussed the appearance of a zero energy edge-state (topological) at the site: $0$, when 2 different topological phases are interfaced, characterized by the coin rotation angles \textcolor{black}{$\gamma=\pi+1.6$ (applied at site 0 with winding number: $\omega = -\frac{1}{2}$) and 
$\gamma= 0.52\pi$} (applied to all other sites with the winding number, $\omega = \frac{1}{2}$) (see Fig.~\ref{c6-f-edge7scss-statcoin}(a)). From the illustrated edge state in the Fig.~\ref{c6-f-edge7scss-statcoin} (a), the walker's probability at the position site $x=0$ remains $\approx 1$ for a large long number of time-steps $t$ for zero disorder. 
Figs.~\ref{c6-f-edge7scss-statcoin}(b) and (c) illustrate the response of the zero energy edge state to static-coin noise (disorder) with $s=0.01$ and $s=0.02$, respectively, relative to the disorder-free case of Fig.~\ref{c6-f-edge7scss-statcoin}(a). Our numerical analysis confirms the robustness of the corresponding topological feature. Our numerical results indicate the resilience of this topological edge state against small static-coin disorder, since the probability at site 0 does not decrease noticeably with disorder.

\subsection{Phase-preserving perturbations}
Phase-preserving perturbations~\cite{pxue} can be introduced by altering the rotation angles $(\gamma)$ by a small amount that define the topological phases via the protocol: SCSS-CQW dynamics, without affecting the winding number ($\omega$). For instance, $\gamma= \pi+1.6$, $\gamma= \pi+1.61$ and $\gamma= \pi+1.64$, corresponds to the same winding number $\omega =-\frac{1}{2}$. By switching the rotation-angle from $\pi+1.6$ to $\gamma=\pi+1.61$ or $\gamma= \pi+1.64$, the topological invariant, namely the winding number, remains unchanged. In Fig.~3 of the manuscript as well as in Fig.~\ref{c6-f-edge7scss-phasepres}(a) referenced above, we demonstrate that a zero energy edge state arises at the lattice site $x=0$ when a phase boundary is engineered by employing two distinct coin parameters: $\gamma = \pi+1.6$ at the site $0$ (corresponding to the winding number, $-\tfrac{1}{2}$) and $\gamma = 0.52\pi$ at the remaining sites (corresponding to winding number $\tfrac{1}{2}$). This results in a pronounced position probability at site 0 that remains stable over a very long time, indicating a long-lived and stable edge state. In Figs.~\ref{c6-f-edge7scss-phasepres}(b)-(c), the perturbation is introduced by changing the coin rotation-angle at site 0 from $\pi+1.6$ to $\pi+1.61$ and $\pi+1.64$, corresponding to winding number $-\frac{1}{2}$. Interestingly, the observed edge state remains almost unaffected under such perturbations (phase preserving), since the probability (at the site $0$), does not decay significantly or vanish over time.

\section{Temporal Stability of the Edge State}
\subsection{Ideal}
We have demonstrated the presence of a localized edge-state in the SCSS-CQW protocol (see, Fig.~3(c) of main). We now analyze the temporal stability of this edge state by examining the time dependence of the survival probability of the edge site.

At each discrete time step $t$, we extract the probability of edge state, as,
\begin{equation}
P(x=0) \equiv \sum_{g=0,1} \left|\langle x=0,g \vert \psi(t) \rangle \right|^2 ,
\end{equation}
which quantifies the probability of finding the walker at site $x=0$ and for coin state $g$. The behavior of $P(x=0)$ serves as a sensitive diagnostic of edge state localization and transport properties of the system.

\begin{figure}
\includegraphics[width = 8.8cm,height=5.5cm]{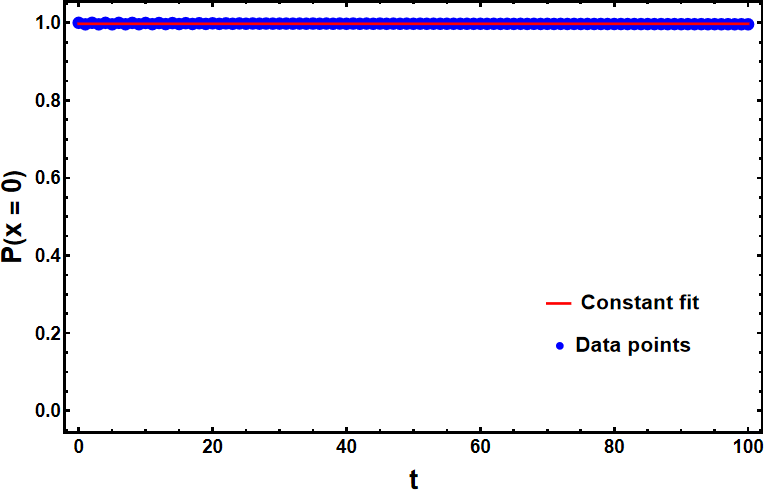}
\caption{\textcolor{black}{Edge-state survival probability $P(x=0)$ vs time-steps $t$ with SCSS-CQW protocol ($D=2$) with coins $\gamma=\pi+1.6$ (at $x=0$, $\omega=\frac{-1}{2}$) and $\gamma=0.52\pi$ (at $x\ne0$, $\omega=\frac{1}{2}$) on a $7$-cycle graph, with disorder free evolution. The numerical data exhibit a constant value $\approx0.99743$, and are well captured by a constant fit (red) with RMSE $\approx 0.0012$, demonstrating a robust edge state.}}
\label{lsq-clean}
\end{figure}

We plot in Fig.~\ref{lsq-clean}, $P(x=0)$ as a function of time. A rapid decay of this quantity would indicate delocalization or leakage of the walker away from the edge, whereas saturation to a finite value would signal robust localization. This plot therefore provides a direct visualization of the persistence and robustness of the zero energy edge state.
Indeed, in Fig.~\ref{lsq-clean}, we see a saturation of $P(x=0)$ at \textcolor{black}{0.99743.} The numerical data is well captured by a constant fit with the mean value of the data.
The constant fit reproduces the numerical data with a root-mean-square error (RMSE) \textcolor{black}{$\approx 0.0012$}, demonstrating excellent agreement over the considered time window.

The absence of any systematic decay term in the data-driven fit confirms that the edge-state survival probability does not deviate considerably from the finite value \textcolor{black}{($\sim 1$)} and remains dynamically stable up to long simulated times. This provides strong evidence that the observed edge state is not transient but represents a robust, long-lived localized mode for ideal SCSS-CQW dynamics.

\begin{figure*}[!t]
         \centering
         \begin{minipage}{\textwidth}
        \centering
     	\includegraphics[width = 17.5cm,height=5cm]{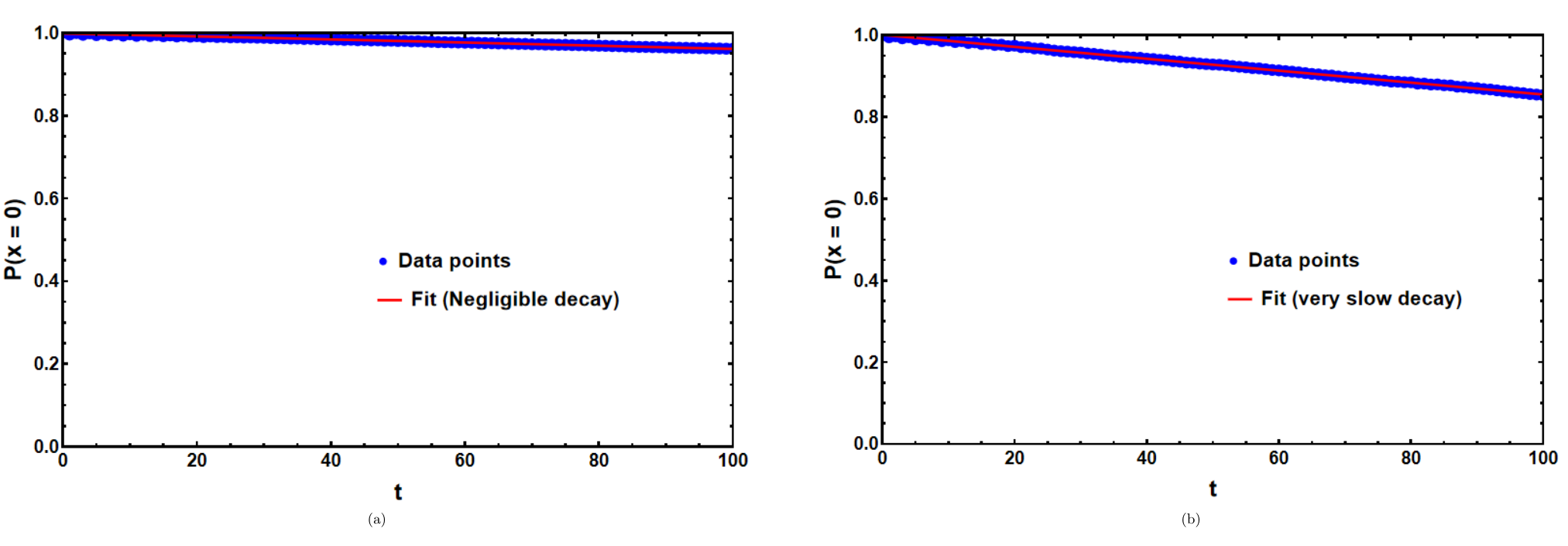}
\caption{\textcolor{black}{Edge-state survival probability $P(x=0)$ vs time-step $t$ using SCSS-CQW protocol (with, $D=2$) with coins $\gamma=\pi+1.6$ (at site 0 with  $\omega=\frac{-1}{2}$) and $\gamma=0.52\pi$ (at remaining sites with $\omega=\frac{1}{2}$) on a $7$-cycle graph, under (a) small \textbf{dynamical disorder} of strength $\alpha=0.01$. The numerical data exhibit a linear fit (red) with very slow decay, demonstrating long-time stability of the localized edge state. (b) under large \textbf{dynamical disorder} of disorder strength: $\alpha=0.02$. The numerical data exhibit a quadratic fit (red) with slow decay, demonstrating long-time stability of the localized edge state.}}
\label{lsq-dyn}
	\end{minipage}
    
    \begin{minipage}{\textwidth}
        \centering
	\includegraphics[width = 17.5cm,height=5cm]{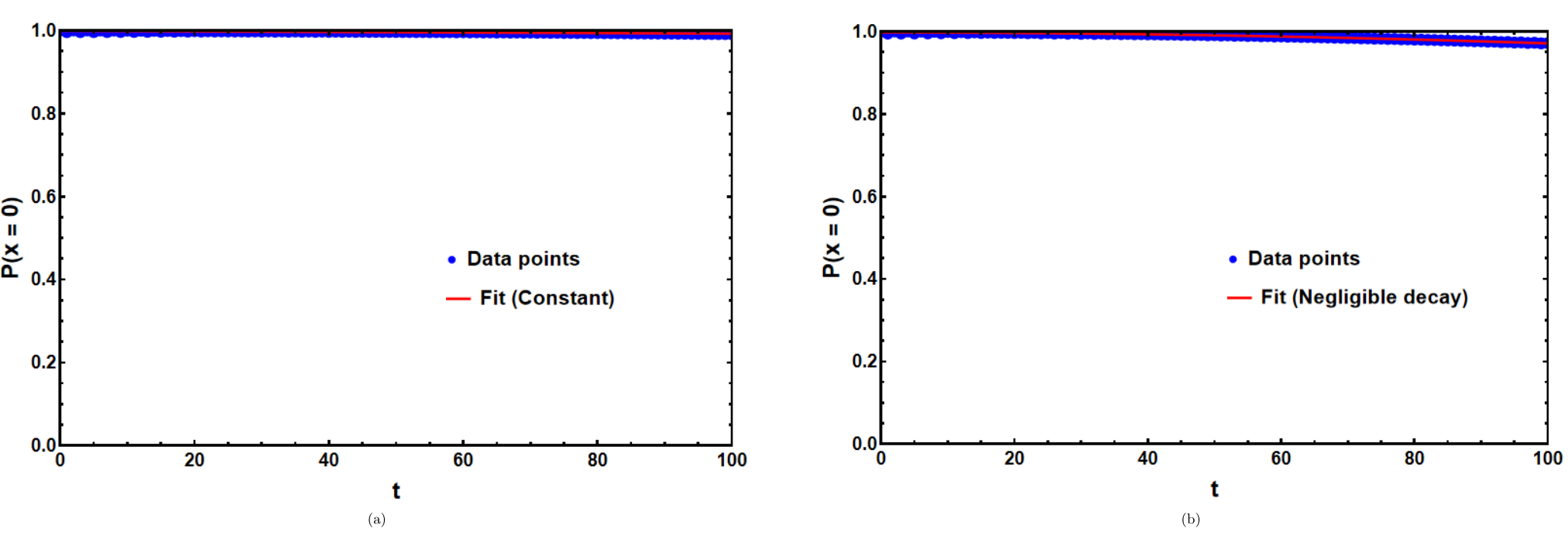}
\caption{\textcolor{black}{Edge-state survival probability $P(0)$ as a function of the time step $t$, using SCSS-CQW protocol ($D=2$) with coin $\gamma=\pi+1.6$ (at $x =0$ with $\omega=\frac{-1}{2}$) and $\gamma=0.52\pi$ (at $x\ne0$ with $\omega=\frac{1}{2}$) on a $7$-cycle graph, in the presence of (a) small \textbf{static-coin disorder} with strength of disorder, $s=0.01$. This numerical data is fitted with a linear fit (red) showing no decay, demonstrating long-time stability of the localized edge state. (b) strong\textbf{ static-coin disorder} with strength $s = 0.02$. This numerical data is fitted with a linear fit (red) and exhibits a very slow negligible decay, which demonstrates long-time stability of the localized edge state.}}
\label{lsq-stat}
\end{minipage}
\begin{minipage}{\textwidth}
        \centering
	\includegraphics[width = 17.5cm,height=5cm]{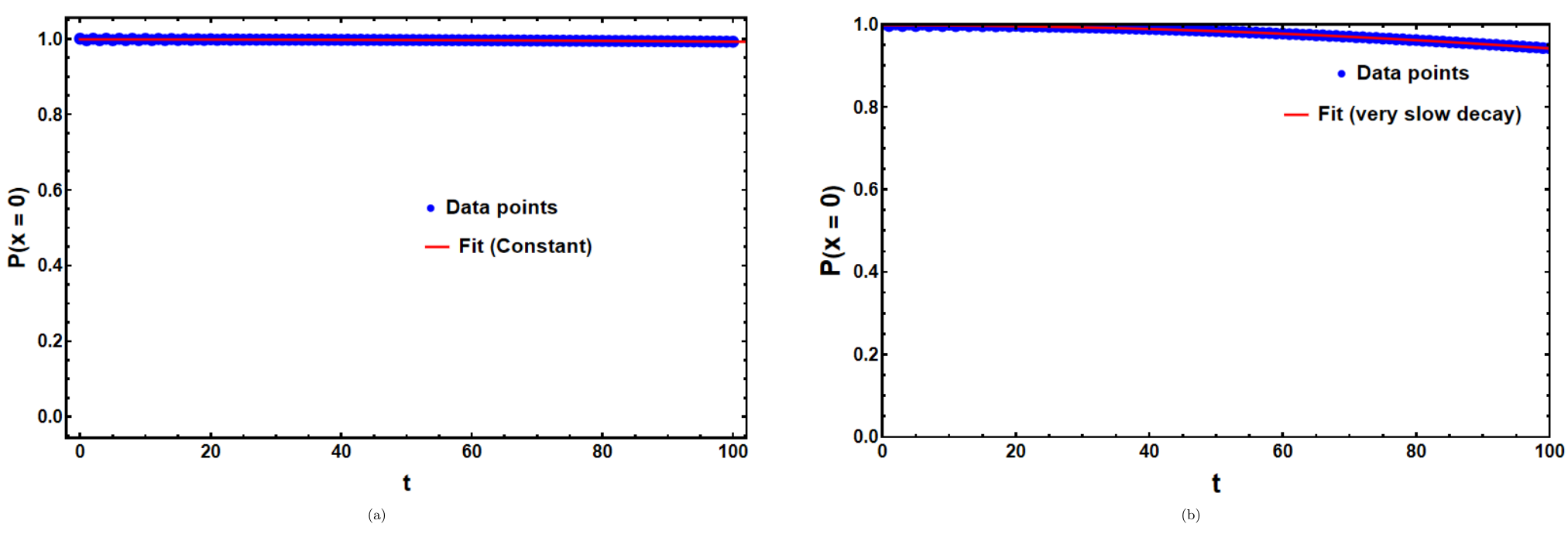}
\caption{\textcolor{black}{(a) Edge-state survival probability $P(x=0)$ vs time-step $t$ using SCSS-CQW protocol (with, $D=2$) with coin $\gamma=\pi+1.61$ (at site 0 with $\omega=\frac{-1}{2}$) and coin $\gamma=0.52\pi$ (at remaining sites with $\omega=\frac{1}{2}$)) on $7$-cycle graph, under phase-preserving perturbations as in Fig.~\ref{c6-f-edge7scss-phasepres}(b). The numerical data is fitted with constant  fit (red), demonstrating long-time stability of the localized edge state. (b) Edge-state survival probability $P(x=0)$ vs time-step $t$ using SCSS-CQW protocol (with, $D=2$) with coin $\gamma=\pi+1.64$ (at site 0 with $\omega=\frac{-1}{2}$) and coin $\gamma=0.52\pi$ (at remaining sites with $\omega=\frac{1}{2}$) on $7$-cycle graph, under phase-preserving perturbations as in Fig.~\ref{c6-f-edge7scss-phasepres}(c). The numerical data is fitted with a cubic polynomial fit (red), demonstrating very slow negligible decay (long-time stability of the localized edge state).}}
\label{lsq-Phase}
\end{minipage}
     \end{figure*}

In  Sec.~II, we studied the resilience of emergent edge states against dynamic coin disorder, then static coin disorder (see, Figs.~\ref{c6-f-edge7scss-dyncoin}, \ref{c6-f-edge7scss-statcoin}), and finally the phase-preserving perturbations (see, Figs.~\ref{c6-f-edge7scss-phasepres}). Below, we provide the time dependence of the edge state under disorder with the same strength of disorder taken in Sec.~II A--C and Figs.~\ref{c6-f-edge7scss-dyncoin}, \ref{c6-f-edge7scss-statcoin} and \ref{c6-f-edge7scss-phasepres}. We utilize Mathematica's built-in \texttt{FindFormula} and \texttt{FindFit} to analyze the behavior of the edge-state survival probability under small to moderate level of disorder.
\subsection{Dynamic disorder}
Under small dynamic-coin disorder of strength $\alpha=0.01$, we analyse the decay of the edge-site survival probability $P(x=0)$, as shown in Fig.~\ref{lsq-dyn}(a). The decay of the edge-state probability is very slow, indicating a high degree of robustness against weak temporal fluctuations in the coin parameters. Applying \texttt{FindFormula} to the data, we find a linear function \textcolor{black}{$0.998558 - 0.00037 t$} fits the data with RMSE \textcolor{black}{$0.0008$}, demonstrating excellent agreement over the entire time window. We conclude that under small dynamical disorder, the observed edge state is robust and long-lived, exhibiting very slow negligible decay.

We repeat the same analysis for a stronger dynamical coin disorder of strength $\alpha=0.02$, shown in Fig.~\ref{lsq-dyn}(b). In this case, applying \texttt{FindFormula} and \texttt{FindFit} to the data, we find the quadratic function, \textcolor{black}{$0.999204-0.0013t-3.09\times 10^{-6}t^2$} fits the data with RMSE \textcolor{black}{$0.0011$}, demonstrating excellent agreement over the entire time window. In this case, the edge-state survival probability shows very slow decay.

\subsection{Static disorder}

In Sec.~II B, we have demonstrated that the generated edge states remain robust under static coin disorder (see Fig.~\ref{c6-f-edge7scss-statcoin}). Here, we analyze the explicit time dependence of the edge-state survival probability for the same disorder strengths considered in Sec.~II. In particular, we study the behavior of the edge-site probability $P(x=0)$ under static disorder using Mathematica’s built-in \texttt{FindFormula} and \texttt{FindFit} to identify effective functional descriptions of the numerical data.

Under small static-coin disorder with degree of disorder $s=0.01$, the resulting edge-site survival probability $P(x=0)$ exhibits no decay over the entire simulated time window, as shown in Fig.~\ref{lsq-stat}(a). Applying \texttt{FindFormula} and \texttt{FindFit} to the data, we find a linear fit, \textcolor{black}{$0.999067 - 0.00007t$} provides the best description, with an RMSE of \textcolor{black}{$\approx 0.0008$}. This confirms that the edge state remains dynamically stable for an appreciable amount of time and exhibits no decay, under small static disorder.

We repeat the same analysis for a stronger static disorder of coin-disorder strength: $s=0.02$, shown in Fig.~\ref{lsq-stat}(b). In this case, the edge-site survival probability $P(x=0)$ exhibits again a negligible decay over the simulated time window compared to the small static disorder case, as shown in Fig.~\ref{lsq-stat}(b). Applying \texttt{FindFormula} and \texttt{FindFit} to the data, we again find that a linear fit, \textcolor{black}{$0.997625 - 0.000016 t$} provides the best description, with an RMSE of \textcolor{black}{$\approx 0.0024$}. These results demonstrate that the zero energy edge state generated in the SCSS-CQW protocol is robust against static coin disorder and remains totally unaffected over long evolution times.

\subsection{Phase-preserving perturbations}

We plot in Figs.~\ref{lsq-Phase}(a) and \ref{lsq-Phase}(b) the time dependence of the edge-site survival probability $P(x=0)$ under two representative phase-preserving perturbations considered in Figs.~\ref{c6-f-edge7scss-phasepres}(b) and \ref{c6-f-edge7scss-phasepres}(c), respectively. As in the previous cases, we analyze the long-time behavior of the edge state by fitting the numerical data using Mathematica’s built-in \texttt{FindFormula} and \texttt{FindFit}. In the plot in Figs.~\ref{lsq-Phase}(a) and \ref{lsq-Phase}(b), the time dependence of the edge-site survival probability $P(x=0)$ can be fitted with a constant fit, \textcolor{black}{$0.999046 - 0.000059t$} (with RMSE $\approx
 0.007$) and a cubic polynomial fit, \textcolor{black}{$0.998162  -8.08324\times10^{-6}t-5.74\times 10^{-6}t^2 + 1.841\times 10^{-9}t^3$ (with RMSE $\approx
0.00077$) respectively}. 
 In the case of Fig.~\ref{lsq-Phase}(a), the survival probability remains finite and exhibits no decay, indicating robustness of the edge state against phase-preserving perturbations, while in Fig.~\ref{lsq-Phase}(b) it exhibits a very slow decay.
In both cases, the edge states shows robustness against phase-preserving perturbations for quite a long time.

\begin{figure*}[!t]
         \centering
         \begin{minipage}{\textwidth}
        \centering
     	\includegraphics[width = 18cm,height=4.3cm]{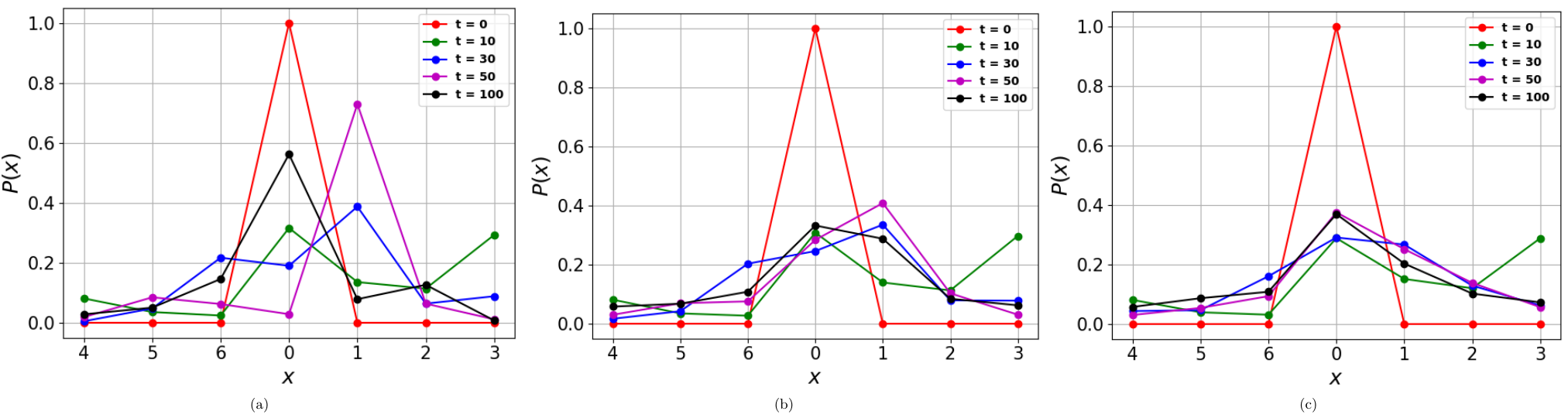}
\caption{\textcolor{black}{(a) The probability profile ($P(x)$) versus position showing a localized or point defect state, generated without a phase boundary using coin $\gamma=\frac{\pi}{3}$ ($\omega=\frac{1}{2}$) at site $x=0$ and coin $\gamma=\tfrac{3\pi}{5}$ ($\omega=\frac{1}{2}$) at other sites, via the SCSS-CQW protocol (for $D=2$) on a 7-cycle graph in absence of disorder. 
(b) Effect of static coin disorder with strength $s=0.01$ on the defect state; see SM Sec.~II above and Ref.~\cite{p5} for more details on how to incorporate different types of disorder into a cyclic quantum walk evolution. 
(c) Corresponding behavior for stronger static-coin disorder $s=0.02$. 
In panels (b) and (c), results are obtained from an ensemble of $1000$ independent disorder realizations, and $P(x)$ represents the ensemble-averaged distribution.}}
\label{point-defect-state}
	\end{minipage}
    
    \begin{minipage}{\textwidth}
        \centering
	\includegraphics[width = 18cm,height=4.3cm]{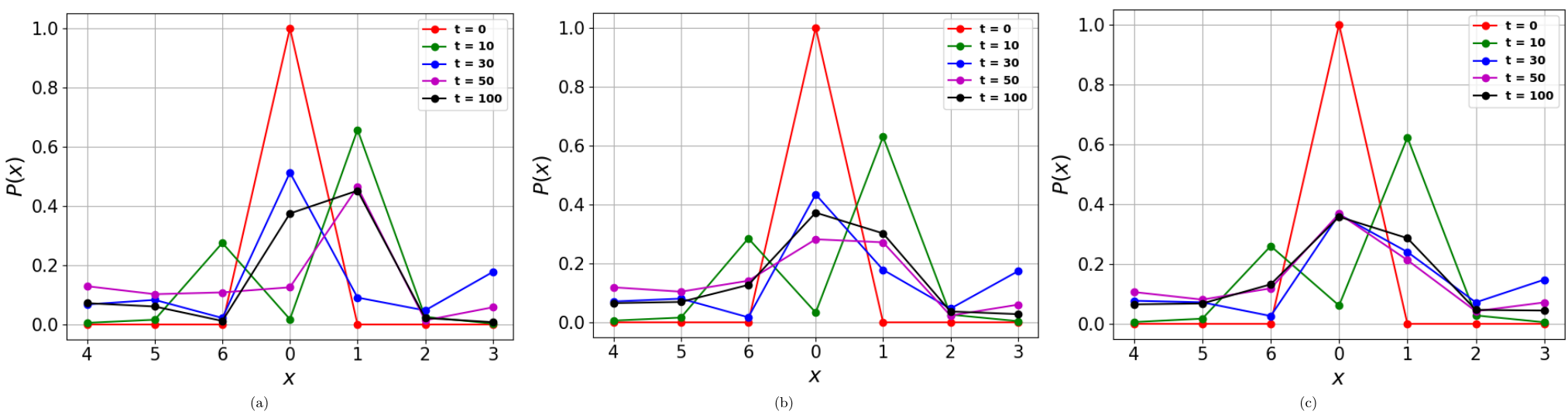}
\caption{\textcolor{black}{(a) The probability profile ($P(x)$) versus position showing a localized or point defect state, generated with a phase boundary using coin $\gamma=\pi+0.2$ ($\omega=\frac{-1}{2}$) at site $x=0$ and coin $\gamma=\tfrac{3\pi}{4}$ ($\omega=\frac{1}{2}$) at other sites, via the SCSS-CQW protocol (for $D=2$) on a 7-cycle graph in absence of disorder. 
(b) Effect of static coin disorder with strength $s=0.01$ on the defect state; see SM Sec.~II above and Ref.~\cite{p5} for more details on how to incorporate different types of disorder into a cyclic quantum walk evolution. 
(c) Corresponding behavior for stronger static-coin disorder $s=0.02$. 
In panels (b) and (c), results are obtained from an ensemble of $1000$ independent disorder realizations, and $P(x)$ represents the ensemble-averaged distribution.}}
\label{point-defect-state-phaseboundary}
\end{minipage}
\begin{minipage}{\textwidth}
        \centering
	\includegraphics[width = 18cm,height=5cm]{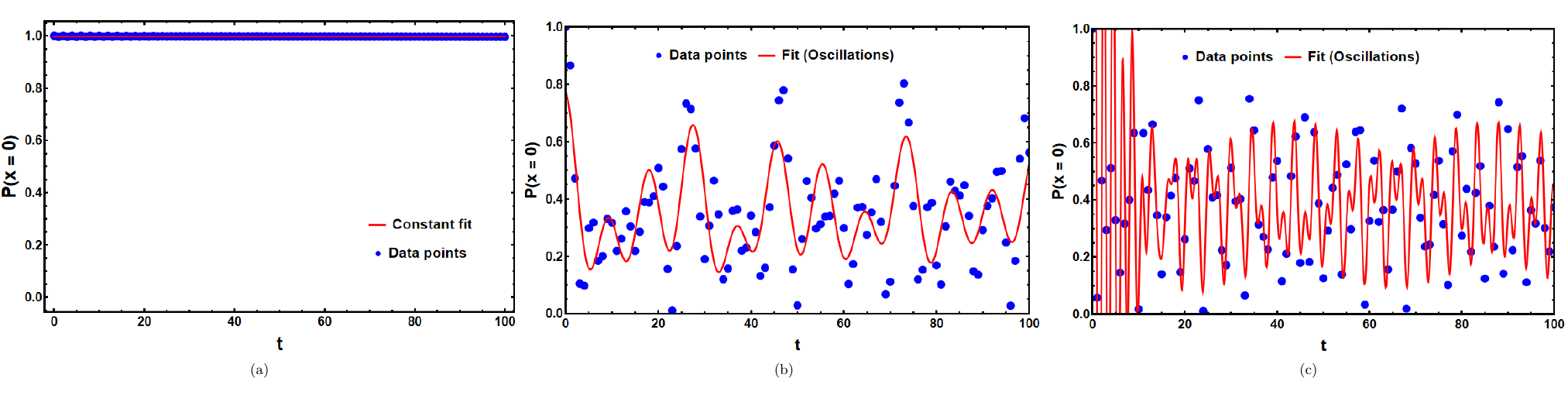}
\caption{\textcolor{black}{(a) Edge-state survival probability $P(x=0)$ vs time-step $t$ using SCSS-CQW protocol (with, $D=2$) using a coin $\gamma=\pi+1.6$ ($\omega=\frac{-1}{2}$) at site 0 and coin with $\gamma=0.52\pi$ ($\omega=\frac{1}{2}$) at sites $x\ne0$, on a $7$-cycle graph, without disorder. The numerical data fit exhibits constant probability ($0.99743$) (red), demonstrating long-time stability of the localized zero energy edge state. (b) Point-defect state survival probability $P(x=0)$ vs time-step $t$ using SCSS-CQW protocol ($D=2$) generated without phase boundary using coin $\gamma=\frac{\pi}{3}$ ($\omega=\frac{1}{2}$) at site 0 and coin $\gamma=\tfrac{3\pi}{5}$ ($\omega=\frac{1}{2}$) at remaining sites $x\ne0$, on a $7$-cycle graph, without disorder. $P(x = 0)$ for the point-defect state shows rapid, large-amplitude oscillations, unlike the stable localized edge state seen in (a). (c) Point-defect state survival probability $P(x=0)$ vs time-step $t$ using SCSS-CQW protocol ($D=2$) generated with phase boundary using coin $\gamma=\pi+0.2$ ($\omega=\frac{-1}{2}$) at site 0 and coin $\gamma=\tfrac{3\pi}{4}$ ($\omega=\frac{1}{2}$) at remaining sites $x\ne0$, on a $7$-cycle graph, without disorder. $P(x = 0)$ for the point-defect state shows rapid, large-amplitude oscillations, unlike the stable localized edge state seen in (a).}}
\label{lsq-clean-defect-state}
\end{minipage}
     \end{figure*}

\section{Comparing topological edge states and localized states}

\subsection{Probability Analysis}
\textcolor{black}{The creation of a phase boundary to generate an edge state requires two distinct coins corresponding to distinct topological winding numbers~\cite{p6,pxue}. The use of two distinct coins is needed to generate a point defect state~\cite{ptdef-1,ptdef-2,ptdef-3} too. A point-defect state, which gives rise to finite energy localized states (see Fig. 6 of Main), can be generated either with two distinct coins forming a topological phase boundary or without any topological phase boundary at the defect site, but in either case, it exhibits substantial spreading into the bulk. A zero energy edge state is not so; it can only be generated with two distinct coins forming a topological phase boundary and exhibits no spreading into the bulk. Below, we systematically analyze both point-defect states and zero energy edge states under disorder to expose the striking differences in their responses, see also Table~\ref{tab-defect-vs-edge}. This analysis dispels any confusion between the edge states generated by our protocol and the point-defect states that may likewise emerge from it.} 

\begin{figure*}[h]
\centering
\includegraphics[width = 18cm,height=5.5cm]{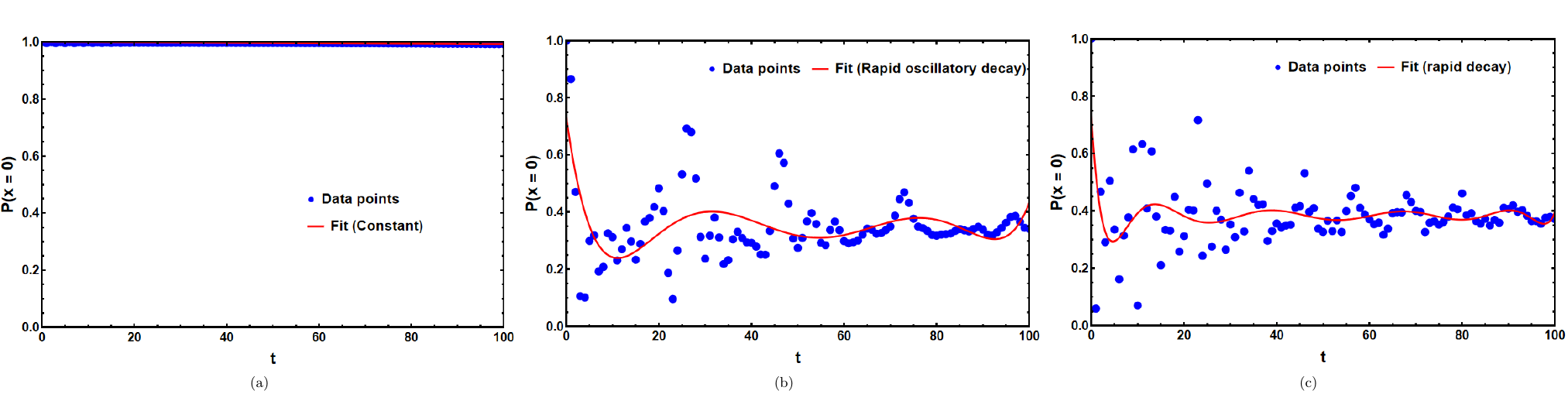}
\caption{\textcolor{black}{(a) Edge-state survival probability $P(0)$ as a function of the time step $t$, using SCSS-CQW protocol ($D=2$) with coin $\gamma=\pi+1.6$ (at site 0 with $\omega=\frac{-1}{2}$) and coin $\gamma=0.52\pi$ (at remaining sites with $\omega=\frac{1}{2}$) on $7$-cycle graph, in presence of \textbf{static-coin disorder} of strength $s=0.01$. The numerical data is fitted with a polynomial fit (red) which depicts a constant fit, that is, negligible decay, demonstrating long-time stability of the localized edge state. (b) Point-defect state survival probability $P(0)$ as a function of the time step $t$, using SCSS-CQW ($D=2$) with coin $\gamma=\frac{\pi}{3}$ (at $x=0$ with $\omega=\frac{1}{2}$) and coin $\gamma=\tfrac{3\pi}{5}$ (at $x\ne0$ with $\omega=\frac{1}{2}$), generated without phase boundary on $7$-cycle graph, in presence of \textbf{static-coin disorder} with strength, $s=0.01$. The numerical data for the point-defect state under static disorder shows a rapid oscillatory decay, unlike the very slow decay seen for an edge state. (c) Point-defect state survival probability $P(0)$ as a function of the time step $t$, using SCSS-CQW (with, $D=2$) using coin $\gamma=\pi+0.2$ (at site 0 with $\omega=\frac{-1}{2}$) and coin $\gamma=\tfrac{3\pi}{4}$ (at remaining sites with $\omega=\frac{1}{2}$), generated with phase boundary on $7$-cycle graph, in presence of \textbf{static-coin disorder} with strength, $s=0.01$. The numerical data for the point-defect state under static disorder shows a rapid oscillatory decay, unlike the very slow decay seen for a zero energy edge state.}}
\label{lsq-stat0025-defect-state}
\end{figure*}

\textcolor{black}{The topological edge states demonstrated in Main and Secs.~I-II of Supplementary Material (SM) are different from point-defect states~\cite{ptdef-1,ptdef-2,ptdef-3} seen in Figs.~\ref{point-defect-state}(a)-(c), \ref{point-defect-state-phaseboundary}(a)-(c), \ref{lsq-clean-defect-state}(b)-(c), \ref{lsq-stat0025-defect-state}(b)-(c). A point-defect state has large spreading into the bulk, while edge states are localized at the topological phase boundary (see, Fig.~5 (b) -(c) of main). For instance, see Fig.~\ref{point-defect-state}(a) where a point-defect state is created at site $x=0$ using coin $\gamma=\frac{\pi}{3}$ at site 0 and coin $\gamma=\frac{3\pi}{5}$ at other sites, both with $\omega=\frac{1}{2}$, i.e., without a phase boundary. For this point-defect state shown in Fig.~\ref{point-defect-state}(a), there is a large spreading into the bulk sites (i.e., $x\ne 0$) both with and without disorder, see Fig.~\ref{point-defect-state}(a)-(c). Further, the point defect state is not stable, even in the absence of disorder: in Fig.~\ref{lsq-clean-defect-state}(b), the  survival probability at site $0$ shows rapid, large-amplitude oscillations about a mean of $0.35$, unlike that of the stable edge state generated via a topological phase boundary, as in Fig.~\ref{lsq-clean-defect-state}(a).}


\textcolor{black}{Further, see Fig.~\ref{point-defect-state-phaseboundary}(a) where a point-defect state is created with a phase boundary, at site $x=0$ using coin $\gamma=\pi+0.2$ (with $\omega=\frac{-1}{2}$) at site 0 and coin $\gamma=\frac{3\pi}{4}$ (with $\omega=\frac{1}{2}$) at other sites. For this point-defect state (Fig.~\ref{point-defect-state-phaseboundary}(a)) again there is a large spreading into the bulk (i.e., $x\ne 0$) both with and without disorder, see Fig.~\ref{point-defect-state-phaseboundary}(a)-(c). Further, the point defect state is not stable, even in the absence of disorder: in Fig.~\ref{lsq-clean-defect-state}(c),  the  survival probability at site $0$ shows rapid, large-amplitude  oscillations about a mean of $0.38$, unlike that of the stable edge state generated via a topological phase boundary, as in Fig.~\ref{lsq-clean-defect-state}(a).} 

\textcolor{black}{An edge state is generated via the creation of a topological phase boundary, i.e., two distinct winding numbers are seen at site 0 and the remaining sites, and is characterised by a pronounced probability amplitude at site $x=0$~\cite{kita-pra,pxue,p6} with almost no spreading into the bulk ($x\ne 0$) and is stable over long times
($t$), see Figs. 3(c) and 5(a) of Main, where the edge-state survival probability does not deviate considerably from 1 $\approx$
0.99743 and remains dynamically stable up to long times, see also Fig.~\ref{lsq-clean} and Fig.~\ref{lsq-clean-defect-state}(a) above. Zero energy edge states generated via the SCSS-CQW protocol are not point defect states, due to the fact that they occur at a topological phase boundary and are stable at site 0, over long times, with negligible spreading into the bulk. Point-defect states are finite energy localized states (see Fig. 6 of Main) and occur irrespective of a phase boundary.}

\textcolor{black}{Furthermore, the numerical data for the point-defect state without phase boundary under static disorder shows a rapid oscillatory decay, see Fig.~\ref{lsq-stat0025-defect-state}(b), approximated by the function, $0.733125-0.116588t+0.009562t^2-0.000333t^3+5.657\times10^{-6}t^4 -4.607\times10^{-8}t^5+1.4439\times10^{-10}t^6$ (RMSE $\approx 0.108$), while the point-defect state with phase boundary under static disorder shows a rapid oscillatory decay and stabilizes to a constant value after the initial decay, see Fig.~\ref{lsq-stat0025-defect-state}(c), its the numerical data is approximated by the function, $0.72956-0.25521t+0.053639t^2-0.00519t^3 +0.000278t^4-9.01216\times10^{-6}t^5+1.8384\times10^{-7}t^6$ (RMSE $\approx 0.1$). On the other hand, an edge state under static disorder exhibits very slow decay, as seen in Fig.~\ref{lsq-stat0025-defect-state}(a) (also see Fig.~\ref{lsq-stat}(a) above). This implies that zero energy edge states show superior resilience against disorder than trivial point defect states. The above-discussed comparative analysis of edge states versus point defect states is juxtaposed in Table~\ref{tab-defect-vs-edge}.}

\subsection{Quasienergy Analysis}
\begin{figure*}[h]
    \centering
    
    \begin{subfigure}[b]{0.32\textwidth}
        \centering
        \includegraphics[width=\textwidth]{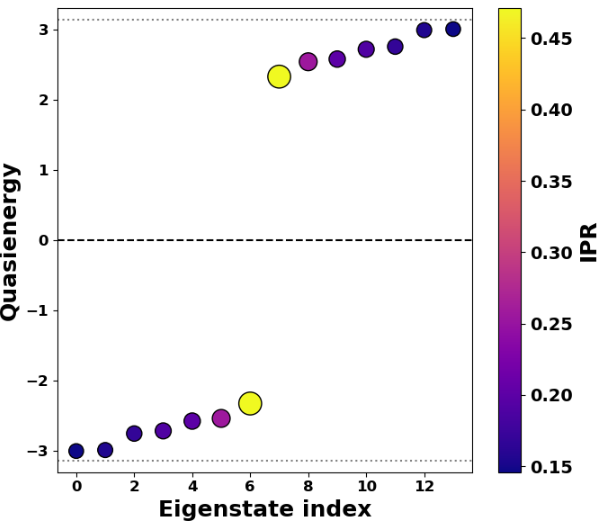}   
        \caption{}
        \label{ipra}
    \end{subfigure}
    \hfill 
    \begin{subfigure}[b]{0.32\textwidth}
        \centering
        \includegraphics[width=\textwidth]{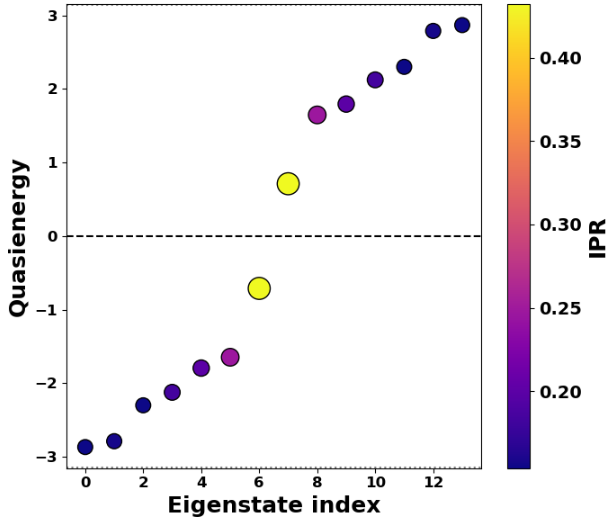}
        \caption{}
        \label{iprb}
    \end{subfigure}
    \hfill 
    \begin{subfigure}[b]{0.32\textwidth}
        \centering
        \includegraphics[width=\textwidth]{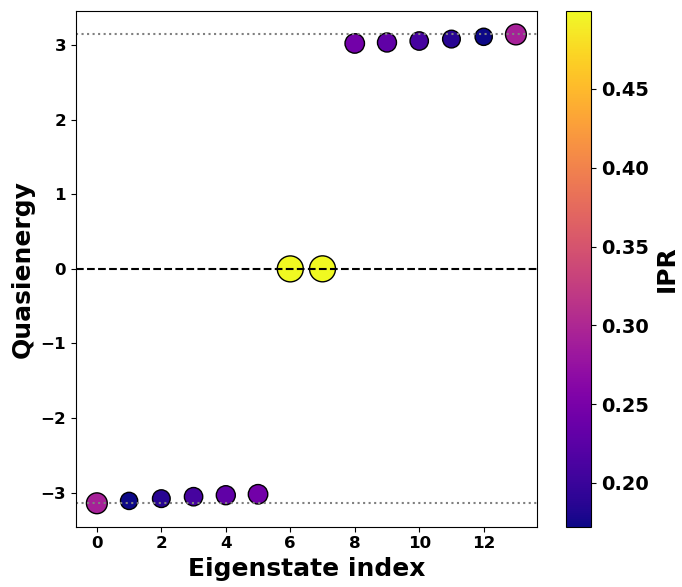}
        \caption{}
        \label{iprc}
    \end{subfigure}

    \caption{\textcolor{black}{Quasienergy spectrum and inverse participation ratio (IPR) for point-defect and topological edge states in the SCSS-CQW protocol on a 7-cycle graph. The quasienergy eigenvalues are obtained by diagonalizing the evolution operator, while the color of each eigenstate represents the corresponding inverse participation ratio (IPR), with larger IPR indicating stronger spatial localization. (a) Localized state generated without a topological phase boundary ($\Delta\omega=0$) using $\gamma=\pi/3$ at $x=0$ and $\gamma=3\pi/5$ elsewhere.  (b) Localized state generated with a topological phase boundary ($\Delta\omega\neq0$) using $\gamma=\pi+0.2$ at $x=0$ and $\gamma=3\pi/4$ at the remaining sites. (c) Topological edge-state configuration generated using $\gamma=\pi+1.6$ at $x=0$ and $\gamma=0.52\pi$ elsewhere.}}
    \label{ipr}
\end{figure*}

\textcolor{black}{To distinguish topological edge states from trivial point-defect states, we analyze the quasienergy spectrum together with the inverse participation ratio (IPR), as shown in Fig.~\ref{ipr}. To characterize the localized eigenstates, we diagonalize the evolution operator $\hat{U}_{evo}$ (see, Eq. (6)). For an $N$-site SCSS-CQW, the site-dependent coin operator as defined in Eq. \ref{cgamma} is, \begin{equation} C=\bigoplus_{x=0}^{N-1} C(\gamma_x),\end{equation} which is a $2N \times 2N$ block diagonal matrix. The evolution operator $\hat{U}_{evo} = S_+CS_-C$, where $S_+$ and $S_-$ are the shift operators defined in Eqs. \ref{s+} and \ref{s-} of the SCSS-CQW is diagonalized to obtain its eigenvalues and eigenvectors,\begin{equation}
    \hat{U}_{evo}\ket{\psi_{i}} = e^{-iE_i}\ket{\psi_{i}},
\end{equation}where $E_i$ denotes the quasienergy of the $i$th eigenstate, where $i \in {1,2...2N}$. Since the eigenvalues of the unitary operator lie on the complex unit circle, the quasienergies are obtained from the phases of the eigenvalues according to, \begin{equation}
    E_i = -\arg(\lambda_i),
\end{equation}where $\lambda_i$ is the corresponding eigenvalue. For the present SCSS-CQW implemented on a 7-cycle graph, the Hilbert space is the tensor product of the two-dimensional coin space and the seven-dimensional position space, resulting in a total Hilbert-space dimension of 14. Consequently, the evolution operator is represented by a $14\times14$ unitary matrix, whose diagonalization yields fourteen quasienergy eigenvalues and their corresponding eigenstates. The probability distribution associated with each site is obtained from the normalized eigenvector, \begin{equation}
    P(x)  = |a_{x, 0_c}|^2+|a_{x,1_c}|^2,
\end{equation} and its degree of localization is quantified through the inverse participation ratio (IPR), defined as,\begin{equation}
    IPR = \sum_x |P(x)|^2.
\end{equation} In periodically driven quantum walks, topological edge states are expected to appear as isolated quasienergy modes inside the bulk spectral gap, typically at quasienergy $E=0$~\cite{asboth}, and are characterized by strong spatial localization, reflected by a large IPR. By contrast, defect-induced localized states may also possess relatively large IPR values but generally do not remain pinned to the topological gap throughout parameter variations and therefore lack topological protection.}
  
\textcolor{black}{For the localized state generated without a topological phase boundary [$\Delta\omega=0$, Fig.~\ref{ipra}], two localized eigenstates with the largest IPR ($\approx0.47$) appear around quasienergies $\pm2.35$, well away from the topological gap at $E=0$. These states coexist with the bulk bands and are therefore identified as trivial localized modes rather than protected zero energy edge states. A similar behavior is observed for the states generated with a phase boundary ($\Delta\omega\neq0$, Fig.~\ref{iprb}), where the localized modes occur near quasienergies $\pm0.7$. Although these states exhibit enhanced localization (large IPR), they are displaced away from the symmetry-protected zero quasienergy centre and, as demonstrated by the real-time dynamics, undergo significant spreading into the bulk and show rapid, large-amplitude oscillations (see, Figs. 20-22). }

\textcolor{black}{In contrast, the topological edge-state (Fig.~\ref{iprc}) exhibits two highly localized eigenstates with the largest IPR ($\approx0.49$) pinned at quasienergy $E=0$, exactly at the centre of the bulk quasienergy gap. Such gap-localized modes are the defining spectral signature of topological edge states predicted by bulk-boundary correspondence~\cite{kita-pra,asboth}. Their localization at the interface between regions of different winding numbers, together with their pinning to the gap center and their long-time dynamical stability, clearly distinguishes them from trivial point-defect states. Therefore, while both point defects and topological edge states may exhibit enhanced IPR, only the latter simultaneously satisfy the three defining characteristics of a topological edge mode: (i) localization at a topological phase boundary, (ii) quasienergy pinned within the topological gap ($E=0$), and (iii) long-time robustness against disorder.}

\begin{figure*}
    \centering
    
    \begin{subfigure}[b]{0.32\textwidth}
        \centering
        \includegraphics[width=\textwidth]{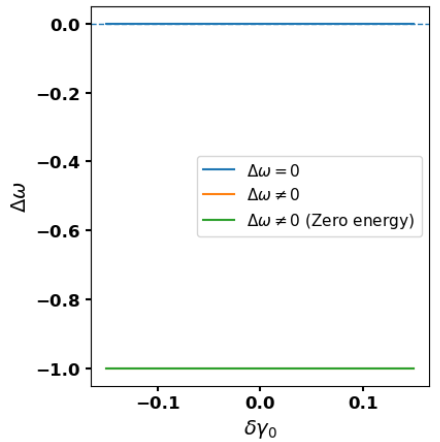}   
        \caption{}
        \label{tracka}
    \end{subfigure}
    \hfill 
    \begin{subfigure}[b]{0.32\textwidth}
        \centering
        \includegraphics[width=\textwidth]{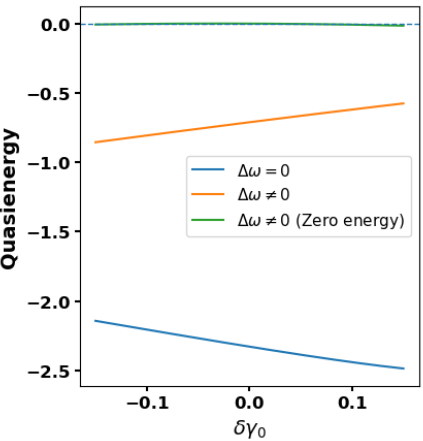}
        \caption{}
        \label{trackb}
    \end{subfigure}
    \hfill 
    \begin{subfigure}[b]{0.32\textwidth}
        \centering
        \includegraphics[width=\textwidth]{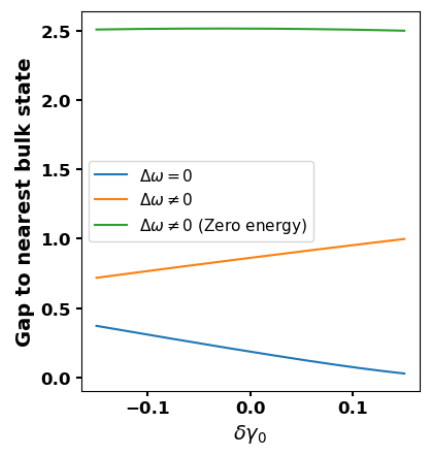}
        \caption{}
        \label{trackc}
    \end{subfigure}
    \caption{\textcolor{black}{Robustness of the tracked eigenstates under local coin perturbation $\delta\gamma_0$. (a) Winding-number difference $\Delta\omega$ as a function of the perturbation for the three representative configurations, $\gamma_{x=0} = \pi/3,\gamma_{x\ne 0} = 3\pi/5$ (blue, trivial FELS), $\gamma_{x=0} = \pi + 0.2,\gamma_{x\ne 0} = 3\pi/4$ (orange, FELS), and $\gamma_{x=0} = \pi + 1.6,\gamma_{x\ne 0} = 0.52\pi$ (green, topological ZEES). Here, the green and orange curves lie on top of each other. (b) Evolution of the tracked quasienergy for the trivial point-defect states and topological edge state. The zero energy edge state remains pinned to the symmetry-protected quasienergy gap at $E=0$, whereas both point-defect states exhibit continuous quasienergy shifts under perturbation. (c) Quasienergy separation between the tracked localized state and the nearest bulk eigenstate. The edge state remains spectrally isolated throughout the perturbation range, whereas the $\Delta\omega=0$ defect state approaches the bulk spectrum and eventually hybridizes with it, while the finite energy ($\Delta\omega=-1$) defect state exhibits a finite separation from the bulk always.}}
    \label{track}
\end{figure*}

\subsection{Robustness of quasienergy under coin perturbations}
\textcolor{black}{To further investigate the robustness of the localized modes, we perturb the coin angle as $\gamma_{x =0} \rightarrow\gamma_{x =0} + \delta \gamma_0 $, with $\delta \gamma_0 \in [-0.15,0.15]$ at site 0, while keeping the bulk coin parameters unchanged, and continuously track the corresponding eigenstate. Of the fourteen eigenstates, we track the localized quasi energy state having the largest IPR at $\delta\gamma_0 =0$. Figure~\ref{track} presents the evolution of the tracked quasienergy and the gap to the nearest bulk state as functions of the perturbation $\delta \gamma_0$. }

\textcolor{black}{Figure~\ref{tracka} shows the winding-number difference $\Delta\omega$ as a function of the local coin perturbation $\delta\gamma_0$. In all three configurations, $\Delta\omega$ remains unchanged throughout the perturbation range, confirming that the local perturbation does not alter the underlying topological configuration. The corresponding tracked quasienergies are shown in Fig.~\ref{trackb}. The topological edge state remains pinned close to the symmetry-protected quasienergy $E=0$ throughout the perturbation range, despite the variation of the local coin parameter. This spectral pinning is the distinguishing feature of the topological edge state. In contrast, the $\Delta\omega=-1$ finite-energy point-defect state exhibits a continuous shift in quasienergy under the same perturbation and is not pinned at a particular energy. The $\Delta\omega=0$ point-defect state also exhibits a strong quasienergy variation and approaches the bulk spectrum as the perturbation increases. These results demonstrate that a winding-number mismatch alone is insufficient to identify a topological edge state; the presence of a pinned zero-quasienergy mode provides the additional necessary spectral criterion.}

\textcolor{black}{The quasienergy separation between the tracked state and the nearest bulk eigenstate is shown in Fig.~\ref{trackc}. For the topological edge state, the separation remains finite and nearly constant throughout the perturbation range, demonstrating that the state remains spectrally isolated from the bulk and does not hybridize with the extended states. The finite-energy point-defect states also remains separated from the bulk over the considered perturbation range, although their separation varies continuously with the perturbation. Localized states occurring at a phase boundary can, depending on their coin parameters and perturbation, either remain spectrally separated from the bulk or hybridize with it, likewise, localized states without a topological phase boundary can, in certain parameter regimes, remain separated from the bulk throughout the perturbation range considered. Thus, the topological edge state is distinguished by the combined presence of a winding-number mismatch, an isolated zero-quasienergy mode pinned to the symmetry-protected gap, and robustness to disorder and phase-preserving perturbations.}

\textcolor{black}{Taken together, these results establish two distinct signatures that differentiate the localized modes. The topological edge state simultaneously exhibits (i) quasienergy pinning near the symmetry-protected gap at $E=0$, and (ii) a finite quasienergy separation from the bulk spectrum throughout the perturbation range. In contrast, the point-defect states fail to satisfy one or more of these criteria. Therefore, while both topological edge states and defect states may initially appear localized, only the former possess simultaneous spectral isolation, localization robustness, and gap protection, providing a clear criterion for distinguishing genuine topological edge modes from trivial defect-induced localized states.}

\begin{table}[H]
    \centering
    \resizebox{\columnwidth}{!}{%
    \renewcommand{\arraystretch}{1.1}
    \begin{tabular}{|c|c|c|}
        \hline
        \textbf{Model~$\downarrow$ / Property$\rightarrow$}
        & \textbf{Topological edge states}
        & \makecell{\textbf{Localized states with/without} \\\textbf{ phase boundary}}
         \\
        \hline

        \makecell{SCSS-CQW\\without disorder}
        &
        \makecell{Stable for long times,\\ negligible spreading into\\ the bulk, and probability\\ at $x=0$ exhibit a\\ constant fit of 0.99743 $\approx 1$.}
        &
        \makecell{Unstable, large spreading into \\the bulk, and probability at $x=0$\\ shows rapid, large-amplitude oscillations.}
        
        \\
        \hline

        \makecell{SCSS-CQW\\with disorder}
        &
        \makecell{Stable for long times\\ with very slow decay.}
        &
        \makecell{Rapid oscillatory decay, followed by \\a saturation to a small constant value ($0.38$).}
        \\
        \hline

        \makecell{Quasienergy\\spectrum\\and IPR of SCSS-CQW \\without disorder}
        &
        \makecell{Highly localized edge modes pinned\\ to the centre of quasienergy gap ($E=0$),\\ exhibiting the largest IPR ($\approx0.49$),\\ consistent with topological \\bulk--boundary correspondence.}
        &
        \makecell{Localized defect modes appear at finite energies,\\ away from the  quasienergy gap,\\ despite relatively large IPR,\\ indicating defect localization.}
        
        \\
        \hline
\makecell{Response to \\coin perturbation\\ ($\delta\gamma_{0}$)}
&
\makecell{Quasienergy remains pinned to\\ $E=0$ and the quasienergy\\ separation from the nearest bulk state\\ remains finite,\\ demonstrating spectral isolation\\ and topological robustness.}
&
\makecell{Quasienergy shifts significantly with\\ perturbation and the quasienergy\\ is not pinned to the topological\\
gap, and the state may hybridize\\
with the bulk spectrum.}
\\
\hline
    \end{tabular}
    }
    \caption{\textcolor{black}{Comparison between topological edge states and point-defect states generated by the SCSS-CQW protocol. Besides their distinct dynamical behavior in clean and disordered systems, the quasienergy spectrum further distinguishes the two classes: topological edge states are pinned to centre of the quasienergy gap and exhibit the largest IPR, whereas point-defect states remain away from the centre of gap despite showing enhanced localization.}}
    \label{tab-defect-vs-edge}
\end{table}

\section{From single-site bulk to finite-width bulk domain}
\textcolor{black}{To demonstrate that the single-site bulk employed throughout this work is not a special or artificial construct, we extend the SCSS-CQW protocol to finite bulk domains. Instead of modifying the coin parameter at only one lattice site, a contiguous domain of width $W$ is assigned the coin angle $\gamma_{\text{bulk}_1}$, while the remaining lattice sites are identified with the bulk coin angle $\gamma_{\text{bulk}_2}$. The corresponding evolution operator is diagonalized for different values of $W$, and the number of localized zero-quasienergy modes, $N_{\text{loc}}$, is evaluated. Here,$N_{\text{loc}}$ denotes the number of localized zero-quasienergy eigenstates. Consequently, $N_{\text{loc}}=0$ indicates the absence of localized zero-energy modes,  while $N_{\text{loc}}=2$ signifies the presence of two localized zero-energy interface modes associated with the two bulk boundaries of the finite-width domain.}

\begin{figure}[H]
    \centering
    \includegraphics[width=0.95\linewidth]{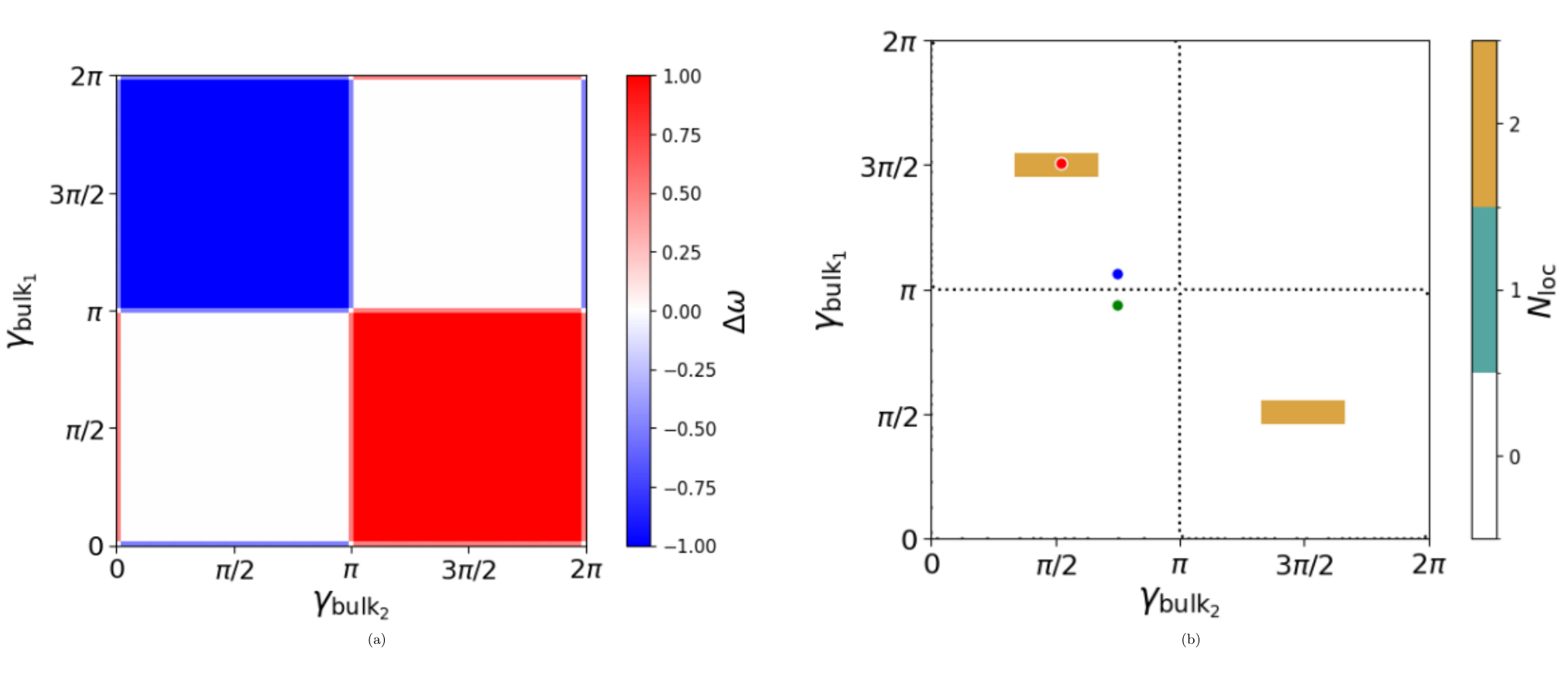}
    \caption{\textcolor{black}{Topological phase diagram and zero-energy modes for a single-site bulk (W=1). (a) Winding-number difference $\Delta\omega$, as a function of the bulk coin parameters. The two nontrivial quadrants $(\Delta \omega = \pm1)$ correspond to topological phase boundaries between the two bulk regions. (b) Number of topological zero-energy modes, $N_{\text{loc}}$. The localized modes appear only within the topologically nontrivial regions,  consistent with the bulk-boundary correspondence; the single-site bulk is the limiting case of a finite-width topological domain, for which these modes occupy only a small patch of the $\Delta\omega\neq 0$ region. The red, blue, and green dots mark the three selected coin-angle pairs $\gamma_{\text{bulk}_1},\gamma_{\text{bulk}_2}=(\pi+1.6,0.52\pi)$, $(\pi+0.2,3\pi/4)$, and $(\pi-0.2,3\pi/4)$, respectively. Topological edge states are observed for the red parameter choice but are absent for the blue and green choices, which lie outside the yellow region corresponding to $N_{\mathrm{loc}}=2$.}}
    \label{width1}
\end{figure}

\textcolor{black}{Fig. \ref{width1} (a) shows the winding-number difference, $\Delta \omega = \omega_{\text{bulk}_1} - \omega_{{\text{bulk}_2}} $, which identifies the topological phase boundaries separating the bulk regions. According to the bulk-boundary correspondence, localized interface states are expected only when the two regions possess different topological invariants ($\Delta \omega \ne 0$). The $N_{\text{loc}}$ map for the single-site bulk ($W=1$) shown in Fig.~\ref{width1} (b) is consistent with this prediction, demonstrating that even the narrowest possible bulk region supports localized zero-quasienergy interface modes only when the two bulks belong to distinct topological phases.}

\begin{figure}[H]
    \centering
    \includegraphics[width=\linewidth]{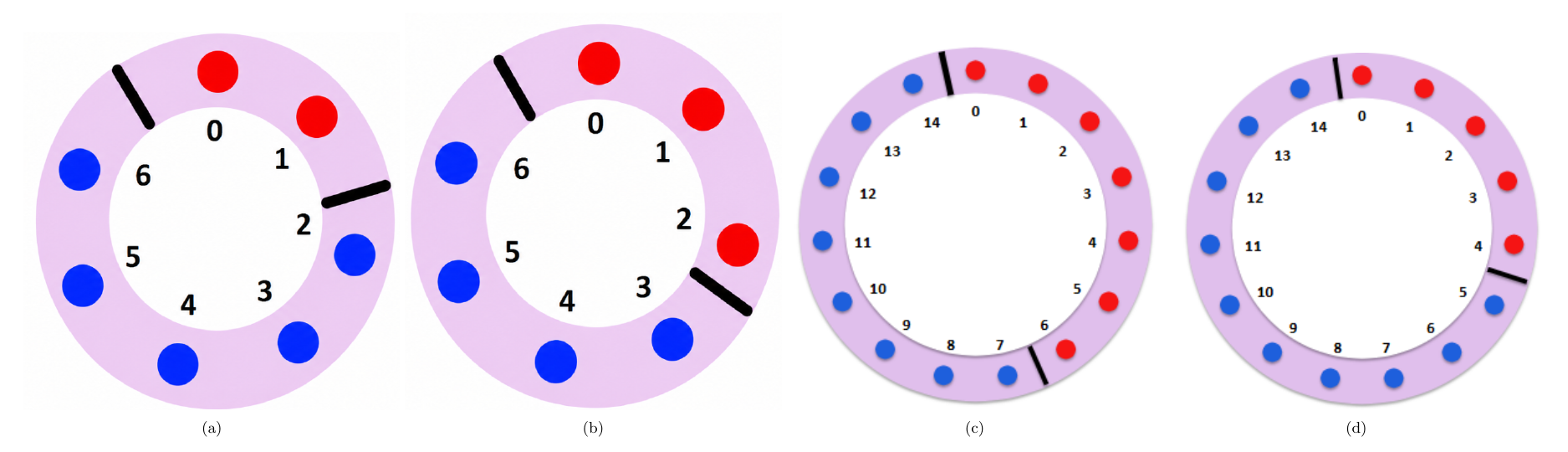}
    \caption{\textcolor{black}{Finite-width domain analysis on a 7-site an 15-site cyclic graphs. (a) 7-site cycle, with sites $0,1$ belonging to bulk 1 in red ($\gamma_{\text{bulk}_1} = \pi + 1.6$, $\omega = -1/2$) and sites $2$--$6$ to bulk 2 in blue ($\gamma_{\text{bulk}_2} = 0.52\pi$, $\omega = 1/2$). (b) 7-site cycle, with sites $0$--$2$ belonging to bulk 1 and sites $3$--$6$ to the bulk 2.(c) 15-site cycle, with sites $0$--$6$ belonging to bulk 1 and sites $7$--$14$ to the other. (d) 15-site cycle, with sites $0$--$4$ belonging to bulk 1  and sites $5$--$14$ to bulk 2.}}
    \label{W23}
\end{figure}
\begin{figure}[H]
    \centering
    \includegraphics[width=0.75\linewidth]{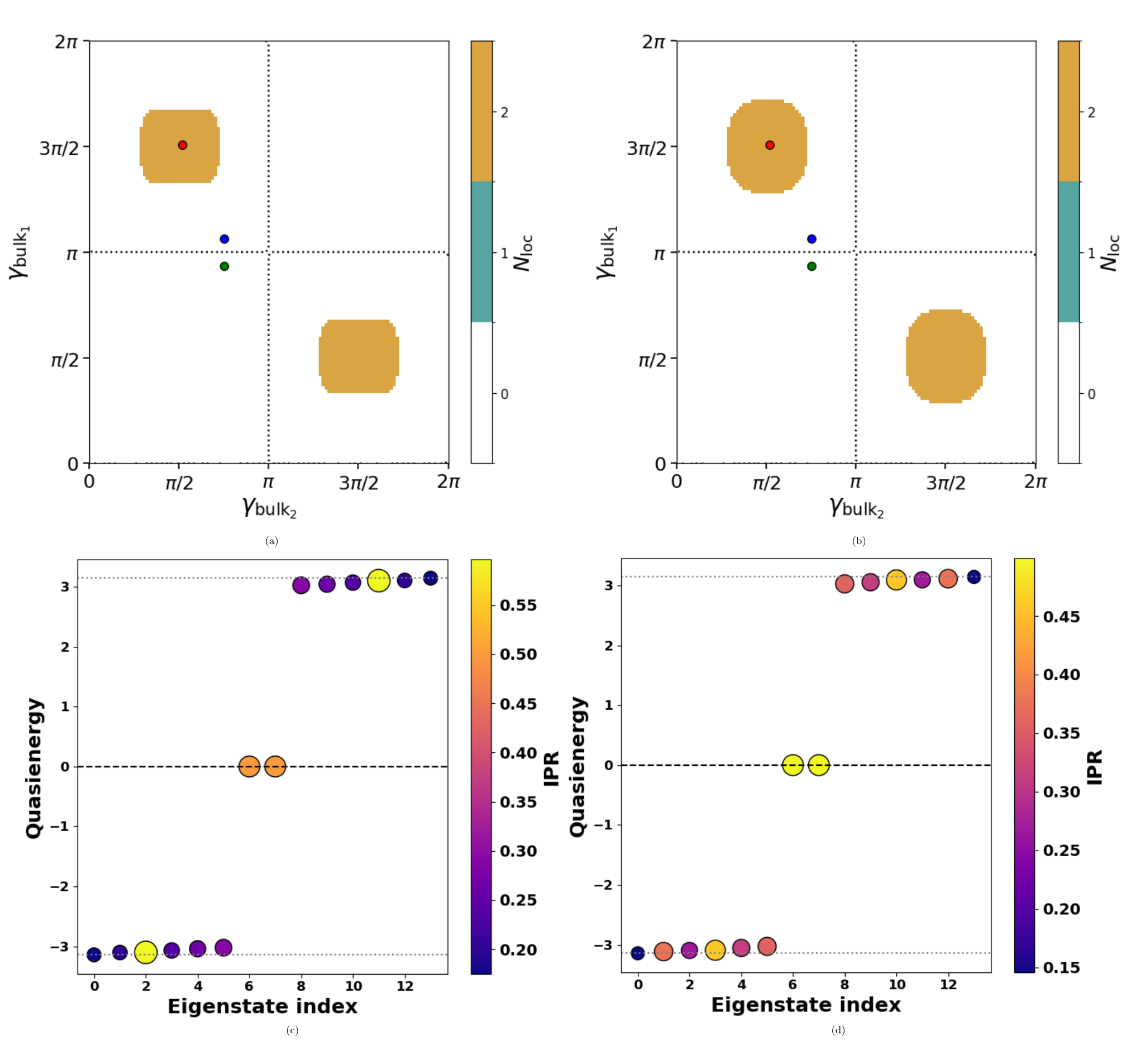}
     \caption{\textcolor{black}{Number of localized zero-quasienergy  modes, $N_{\text{loc}}$, for finite-width bulk domains of width (a) $W=2$, and (b) $W=3$,  on a 7-site cyclic graph. In both cases, topological zero energy modes remain confined to the topologically nontrivial regions. (c-d) Quasienergy spectra for representative parameter values selected from the corresponding topological regions in (a)-(b) (for coin parameters specified in red). The color of each eigenstate denotes its inverse participation ratio (IPR), with larger IPR indicating stronger localization. In both cases, two isolated eigenstates are pinned at $E=0$, confirming that the localized modes are genuine topological edge states. The red, blue, and green dots mark the three selected coin-angle pairs $\gamma_{\text{bulk}_1},\gamma_{\text{bulk}_2}=(\pi+1.6,0.52\pi)$, $(\pi+0.2,3\pi/4)$, and $(\pi-0.2,3\pi/4)$, respectively. Topological edge states are observed for the red parameter choice but are absent for the blue and green choices, which lie outside the yellow region corresponding to $N_{\mathrm{loc}}=2$. As the bulk domain width increases from a single site to a finite number of sites, the range of coin parameters supporting topological ZEES expands in the presence of a winding-number mismatch.}}
    \label{width21}
\end{figure}

\textcolor{black}{We first consider a 7-site cyclic graph to examine the evolution from a single-site bulk to a finite-width domain. For $W=2$, sites $0$ and $1$ are with the coin angle $\gamma=\pi+1.6$, corresponding to $\omega=-1/2$, while sites $2$--$6$ are with coin angle $\gamma=0.52\pi$, corresponding to $\omega=1/2$. Increasing the bulk width to $W=3$ extends the first domain to sites $0$--$2$, while sites $3$--$6$ remain in the second topological phase.}
\textcolor{black}{Figs. \ref{width21}(a) and \ref{width21}(b) present the corresponding phase diagrams as the bulk width is increased to $W = 2$ and $W = 3$ on a $7$--site cyclic graph. The overall topological phase diagram remains unchanged, and the localized modes continue to appear exclusively within the regions where $\Delta \omega \ne 0$. The principal effect of increasing the bulk width is that the coin parameter region supporting localized zero-energy modes gradually expands. This occurs because a wider bulk domain more closely approximates an extended topological phase, reducing the hybridization between the interface-localized states and making them observable over a broader range of coin parameters. The corresponding spectra are shown in Figs.~\ref{width21}(c) and \ref{width21}(d), where the color of each eigenstate represents its inverse participation ratio (IPR). In both cases, two eigenstates remain pinned at the symmetry-protected quasienergy $ E=0,$ confirming the existence of topological edge modes. This indicates that the defining signature is the simultaneous presence of strong localization and spectral pinning at the symmetry-protected quasienergy $E=0$. The remaining eigenstates form the bulk quasienergy bands and are distributed away from the gap.}

\begin{figure}
    \centering
    
   \begin{subfigure}[b]{0.24\textwidth}
        \centering
        \includegraphics[width=\textwidth]{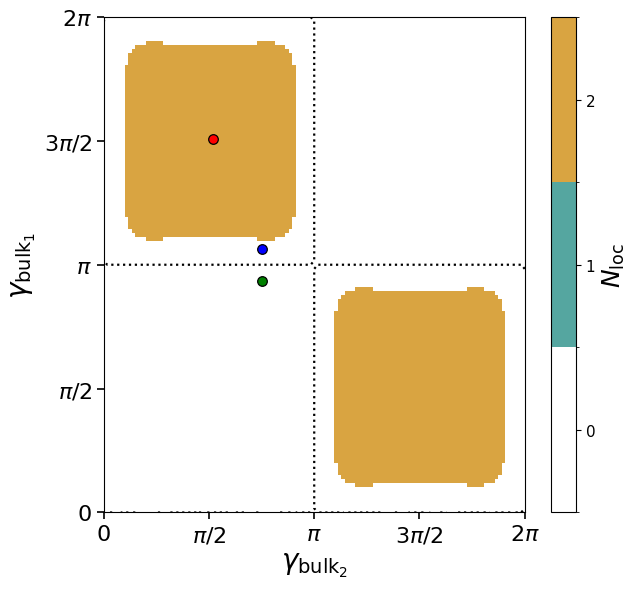}   
        \caption{}
        \label{15a}
    \end{subfigure}
    \hfill 
    \begin{subfigure}[b]{0.24\textwidth}
        \centering
        \includegraphics[width=\textwidth]{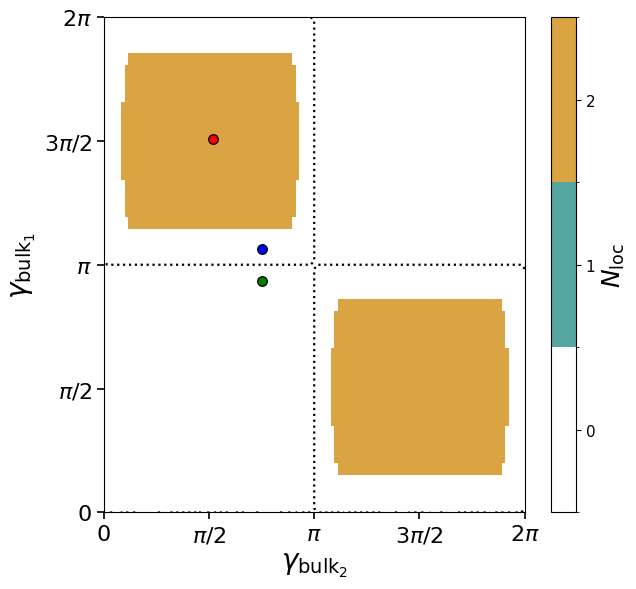}   
        \caption{}
        \label{15b}
    \end{subfigure}
    \hfill
    \begin{subfigure}[b]{0.24\textwidth}
        \centering
        \includegraphics[width=\textwidth]{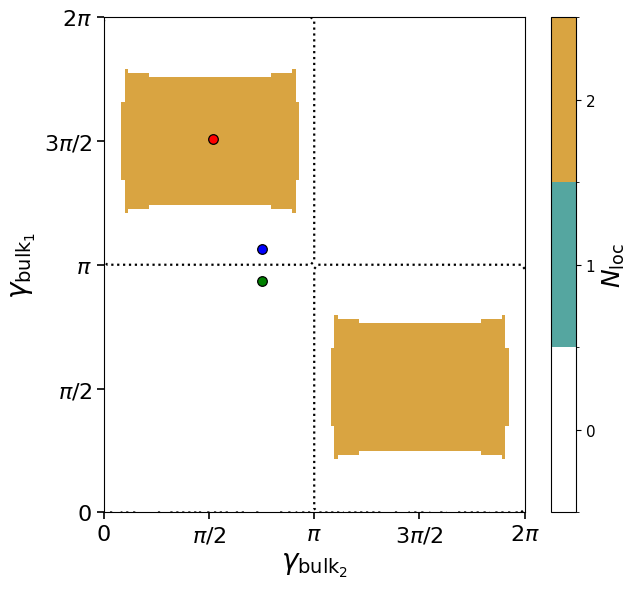}
        \caption{}
        \label{15c}
    \end{subfigure}
    \hfill 
    \begin{subfigure}[b]{0.24\textwidth}
        \centering
        \includegraphics[width=\textwidth]{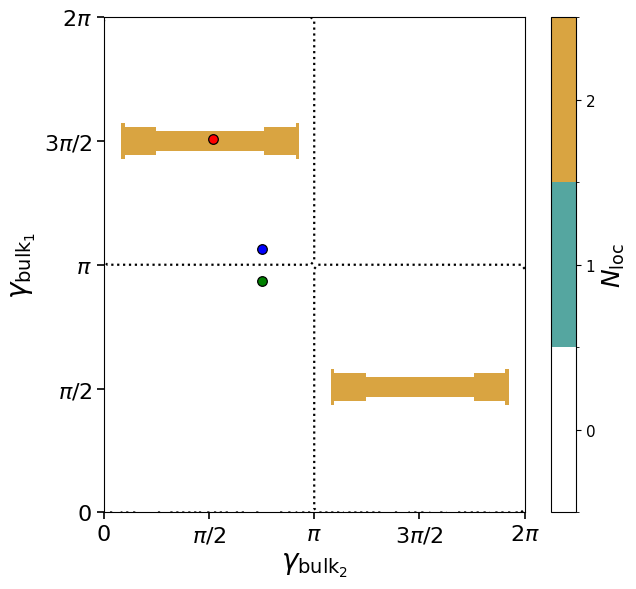}
        \caption{}
        \label{15d}
    \end{subfigure}
    
    \begin{subfigure}[b]{0.24\textwidth}
        \centering
        \includegraphics[width=\textwidth]{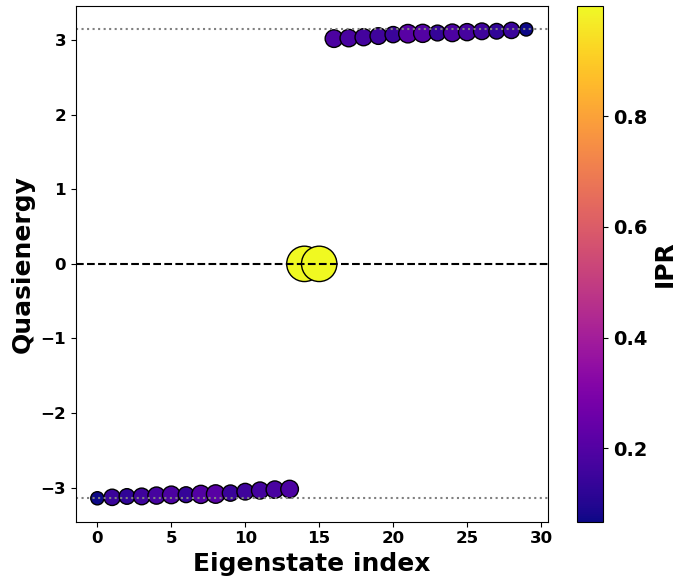}   
        \caption{}
        \label{15e}
    \end{subfigure}
    \hfill 
    \begin{subfigure}[b]{0.24\textwidth}
        \centering
        \includegraphics[width=\textwidth]{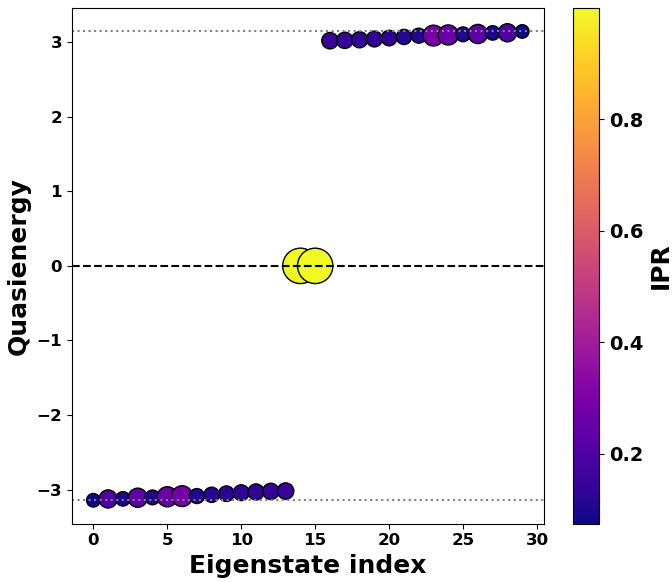}   
        \caption{}
        \label{15f}
    \end{subfigure}
    \hfill
    \begin{subfigure}[b]{0.24\textwidth}
        \centering
        \includegraphics[width=\textwidth]{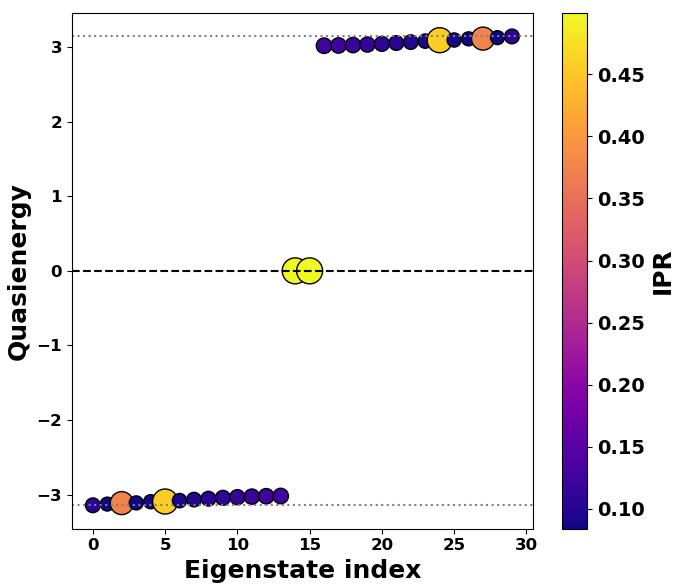}
        \caption{}
        \label{15g}
    \end{subfigure}
    \hfill 
    \begin{subfigure}[b]{0.24\textwidth}
        \centering
        \includegraphics[width=\textwidth]{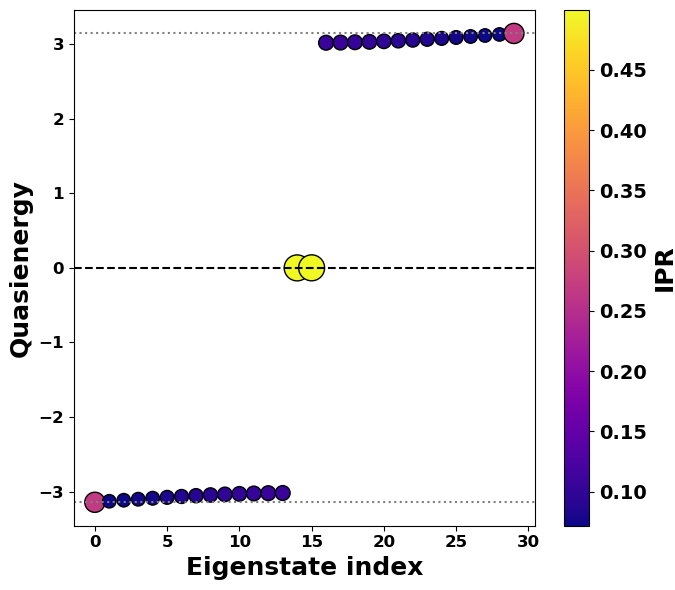}
        \caption{}
        \label{15h}
    \end{subfigure}
    
    \caption{\textcolor{black}{Number of localized zero-quasienergy modes, $N_{\text{loc}}$, for finite-width bulk domains of width (a) $W=7$, (b) $W=5$, (c) $W = 3$ and (d) $W = 1$ on a $15$-site cyclic graph. In all cases, localized modes remain confined to the topologically nontrivial regions. (e-h) Quasienergy spectra for representative parameter values selected from the corresponding topological regions in (a-d) (for coin parameters specified in red) respectively. The color of each eigenstate denotes its inverse participation ratio (IPR), with larger IPR indicating stronger localization. In all cases, two isolated eigenstates are pinned at $E=0$, confirming that the localized modes are genuine topological edge states. The red, blue, and green dots mark the three selected coin-angle pairs $\gamma_{\text{bulk}_1},\gamma_{\text{bulk}_2}=(\pi+1.6,0.52\pi)$, $(\pi+0.2,3\pi/4)$, and $(\pi-0.2,3\pi/4)$, respectively. Topological edge states are observed for the red parameter choice but are absent for the blue and green choices, which lie outside the yellow region corresponding to $N_{\mathrm{loc}}=2$. As the bulk domain width increases from a single site to a finite number of sites, the range of coin parameters supporting topological ZEES expands in the presence of a winding-number mismatch.}}
    \label{width15}
\end{figure}

\textcolor{black}{To further establish the connection between a single-site bulk and an extended topological domain, we consider larger cyclic graphs with $N=15$ and $N=51$ sites. For $N=15$, we consider bulk domain widths $W=1,3,5$, and $7$. Specifically, for a given $W$, sites $0,\ldots,W-1$ belong to the first bulk domain, employed with coin angle $\gamma=\pi+1.6$, while sites $W,\ldots,14$ belong to the second domain, with coin angle $\gamma=0.52\pi$. The phase diagrams for $W=5$ and $W=7$ (see Fig. (\ref{width15})) show that the region supporting $N_{\rm loc}=2$ expands as the width of the finite domain increases. The corresponding quasienergy spectra contain two isolated states pinned at the symmetry-protected quasienergy (E=0), with large inverse participation ratios (IPRs), confirming their localized character. Thus, increasing the width of the topological domain does not change the underlying topological criterion; rather, it reduces the finite-size hybridization between the two interfaces, allowing the two interface modes to become increasingly well resolved.}

\begin{figure}
    \centering
    \includegraphics[width=0.95\linewidth]{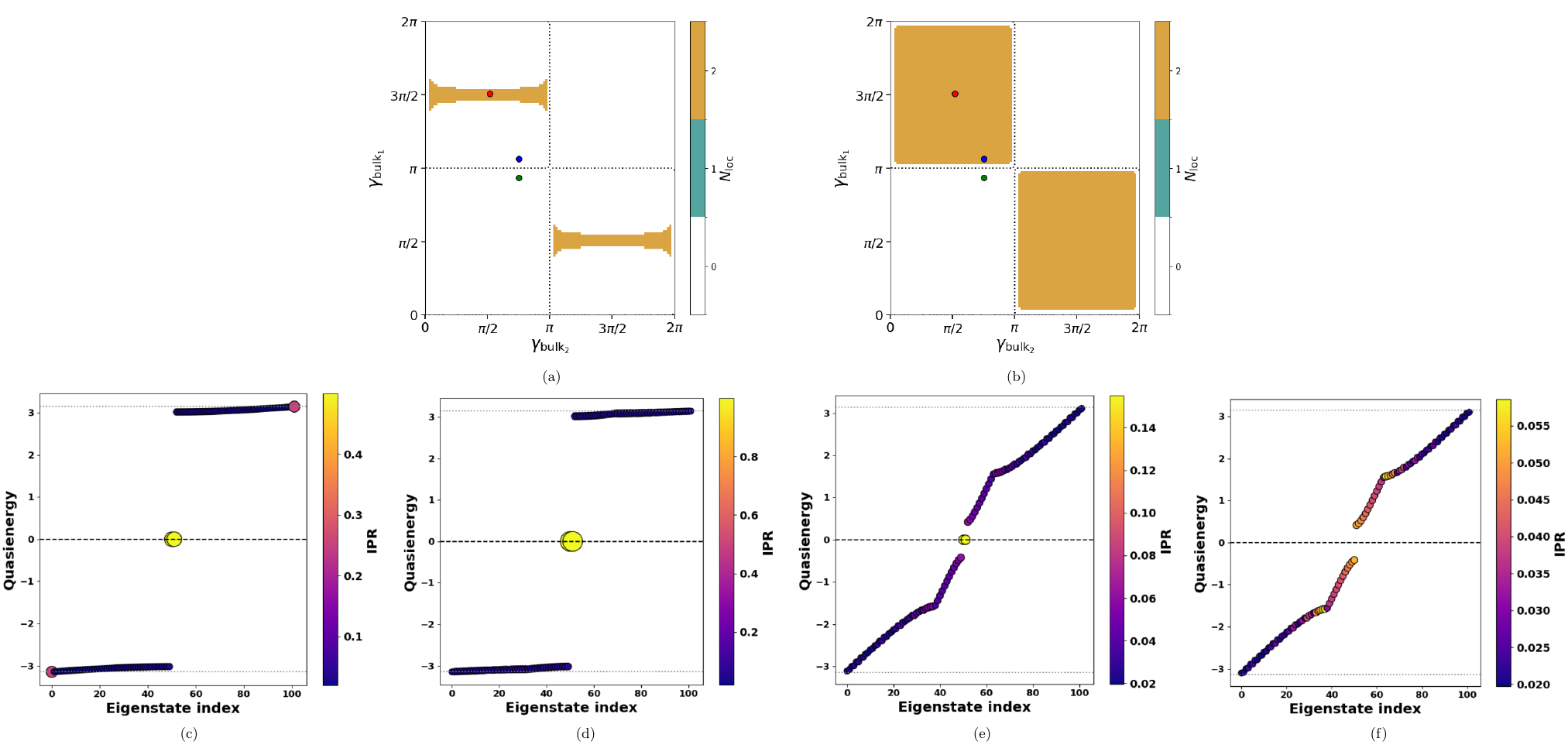}
    \caption{\textcolor{black}{Topological phase diagram showing the no. of zero energy modes for (a) single-site bulk (W=1) and (b) finite-width bulk (W = 25). The ZEES appear only within the topologically nontrivial regions, covering a small patch of them for $W = 1$ and nearly all of them for $W = 25$, showing that the single-site bulk is the limiting case of a finite-width topological domain. (c-f) Quasienergy spectra for representative parameter values selected from the corresponding regions in (a)-(b). The color of each eigenstate denotes its inverse participation ratio (IPR), with larger IPR indicating stronger localization. The red, blue, and green dots mark the three selected coin-angle pairs $\gamma_{\text{bulk}_1},\gamma_{\text{bulk}_2}=(\pi+1.6,0.52\pi)$, $(\pi+0.2,3\pi/4)$, and $(\pi-0.2,3\pi/4)$, respectively.  Topological edge states are observed for both the red and blue parameter choices, whereas they are absent for the green choice. Panels (c) and (d) show the quasienergy spectra for the red parameter choice for (W=1) and (W=25), respectively. Panels (e) and (f) show the spectra for (W=25) for the blue and green parameter choices, respectively. }}
    \label{width51}
\end{figure}

\textcolor{black}{We observe the same trend for the larger cyclic graph with $N=51$, where we consider a single-site bulk domain ($W=1$) and a finite-width domain ($W=25$), as shown in Figs.~\ref{width51}(a) and ~\ref{width51}(b), respectively. For $W=25$, the region of coin parameters supporting $N_{\rm loc}=2$ encompasses nearly the entire region in which the winding-number mismatch occurs. The corresponding quasienergy spectra, shown in Figs. (\ref{width51})(c)–(f), further illustrate the presence or absence of localized zero-energy edge states for representative coin-angle configurations. In particular, the comparison between the single-site and finite-width domains demonstrates that increasing the domain width broadens the parameter region supporting topological zero-energy edge states when the two bulk regions have different winding numbers. In the finite-width bulk domain, a topological ZEES implies that there is a winding number mismatch, and the reverse also holds true for nearly all coin angles, i.e., winding number mismatch almost always implies the presence of a topological zero-energy edge state.}

\textcolor{black}{Therefore, the single-site bulk considered in the main text should be viewed as the limiting case ($W=1$) of a finite-width topological domain. The wider-domain calculations do not alter the underlying topological physics or the bulk-boundary correspondence; instead, they demonstrate that the localized zero energy edge states in the Main remain pinned to the zero quasienergy regardless of the bulk width. Consequently, the conclusions drawn from the single-site bulk remain physically well-founded and are representative of the more general finite-width interface geometry.}

\section{Chiral Signature of topological ZEES}
\textcolor{black}{A topological ZEES is identified by a winding-number mismatch together with a quasienergy mode pinned at $E=0$. This pinning follows from a chiral symmetry of the evolution operator~\cite{asboth}, which defines a local chiral density that is nonzero only at the interfaces hosting the ZEES. Herein, we first establish the chiral symmetry in the time frame used throughout the manuscript, then we consider another symmetric time frame, where the chiral operator takes a different form, and we show that the local chiral density is same in both the cases. To further characterize the topological nature of the edge states, we investigate the local chiral density at every site of the cyclic graph for different bulk domain widths.} \\

\subsection{Chiral symmetry in the time frame $\hat{U}_{evo} = \hat{S}_+\hat{C_\gamma}\hat{S}_-\hat{C}_\gamma$}
\textcolor{black}{The winding number of Eq.~(\ref{eq:zak_phase}) is defined with the unit vector $\hat A(\gamma,D)=(\cos\Delta\gamma,\,0,\,\sin\Delta\gamma)$, $\Delta=D/2$, which does not depend on $k$. It can therefore be applied on each site, where the coin angle $\gamma$ is applied depending upon the bulk width chosen in the cycle. This begets the chiral operator as,
\begin{equation}
\hat\Gamma_U=\sum_{x=0}^{N-1}\ket{x}\bra{x}\otimes\hat A(\gamma)\!\cdot\!\vec\sigma
=\sum_{x=0}^{N-1}\ket{x}\bra{x}\otimes\left[\cos(\Delta\gamma)\,\sigma_x+\sin(\Delta\gamma)\,\sigma_z\right],
\label{gamma-U}
\end{equation}
which is Hermitian, satisfies $\hat\Gamma_U^2=\mathbb 1$, and obeys
\begin{equation}
\hat\Gamma_U\,\hat U_{evo}\,\hat\Gamma_U=\hat U_{evo}^{\dagger}.
\label{chiral-sym-U}
\end{equation}
Consequently as $\hat U_{evo}\ket{\psi_n}=e^{-iE_n}\ket{\psi_n}$, where $\ket{\psi_n}$ is the eigenstate corresponding to eigenvalue $E_n$, thus, 
\begin{equation}
\hat U_{evo}\big(\hat\Gamma_U\ket{\psi_n}\big)=\hat\Gamma_U\,\hat U_{evo}^{\dagger}\ket{\psi_n}=e^{+iE_n}\,\hat\Gamma_U\ket{\psi_n},
\end{equation}
so $\hat\Gamma_U$ maps an eigenstate at $E_n$ to one at $-E_n$.}

\subsection{Chiral symmetry in the symmetric time frame}
\textcolor{black}{Eq.~(\ref{chiral-sym-U}) is not the form in which chiral symmetry is usually defined for quantum walks. It becomes the standard one in the symmetric time frame ~\cite{asboth, kita-pra}, which is, 
\begin{equation}
\hat F=\hat C^{1/2}\,\hat S_+\,\hat C\,\hat S_-\,\hat C^{1/2}=\hat C^{1/2}\,\hat U_{evo}\,\hat C^{-1/2},
\label{F-frame}
\end{equation}
where $\hat C=e^{-i\Delta\gamma\sigma_y}$ and $\hat C^{1/2}=e^{-i\Delta\gamma\sigma_y/2}$. Since $\hat F$ and $\hat U_{evo}$ related via a similarity tranformation, the two have the same quasienergies. In this frame the relevant unit vector is $\hat A_F=(1,0,0)$ for every site, i.e. the chiral operator is
\begin{equation}
\hat\Gamma_F=\mathbb 1_N\otimes\sigma_x,\qquad \hat\Gamma_F\,\hat F\,\hat\Gamma_F=\hat F^{\dagger},
\label{chiral-sym-F}
\end{equation}
independent of coin angles. The chiral symmetry in Eq. (\ref{chiral-sym-F}) follows from $\sigma_x\hat C^{1/2}\sigma_x=\hat C^{-1/2}$, $\sigma_x\hat C\sigma_x=\hat C^{-1}$, $\sigma_x\hat S_+\sigma_x=\hat S_-^{\dagger}$ and $\sigma_x\hat S_-\sigma_x=\hat S_+^{\dagger}$. Thus, 
\begin{equation}
\sigma_x\hat F\sigma_x=\hat C^{-1/2}\hat S_-^{\dagger}\hat C^{-1}\hat S_+^{\dagger}\hat C^{-1/2}=\hat F^{\dagger},
\end{equation}
since $\hat C$ is a real orthogonal matrix. The two chiral operators are related by the same transformation as the evolution operators,
\begin{equation}
\hat\Gamma_U=\hat C^{-1/2}\,\hat\Gamma_F\,\hat C^{1/2}=\sum_x\ket{x}\bra{x}\otimes\left[\cos(\Delta\gamma_x)\sigma_x+\sin(\Delta\gamma_x)\sigma_z\right],
\end{equation}
so Eq.~(\ref{gamma-U}) is the chiral operator of the frame used in the paper, with a vector $\hat A(\gamma)$, while in the symmetric frame it is $\hat A_F$. The same argument as above shows that $\hat\Gamma_F$ maps an eigenstate of $\hat F$ at $E$ to one at $-E$.}

\subsection{Local chiral density}
\textcolor{black}{To determine the chirality of the zero-energy modes, one needs to evaluate $\langle\hat\Gamma\rangle$ for each eigenstate. This is ambiguous when two pinned modes of opposite chirality hybridize, the resulting eigenstates are their sum and difference, and each has $\langle\hat\Gamma\rangle=0$. We therefore use all the eigenstates in the zero-quasienergy gap together. Here the zero-quasienergy gap is the range of quasienergies around $E=0$ that contains no bulk states. Let $\hat P_{\rm gap}$ be the projector onto the eigenstates of the evolution operator with quasienergy in a window $|E|<E_c$ around $E=0$, where $E_c$ only needs to separate the pinned states from the bulk bands. Since $\hat\Gamma$ maps $E\to-E$, this window contains every state together with its chiral partner, and $\hat P_{\rm gap}$ commutes with $\hat\Gamma$. The local chiral density can thus be defined as,
\begin{equation}
\rho_\Gamma(x)=\operatorname{Tr}_{c}\bra{x}\hat P_{\rm gap}\hat\Gamma\ket{x},
\label{rho-def}
\end{equation}}
\textcolor{black}{where $\operatorname{Tr}_c$ is the trace over the coin and $x$ labels the site of the cyclic graph. Because $\hat P_{\rm gap}$ commutes with $\hat\Gamma$, the in-gap subspace is spanned by states with $\hat\Gamma=\pm1$, and then $\rho_\Gamma(x)=P_+(x)-P_-(x)$, where $P_\pm(x)$ is the probability density at site $x$ of the zero-energy states with chirality $\pm1$. The density is therefore positive where a $\hat\Gamma=+1$ mode is localized and negative where a $\hat\Gamma=-1$ mode is localized, and it does not depend on how the eigenstates are chosen inside the gap. Bulk states and FELS lie outside the zero-energy window and do not contribute.}

\textcolor{black}{The local chiral density is the same in the two time frames. The eigenstates of $\hat F$ are $\hat C^{1/2}\ket{\psi_n}$ with the same energies, so $\hat P^{F}_{\rm gap}\hat\Gamma_F=\hat C^{1/2}\big(\hat P^{U}_{\rm gap}\hat\Gamma_U\big)\hat C^{-1/2}$. Since $\hat C^{1/2}$ acts on the coin at each site separately, the coin trace at every site is unchanged, and
\begin{equation}
\rho^{F}_\Gamma(x)=\rho^{U}_\Gamma(x).
\end{equation}}
\textcolor{black}{To calculate $\rho_\Gamma(x)$ for coin angles $\{\gamma\}$, we diagonalize the evolution operator, build $\hat P_{\rm gap}$ from the eigenstates in the zero-quasienergy gap, and evaluate Eq.~(\ref{rho-def}) for every site.}\\

\textcolor{black}{The local chiral density for the 7-cycle graph, in case of a single-site ($W=1$) as well as a finite-width ($W=3$) bulk domain, with $D=2$ is calculated. For $W=1$, the coin rotation angle is $\gamma_0=\pi+1.6$ at site $x=0$, while the at remaining six sites the coin angle $\gamma_x=0.52\pi$ is applied, so the two domain walls are at the junctions $x=6|0$ and $x=0|1$. For $W=3$, $\gamma_x=\pi+1.6$ is applied to sites $x=0,1,2$ and $\gamma_x=0.52\pi$ is applied at the remaining sites, so the two interfaces are at $x=6|0$ and $x=2|3$. We take $E_c=0.1$ in both cases only two eigenstates lie below it, so the projector selects exactly the pinned pair. All results are collated in Table~\ref{tab-chiral-density}.}

\begin{table}[H]
\centering
\renewcommand{\arraystretch}{1.2}
\begin{tabular}{|c|c|c|}
\hline
 & $W=1$ & $W=3$ \\ \hline
Phase boundary & $x=6|0,\ 0|1$ & $x=6|0,\ 2|3$ \\ \hline
Number of zero-energy edge states & 2 & 2 \\ \hline
$\|\hat\Gamma_U\hat U_{\rm evo}\hat\Gamma_U-\hat U_{\rm evo}^{\dagger}\|$ & $1.21\times10^{-16}$ & $2.52\times10^{-16}$ \\ \hline
$\|\hat\Gamma_F\hat F\hat\Gamma_F-\hat F^{\dagger}\|$ & $7.9\times10^{-17}$ & $2.4\times10^{-16}$ \\ \hline
$\max_x|\rho^{F}_\Gamma(x)-\rho^{U}_\Gamma(x)|$ & $2.22\times10^{-16}$ & $4.44\times10^{-16}$ \\ \hline
\multicolumn{3}{|c|}{Local chiral density $\rho_\Gamma(x)$} \\ \hline
$x=0$ & $-0.99880$ & $-0.99901$ \\ \hline
$x=1$ & $+0.99979$ & $-0.00000$ \\ \hline
$x=2$ & $+0.00000$ & $+0.00021$ \\ \hline
$x=3$ & $0.00000$ & $+0.99979$ \\ \hline
$x=4$ & $0.00000$ & $+0.00000$ \\ \hline
$x=5$ & $0.00000$ & $0.00000$ \\ \hline
$x=6$ & $-0.00099$ & $-0.00099$ \\ \hline

\end{tabular}
\caption{\textcolor{black}{Local chiral density of the zero-energy states for the 7-cycle ($D=2$), with $\gamma=\pi+1.6$ on the first $W$ sites and $\gamma=0.52\pi$ on the remaining sites. $\rho_\Gamma(x)$ is computed from the projector onto the eigenstates with $|E|<E_c=0.1$.}}
\label{tab-chiral-density}
\end{table}

\textcolor{black}{For $W=1$, the local chiral density has two pronounced peaks of opposite sign, $\rho_\Gamma(0)=-0.9988$ and $\rho_\Gamma(1)=+0.9998$, with a small tail $\rho_\Gamma(6)=-0.0010$ and negligible values at all other sites. The chiral character of the zero-energy states is therefore concentrated at the two domain walls. For $W=3$, the negative peak remains at $x=0$, with $\rho_\Gamma(0)=-0.9990$, whereas the positive peak moves to $x=3$, with $\rho_\Gamma(3)=+0.9998$, and the density at the intermediate and remaining bulk sites is essentially zero. Increasing the domain width from $W=1$ to $W=3$ thus moves the positive peak from $x=1$ to $x=3$, following the displacement of the corresponding phase boundary, while the negative peak stays at $x=0$.} \\

\textcolor{black}{Similarly, we can calculate the local chiral density for the 15-cycle graph, for a single-site ($W=1$) and a finite-width ($W=7$) bulk domain, with $D=2$. For $W=1$, the coin rotation angle $\gamma_0=\pi+1.6$ is applied at site $x=0$, while at the remaining $14$ sites $\gamma_x=0.52\pi$ is applied, so the two phase boundaries are at the junctions $x=14|0$ and $x=0|1$. For $W=7$, $\gamma_x=\pi+1.6$ is applied at sites $x=0-6$ and $\gamma_x=0.52\pi$ is applied at the remaining sites, so the two interfaces are at $x=14|0$ and $x=6|7$. Herein, we also take $E_c=0.1$ in both cases only two eigenstates lie below it, thus the projector selects exactly the pinned pair. The findings are juxtaposed in Table~\ref{tab-chiral-density15}.}
\begin{table}[H]
\centering
\renewcommand{\arraystretch}{1.2}
\begin{tabular}{|c|c|c|}
\hline
 & $W=1$ & $W=7$ \\ \hline
Phase boundary & $x=14|0,\ 0|1$ & $x=14|0,\ 6|7$ \\ \hline
Number of zero-energy states & 2 & 2 \\ \hline
$\|\hat\Gamma_U\hat U_{\rm evo}\hat\Gamma_U-\hat U_{\rm evo}^{\dagger}\|$ & $1.33\times10^{-16}$ & $4.04\times10^{-16}$ \\ \hline
$\|\hat\Gamma_F\hat F\hat\Gamma_F-\hat F^{\dagger}\|$ & $7.86\times10^{-17}$ & $3.92\times10^{-16}$ \\ \hline
$\max_x|\rho^{F}_\Gamma(x)-\rho^{U}_\Gamma(x)|$ & $5.55\times10^{-16}$ & $3.33\times10^{-16}$ \\ \hline
\multicolumn{3}{|c|}{Local chiral density $\rho_\Gamma(x)$} \\ \hline
$x=0$ & $ -0.99879$ & $-0.99901$ \\ \hline
$x=1$ & $+0.99978$ & $0.00000$ \\ \hline
$x=2$ & $+0.00000$ & $0.00000$ \\ \hline
$x=3$ & $0.00000$ & $0.00000$ \\ \hline
$x=4$ & $0.00000$ & $0.00000$ \\ \hline
$x=5$ & $0.00000$ & $0.00000$ \\ \hline
$x=6$ & $0.00000$ & $+0.00021$ \\ \hline
$x=7$ & $0.00000$ & $+0.99978$ \\ \hline
$x=8$ & $0.00000$ & $0.00000$ \\ \hline
$x=9$ & $+0.00000$ & $0.00000$ \\ \hline
$x=10$ & $0.00000$ & $0.00000$ \\ \hline
$x=11$ & $0.00000$ & $+0.00000$ \\ \hline
$x=12$ & $0.00000$ & $0.00000$ \\ \hline
$x=13$ & $0.00000$ & $0.00000$ \\ \hline 
$x=14$ & $-0.00098$ & $-0.00098$ \\ \hline
\end{tabular}
\caption{\textcolor{black}{Local chiral density of the zero-energy states for the 15-cycle ($D=2$), with $\gamma=\pi+1.6$ on the first $W$ sites and $\gamma=0.52\pi$ on the remaining sites. $\rho_\Gamma(x)$ is computed from the projector onto the eigenstates with $|E|<E_c=0.1$.}}
\label{tab-chiral-density15}
\end{table}
\textcolor{black}{For $W=1$, the local chiral density has two pronounced peaks of opposite sign, $\rho_\Gamma(0)=-0.99879$ and $\rho_\Gamma(1)=+0.99978$, with a small tail $\rho_\Gamma(14)=-0.00098$ and negligible values at all other sites. The chiral character of the zero-energy states is therefore concentrated at the two phase boundaries. For $W=7$, the negative peak remains at $x=0$, with $\rho_\Gamma(0)=-0.99901$, whereas the positive peak now moves to $x=7$, with $\rho_\Gamma(7)=+0.99978$, and the density at the intermediate and remaining bulk sites is zero.} \\
\textcolor{black}{ In both frames, for both cycle sizes and both domain widths, the chiral-symmetry relations hold to numerical precision. For the 7-cycle, $\|\hat\Gamma_U\hat U_{\rm evo}\hat\Gamma_U-\hat U_{\rm evo}^{\dagger}\|=1.21\times10^{-16}$ ($W=1$) and $2.52\times10^{-16}$ ($W=3$), and $\|\hat\Gamma_F\hat F\hat\Gamma_F-\hat F^{\dagger}\|=7.9\times10^{-17}$ and $2.4\times10^{-16}$. For the 15-cycle, the corresponding values are $1.33\times10^{-16}$ ($W=1$) and $4.04\times10^{-16}$ ($W=7$) for $\hat U_{\rm evo}$, and $7.86\times10^{-17}$ and $3.92\times10^{-16}$ for $\hat F$. The local chiral densities computed in the two frames agree at every site, confirming that $\rho_\Gamma(x)$ does not depend on the time frame. The total chiral density vanishes in both cases, consistent with two interface modes of opposite chirality. These results show that the projector-based local chiral density is a spatial diagnostic of the zero-energy topological states as it is concentrated at the domain walls and negligible across the remaining sites. }

\section{Resource overhead}
The total experimental cost, or resource overhead of the CQW protocol is determined by the number of coin and shift operators required, together with the number of detectors used for readout. In this letter, we quantify this overhead for the SCSS-CQW protocol and compare them with the \textbf{scattering junction-based protocol }of Refs.~\cite{frac-asboth,frac2-asboth,frac3-asboth} with 1D lattices, which can be used to generate fractional topological phases, and edge-states in photon-based experiments~\cite{bian,pxue,barry-pra}. The SCSS-CQW protocol substantially lowers the required experimental resources, as it uses fewer coin-shift operations and only a small number of single-photon detectors, unlike the scattering junction-based protocols of Refs.~\cite{frac-asboth,frac2-asboth,frac3-asboth}. Specifically, the SCSS-CQW protocol reduces the total resource count by approximately a factor of $3.4$, with the operator count reduced by a factor of $2$, while the detector requirement scales far more favorably than the O$(6\tau)$ detector scaling of the scattering junction-based protocol of Refs.~\cite{frac-asboth,frac2-asboth,frac3-asboth} for a walk of $\tau$ steps. This significant reduction improves the practicality of implementing the SCSS-CQW protocol for realizing topological phases and edge states across photonic and other quantum platforms~~\cite{per, peruzzo, schreiber, broome, tamura, schmitz, zaehringer, ryan}. A detailed comparison of resource costs is provided in Table~\ref{c6-tab-reso}.

The time-evolution of the quantum particle (walker) via our SCSS-CQW is governed by the following evolution operator, 
$\hat{U}_{evo}= \hat{S}_{+} \hat{C}_\gamma \hat{S}_{-} \hat{C}_\gamma$,
where the shift operators are: $\hat{S}_{+} = \sum_{x=0}^{N-1} \left( \ket{0}_c \bra{0}_c \otimes \ket{(x+1)\mod N} \bra{x} + \ket{1}_c \bra{1}_c \otimes \ket{x} \bra{x} \right), \hat{S}_{-} = \sum_{x=0}^{N-1} \left( \ket{0}_c \bra{0}_c \otimes  \ket{x} \bra{x} + \ket{1}_c \bra{1}_c \otimes \ket{(x-1)\mod N} \bra{x} \right),$ and the 
coin operator has the form, $
\hat{C}_{\gamma} = e^{-i \frac{D \gamma}{2} \sigma_y}$.

In Refs.~\cite{frac-asboth,frac2-asboth,frac3-asboth}, the authors give a more complex unitary protocol (scattering junction-based) to generate fractional topological invariants via a scattering junction composed of 3 independent regions: two semi-infinite 1D lattices (left and right leads) wherein QW evolution is via $U_{lead}=\hat{S}_{\downarrow} \hat{S}_{\uparrow}$ and a finite 1D lattice (scattering region) wherein evolution is via $U_{SS} = \hat{S}_{\downarrow} \hat{C}_2 \hat{S}_{\uparrow} \hat{C}_1 $ which is a double coin split-step quantum walk (SS-QW) and requires 2 coin operators $C_1, C_2$. The shift operators, $\hat{S}_{\downarrow}=\sum_{x=-\infty}^{\infty} \ket{0}_c \bra{0}_c \otimes \left( \ket{x} \bra{x} + \ket{1} \bra{1} \otimes \ket{x-1} \bra{x} \right)$ and $ \hat{S}_{\uparrow}=\sum_{x=-\infty}^{\infty} \ket{0}_c \bra{0}_c \otimes \left( \ket{x+1} \bra{x} + \ket{1} \bra{1} \otimes \ket{x} \bra{x} \right)$.

When the 3 regions are combined, fractional topological invariants can be derived from the reflection amplitude. Fractional topological invariants in
Refs.~\cite{frac-asboth,frac2-asboth,frac3-asboth} are not intrinsic to the evolution of quantum walker, but are a derivative of the scattering considered.\\
If $\ket{\psi(0)}$ represents walker's initial state, then the quantum mechanical state at the later timestep $t$ becomes, $\ket{\psi(t)} = U^t (\ket{\psi(0)})$.
The evolution operators of our SCSS-CQW protocol, and scattering junction-based protocol (for the scattering region with the SS-QW evolution, $U_{SS}$ and left/right lead with evolution $U_{lead}$), at time $t=\tau$, have the forms,
\begin{equation}
\begin{split}
    & \hat{U}_{evo}^\tau= (\hat{S}_{+} \hat{C}_\gamma \hat{S}_{-} \hat{C}_\gamma)^\tau \;, U_{SS}^\tau = (\hat{S}_{\downarrow} \hat{C}_2 \hat{S}_{\uparrow} \hat{C}_1)^\tau.\\ &
    U_{lead}^\tau=(\hat{S}_{\downarrow} \hat{S}_{\uparrow})^\tau
\end{split}
\label{c6-e-att}
\end{equation}

 One can explicitly determine the resource overhead, namely the scaling of the total number of operators required and the number of detectors involved. The operators here refer to both coin and shift operators. In our SCSS-CQW protocol, as well as in the scattering region (SS-QW), each time step consists of four such operators (two coin and two shift operators), as given in Eq.~(\ref{c6-e-att}). Consequently, after $\tau$ time steps, the operator count scales as, $2\tau+2\tau=4\tau$. In the scattering junction-based protocol (besides the SS-QW), each of the right and left leads requires $2\tau$ number of shift operators. Thus, the total operator count in the scattering junction-based protocol scales as $4\tau+2(2\tau)=8\tau,$ which is 2 times larger than our SCSS-CQW protocol.

 However, our SCSS-CQW protocol offers a dramatic reduction in the number of particle detectors required. In each case, the number of detectors is fixed and equals to the count of lattice sites in the small cyclic (periodic) graph ($N = 7, 8, \dots$), and it does not depend on the number of time steps in the walk. By contrast, the scattering junction-based protocol (combination of left lead, right lead and SS-QW of the scattering region) requires $3*(2\tau+1)=(6\tau + 3)$ detectors for a walk of $\tau$ steps, leading to a detector overhead that grows linearly with time. Consequently, our protocol exhibits an $O(1)$ detector scaling in contrast to the $O(6\tau)$ scaling of the scattering junction-based protocol of Refs.~\cite{frac-asboth,frac2-asboth,frac3-asboth}, to generate edge states. Thus, our protocol not only generates fractional topological phases it also greatly improves the scalability and practical implementation of topological phase detection and edge-state generation on photonic and other quantum platforms.
Furthermore, our SCSS-CQW protocol needs only two distinct coin operators (i.e., two choices of $\hat{C}_\gamma$) to engineer a phase boundary in real position space, an essential component for producing edge states. In contrast, the scattering junction-based protocol (SS-QW) relies on three distinct coins (two versions of $\hat{C}_1$ and a fixed $\hat{C}_2$) to create the same boundary.

\begin{table}[!]
    \centering

    \renewcommand{\arraystretch}{2}
    \begin{tabular}{|c|c|c|c|c|c|}
       \hline        
       \textbf{Protocol} & \textbf{\makecell{Distinct \\coins\\($\#_{\theta}$)} }& \textbf{\makecell{Detectors\\($\#_{d}$)}}& \textbf{\makecell{Shift \\operators\\($\#_{s}$)}} &\textbf{\makecell{Coin \\operators\\($\#_{c}$)}}& \textbf{\makecell{Resource overhead \\($R=\#_{d}+\#_{s}+\#_{c}+\#_{\theta}$)}}\\
        \hline        \rowcolor{gray!15}\multicolumn{6}{|c|}{\textbf{This letter: SCSS-CQW protocol (for $N$, e.g., 7, 8,...)}} \\ \hline
         SCSS-CQW &$2$ & $N$ & $2\tau$& $2\tau$& $R_{SCSS}=4\tau+N+2$ \\ \hline
         \multicolumn{6}{|c|}{\textbf{Example: This letter: SCSS-CQW protocol ($N=7,\; \tau=100$)}} \\ \hline
         SCSS-CQW &$2$ & $7$ & $200$& $200$& $R_{SCSS}=409$ \\ \hline        \rowcolor{gray!15}\multicolumn{6}{|c|}{\textbf{Scattering junction-based protocol (for $N_s, N_l$, e.g., $200, 1000,...$)}} \\ \hline
        SS-QW& $3$ & $2\tau+1$ & $2\tau$& $2\tau$& $R_{SS}=4\tau+(2\tau+1)+3$\\
        Left(Right) Lead & $0$ & $2(2\tau+1)$ & $2(2\tau)$& $0$& \makecell{\underline{$2R_{lead}=2(4\tau+1)\;\;$} \\ $R_{total}=14\tau+6$}\\ \hline
         \multicolumn{6}{|c|}{\textbf{Example: Scattering junction-based protocol (for $N_s=200, N_l=200, \tau=100$)}} \\ \hline
        SS-QW  & $3$ & $201$ & $200$&$200$& $R_{SS}=604$ \\ 
       Left (Right) Lead & $0$ & $201$ & $200$&$0$& \makecell{\underline{$2R_{lead}=802$\;\;}\\  $R_{total}=1406$}   \\ \hline 
    \end{tabular} 
    \caption{Comparison of resource overhead in implementing different unitary quantum walk protocols, i.e., the SCSS-CQW protocol on finite cyclic graphs vs. existing QW on infinite 1D lattice with scattering junction (with SS-QW in the scattering region), to generate fractional topological phases, and protected edge-states. \small{Note: Here $N_s$ ($N_l$) denotes the number of sites on the scattering region (left/right lead) and $N$ denotes the number of position sites in a cyclic graph, while $\tau$ represents the number of time steps in the quantum walk.}}
    \label{c6-tab-reso}
\end{table}
Taking all resources into account, the total overhead for our SCSS-CQW protocol is, $R_{SSCQW}=4\tau+N+2$, as compared to the significantly larger overhead of the scattering junction-based protocol, $R_{total}=R_{SS}+2*R_{lead}=8\tau+(6\tau+3)+3=14\tau+6$, as juxtaposed in the Table~\ref{c6-tab-reso}.
\\
\underline{Example:} If we consider a cyclic quantum walk with $N = 7$ sites and $\tau = 100$ time steps. Our SCSS-CQW protocol requires a total resource overhead $R_{\text{SSCQW}} = 4\tau + N + 2 = 409$.
In comparison, the scattering junction-based protocol requires total resource overhead: $R_{total} = 14\tau + 6 = 1406.$ 
As is evident, our protocol lowers the total resource overhead by more than a factor of \text{3.4}. In absolute terms, this corresponds to a reduction of $997$ components (detectors plus operators), which amounts to a relative saving of approximately \textbf{70.9\%}, without compromising on the ability to generate topological phase boundaries and edge states. Such a substantial decrease in operational and detection requirements highlights the enhanced experimental efficiency and practical implementability of the SCSS-CQW protocol.

\section{Algorithm for edge state generation via SCSS-CQW}

\subsection{Generating QW dynamics via SCSS-CQW on N-cyclic graph}
Below, we outline the algorithm, i.e., steps to generate quantum-walk dynamics and edge-state using a \emph{single-coin split-step cyclic quantum walk} (SCSS-CQW) on a $N$-cycle graph (see, Fig.3 (c) of Main).


\begin{algorithmic}[1]

\Require Total number of time steps $T$

\Require Number of lattice sites $N$ (cyclic graph: $x=0,1,\dots,N-1$)
\Require Single-coin operator $\hat{C}_\gamma$. Choosing coin $\hat{C}_{\gamma_{x=0}}$  at site $x=0$ and $\hat{C}_{\gamma_{x\ne0}}$  at sites $x\in\{1, \dots, N-1\}$ with two different winding numbers
\Require Conditional shift operators $\hat{S}_{+}$ and $\hat{S}_{-}$  
\Require Initial walker state $\ket{\psi(0)}$ 

\Ensure Position probability distribution: $\texttt{prob}[n][x]$ at time-step $\tau=0,1,2,...,T$ and position $x=0, 1,2,...,N-1$
\Ensure Edge-state survival probability at $x=0$

\State Initialize the array storing the position-probability: $\texttt{prob} \leftarrow$ 0s of the shape $T+1 \times N$

\State Prepare initial state:
\[
\ket{\psi(0)} = \ket{0} \otimes \ket{0_c}.
\]

\State Define the SCSS-CQW single-step evolution operator:
\[
\hat{U}_{evo} = \hat{S}_{+}\,\hat{C}_\gamma\,\hat{S}_{-}\,\hat{C}_\gamma
\]

\For{time step $n = 1$ to $n=T$,}
    \State Update the quantum state of the quantum walker
    \[
    \ket{\psi(n)} = \hat{U}_{evo} [\ket{\psi(n - 1)}]
    \]
    
    \For{ the position site $0$ up to site $(N-1)$,}
        \State Evaluate site specific probability value (i.e., $P(x,n)$):
        $$\sum_{g=0}^{1} [(\bra{x,g_c}\psi(n)\rangle)*(\bra{x,g_c}\psi(n)\rangle)]$$
        
        \State Save the probability:
        $\texttt{prob}[n][x]$ from the computed probability, $[P(x,n)]$
    \EndFor.
\EndFor.

\State Compute edge-state survival probability:
\[
P(x=0) = \texttt{prob}[n][0]
\]
A pronounced value of $P(x=0)$ serves as a hallmark of the generated edge state.
\State \Return $\texttt{prob}$: full spatio-temporal probability distribution of the SCSS-CQW.
\end{algorithmic}

The survival probability of the edge state at each time step is obtained as $P(x=0)$ from the stored distribution, also see Python code~\cite{AR-git}. This is used to analyze localization and robustness of the generated edge state via the SCSS-CQW protocol.

\subsection{Quasienergy spectrum}

\textcolor{black}{The quasienergy spectrum of the SCSS-CQW (see, Fig. \ref{ipr})
are obtained by diagonalizing the evolution operator $U_{evo}$.
For an $N$-site cyclic graph, the combined position and coin Hilbert space has dimension $2N$, resulting in $2N$ eigenstates. The quasienergy is calculated for each eigenstate as follows,}

\begin{algorithmic}[1]

\Require Number of lattice sites $N$
\Require Single-coin operator $\hat{C}_\gamma$. Choosing coin $\hat{C}_{\gamma_{x=0}}$  at site $x=0$ and $\hat{C}_{\gamma_{x\ne0}}$  at sites $x\in\{1, 2, \dots, N-1\}$ with two different winding numbers
\Require Conditional shift operators
$\hat{S}_{+}$ and $\hat{S}_{-}$
\Ensure Quasienergy spectrum $\{E_i\}_{i=1}^{2N}$
\Ensure Eigenstates $\{\ket{\psi_i}\}_{i=1}^{2N}$

\State Define the evolution operator:
\[
\hat{U}_{\mathrm{evo}}
=
\hat{S}_{+}\hat{C}_{\gamma}
\hat{S}_{-}\hat{C}_{\gamma}.
\]

\State Diagonalize the evolution operator:
\[
\hat{U}_{\mathrm{evo}}\ket{\psi_i}
=
\lambda_i\ket{\psi_i},
\qquad
i=1,2,\ldots,2N.
\]

\State Extract the quasienergy from each eigenvalue:
\[
E_i=-\arg(\lambda_i),
\qquad
E_i\in[-\pi,\pi].
\]

\State Sort $\{E_i\}$ and the corresponding eigenvectors
$\{\ket{\psi_i}\}$ in ascending order of quasienergy.

\State Plot the quasienergy spectrum:
\[
i \longrightarrow E_i.
\]

\State \Return
$\{E_i,\ket{\psi_i}\}$.

\end{algorithmic}
\textcolor{black}{Ref. \cite{AR-git} provides the Python code for obtaining the quasienergy spectra for the topological edge states.} 
\subsection{Inverse Participation ratio}
\textcolor{black}{The localization properties
are characterized by the inverse participation ratio (IPR) calculated from the corresponding eigenstates (see, Fig.~\ref{ipr}). The procedure is summarized below.}
\begin{algorithmic}[1]
\Require Quasienergy eigenstates $\{\ket{\psi_i}\}_{i=1}^{2N}$
\Require Number of lattice sites $N$
\Ensure IPR $\{\mathrm{IPR}_i\}_{i=1}^{2N}$
\For{each eigenstate $i=1,\ldots,2N$}

    \State Express the eigenstate in the
    position--coin basis:$\ket{\psi_i}
    =
    \sum_{x=0}^{N-1}
    \sum_{g=0}^{1}
    \psi_n(x,g)\ket{x,g}.$

    \State Calculate the position-resolved probability:
    \[
    P(x)
    =
    \sum_{g=0}^{1}
    |\psi_i(x,g)|^2.
    \]

    \State Normalize the position probability:
    \[
    P(x)
    \leftarrow
    \frac{P(x)}
    {\displaystyle\sum_{x=0}^{N-1}P(x)}.
    \]

    \State Calculate the inverse participation ratio:
    $
    \mathrm{IPR}_i
    =
    \sum_{x=0}^{N-1}[P(x)]^2.
    $

    \State Store $\mathrm{IPR}$.

\EndFor

\State Identify eigenstates with enhanced IPR as
localized eigenstates.

\State Encode $\mathrm{IPR}_n$ using the color scale in the
quasienergy spectrum.

\State \Return
$\{\mathrm{IPR}_n\}$.

\end{algorithmic}
\textcolor{black}{Ref. \cite{AR-git} provides the Python code for obtaining the IPR for the corresponding quasienergy eigenstates.}

\clearpage

\end{document}